\documentclass[%
 aip,pof,
 amsmath,amssymb,
preprint,%
]{revtex4-1}
\usepackage{color}
\usepackage{hyperref}
\usepackage{graphicx}
\usepackage{dcolumn}
\usepackage{bm}
\usepackage[mathlines]{lineno}
\usepackage{mathrsfs}
\usepackage[utf8]{inputenc}
\usepackage[T1]{fontenc}
\usepackage{mathptmx}
\usepackage{subfigure}
\usepackage{tabularx}
\usepackage{booktabs}
\usepackage{bm}
\usepackage{amsmath}
\usepackage{mathrsfs}
\usepackage{float}
\usepackage{multirow}
\newcommand{\Ch}{{\it C\hspace{-1.7pt}h} } 
\newcommand{\We}{{\it W\hspace{-2.5pt}e} } 
\newcommand{\CFL}{{\it C\hspace{-1.7pt}F\hspace{-1.7pt}L}} 
\begin{document}

\preprint{AIP/POF}
\title{Study of a droplet breakup process in decaying homogeneous isotropic turbulence based on the phase-field DUGKS approach}

\author{Jun Lai
}
\affiliation{
State Key Laboratory for Turbulence and Complex Systems, College of engineering, Peking University, Beijing 100871, P.R. China
}
\affiliation{
Guangdong Provincial Key Laboratory of Turbulence Research and Applications, Center for Complex Flows and Soft Matter Research and Department of Mechanics and Aerospace Engineering, Southern University of Science and Technology, Shenzhen 518055, P.R. China
}
\author{Tao Chen
}
\affiliation{
	Guangdong Provincial Key Laboratory of Turbulence Research and Applications, Center for Complex Flows and Soft Matter Research and Department of Mechanics and Aerospace Engineering, Southern University of Science and Technology, Shenzhen 518055, P.R. China
}
\author{Shengqi Zhang
}
\affiliation{
	State Key Laboratory for Turbulence and Complex Systems, College of engineering, Peking University, Beijing 100871, P.R. China
}
\author{Zuoli Xiao
}
\affiliation{
	State Key Laboratory for Turbulence and Complex Systems, College of engineering, Peking University, Beijing 100871, P.R. China
}
\author{Shiyi Chen
}
\affiliation{
	State Key Laboratory for Turbulence and Complex Systems, College of engineering, Peking University, Beijing 100871, P.R. China
}
\affiliation{
	Guangdong Provincial Key Laboratory of Turbulence Research and Applications, Center for Complex Flows and Soft Matter Research and Department of Mechanics and Aerospace Engineering, Southern University of Science and Technology, Shenzhen 518055, P.R. China
}
\affiliation{Guangdong-Hong Kong-Macao Joint Laboratory for Data-Driven Fluid Mechanics and
	Engineering Applications, Southern University of Science and Technology, Shenzhen
	518055, China}
\author{Lian-Ping Wang
}
\email{wanglp@sustech.edu.cn}
\affiliation{
	Guangdong Provincial Key Laboratory of Turbulence Research and Applications, Center for Complex Flows and Soft Matter Research and Department of Mechanics and Aerospace Engineering, Southern University of Science and Technology, Shenzhen 518055, P.R. China
}
\affiliation{Guangdong-Hong Kong-Macao Joint Laboratory for Data-Driven Fluid Mechanics and
	Engineering Applications, Southern University of Science and Technology, Shenzhen
	518055, China}


%
%

\date{\today}

\begin{abstract}
\centerline{\bf Abstract}
The breakup of a spherical droplet in a decaying homogeneous isotropic turbulence 
is studied by solving the Cahn-Hilliard-Navier-Stokes equations,
using the discrete unified gas kinetic scheme combined with the free-energy-based phase-field model. 
We focus on the combined effects of turbulence and surface tension on the breakup process 
by assuming that the two fluid phases have the same density and same viscosity. 
The key physical parameters of the system include
the volume fraction ($\varphi=6.54\%$), the initial Weber number ({\it We} $=21.7$), and the initial Taylor microscale Reynolds number ($Re_\lambda=58$).  Due to the turbulence decay, the Weber number decreases monotonically in time to a value
of less than 0.01, providing a great opportunity to study the competing effects of turbulent kinetic energy and interfacial
free energy on the dynamics of the two-phase system.  
Three distinct stages of droplet evolution are identified, namely, 
the deformation stage when the initially spherical droplet evolves into an irregular geometric shape with complex structures, 
the breakup stage when many daughter droplets are formed,
and the restoration stage when the droplets relax towards spherical shape.
These three stages are analyzed systematically from several perspectives:
(1) a geometric perspective concerning the maximum equivalent diameter, the total number of droplets, total interface area, and 
probability distribution of droplet diameters, (2) a dynamic perspective concerning the evolution of local velocity
and vorticity at the fluid-fluid interface, (3) a global perspective concerning the evolution of 
average kinetic energy / dissipation rate and their Fourier spectra, (4) spherical harmonics based energetics concerning
simultaneous  transfer of
kinetic energy  
across different length scales and different radii relative to initial droplet center,
and (5) the time evolution of global kinetic energy and free energy of the system.
It is found that the ending time of the breakup stage can be estimated by the Hinze criterion.
The kinetic energy of the two-phase flow during the breakup stage is found to have a power-law decay with an exponent $-1.76$, 
compared to the exponent ($-1.65$) for the single-phase flow during the same time period, 
mainly due to the enhanced viscous dissipation generated by the daughter droplets. 
Energy spectra of the two-phase flow show power-law decay, with a slope between $-4$ and $-3$,  at high wave numbers, both in the usual Fourier spectral space
and in the spherical harmonics space.
\end{abstract}


\maketitle


\section{Introduction}
Liquid-liquid or gas-liquid two-phase turbulent flows appear in many natural processes and engineering applications, such as wind-cloud interaction, rain formation, air-sea interaction, spray combustion, hydrocarbon separation, emulsion polymerization, fire extinction, irrigation, {\it etc.} Understanding the evolution and properties of two-phase flows is therefore of major technological and scientific interest.~\cite{2016Multiphase} 
These flows could be governed by a number of parameters including density ratio, viscosity ratio, volume fraction, Weber number  $\We$ 
(the ratio of inertial force to surface tension force), and Reynolds number $Re$ (the ratio of inertial force to viscous force). The interaction between the 
evolving interface topology  and the turbulent flow is the most important and difficult issue in these flows.

With the advances in computer science
and numerical methods, direct numerical simulation (DNS) has been developed to address
the dynamics of two-phase turbulent flows at the interface scale.~\cite{scarbolo2015coalescence,albernaz_do-quang_hermanson_amberg_2017,rosti_ge_jain_dodd_brandt_2019,mukherjee_2019,soligo_roccon_soldati_2019,lu2005effect,qian2006simulation,FENG201792} There are different ways of classifying these flows.~\cite{2019DNS,rosti_ge_jain_dodd_brandt_2019} By virtue of the properties of dispersed phase, they can be classified into droplet-~\cite{dodd2016interaction,mukherjee_2019,roccon2017viscosity} and bubble-laden~\cite{1998Direct,qian2006simulation,lu2005effect} flows, mainly depending on the density of dispersed phase compared to the continuous phase. The dispersed phase can be classified into non-deformable droplets/bubbles ({\it e.g.}, the point-particle model)~\cite{2004On,2014Water,2015Effect} or deformable droplets/bubbles~\cite{lu2005effect,Lohse2016Deformation,roccon2017viscosity}, depending on the deformability of the dispersed phase. 
The  two-phase turbulent flows studied in DNS so far include mostly decaying homogeneous isotropic turbulence (DHIT),~\cite{1998Direct,dodd2016interaction} forced homogeneous isotropic turbulence (FHIT),~\cite{qian2006simulation,2007Multi,albernaz_do-quang_hermanson_amberg_2017} and turbulent channel flows.~\cite{lu2005effect,scarbolo2013turbulence,scarbolo2015coalescence,roccon2017viscosity}

Interface-resolved DNS methods are divided into macroscopic methods and mesoscopic methods.~\cite{PhysRevE.85.046309,2015Multiphase} The macroscopic methods solving the continuum mechanics
equations include the front-tracking method,~\cite{1996A} volume-of-fluid (VOF) method,~\cite{Scardovelli1999DIRECT,2021A} level-set method,~\cite{SUSSMAN1998663,2021A} {\it etc.} These are the traditional multiphase computational fluid dynamics (CFD) methods. It remains challenging for the front-tracking method to model interface breakup and coalescence, because the interface needs to be artificially  ruptured.~\cite{1996A,PhysRevE.85.046309} For the VOF and level-set methods, an interface reconstruction step is required, which would be 
a complex task to implement.~\cite{Scardovelli1999DIRECT,PhysRevE.85.046309} 

The mesoscopic approaches are formulated based on properly-designed model Boltzmann equations.
The widely used mesoscopic approaches, developed mainly over the past 30 years, include
the color-gradient model,~\cite{1991Lattice} pseudo-potential model,~\cite{shan1993lattice,1994Simulation} free-energy model,~\cite{swift1996lattice} {\it etc.} Gunstensen~{\it et al.}~(1991)~\cite{1991Lattice} developed the first multicomponent lattice Boltzmann method,~\cite{chen1998lattice} which is usually called color-gradient model or color-fluid model nowadays. The perturbation step in this model would cause an anisotropic surface tension that induces unphysical velocity and vorticity near the fluid-fluid interface.~\cite{chen1998lattice,PhysRevE.85.046309} Shan \& Chen~(1993) \cite{shan1993lattice} presented a pseudo-potential model, known as the Shan-Chen model, which can separate fluid phases or components automatically and improve the isotropy of the surface tension as well,~\cite{chen1998lattice} while the spurious velocity is typically observed near the interface.~\cite{PhysRevE.85.046309} Swift~{\it et al.}~(1996)~\cite{swift1996lattice} proposed the free-energy model. 
Compared to the previous models, the free-energy-based phase-field model is thermodynamically consistent,~\cite{chen1998lattice,PhysRevE.85.046309} which serves the physical basis for the simulations in this paper.
It is noted that the phase field equations can also be solved directly using the traditional CFD methods.~\cite{LIU2021110659}

In recent years, homogeneous isotropic turbulence (HIT) laden with deformable droplets is an active research area.
For droplets evolution in a forced HIT (FHIT), Derksen \& Akker~(2007)~\cite{2007Multi} have reported one of the first DNS results on liquid-liquid dispersions. Using a lattice Boltzmann method (LBM), {\it i.e.}, the He-Chen-Zhang (HCZ) model,~\cite{HE1999642} they found that the droplets 
increased the turbulence energy at small scales because of the generation of small-scale turbulence in the continuous phase. 
Perlekar~{\it et al.}~(2012)~\cite{perlekar2012droplet} studied the droplet size distribution in FHIT based on the multicomponent Shan-Chen LBM. They found that the droplet emulsion can be maintained for a long time, and, for the small volume fraction case, the 
average droplet diameter is in agreement with the Hinze criterion.~\cite{hinze1955fundamentals} Komrakova~{\it et al.}~(2015)~\cite{komrakova2015numerical} simulated liquid-liquid dispersions in FHIT based on the phase-field model combined with LBM. They observed that it is impossible to form a dispersion if the volume fraction of the initial droplet is higher than 0.05, because a large portion of dispersed phase would remain connected. Albernaz~{\it et al.}~(2017)~\cite{albernaz_do-quang_hermanson_amberg_2017} used pseudo-potential LBM with the multi-relaxation time (MRT) collision operator, to study the heat transfer of droplets in FHIT.
They found that temperature fluctuations, surface tension variation and turbulence intensity influence the occurrence of evaporation and condensation.
Shao~{\it et al.}~(2018)~\cite{shao2018direct} used the level-set method to study
the effects of the Weber number on  a droplet breakup in FHIT. They found that the initial spherical droplet tended to break down into small droplets with increasing Weber number.
Furthermore, at the statistically stationary state, the local topology of the bi-axial strain is suppressed inside the droplet region compared to the outside carrier-phase.
Mukherjee~{\it et al.}~(2019)~\cite{mukherjee_2019} studied droplet-turbulence interactions and quasi-equilibrium dynamics in FHIT, based on the pseudo-potential LBM. They observed that droplet breakup extracts the kinetic energy from the large scales and injects into the small scales, and turbulent emulsions would evolve into a quasi-equilibrium cycle of alternating coalescence and breakup dominated processes.

While FHIT produces a statistically stationary flow field which is beneficial for studying droplet size distribution,~\cite{2007Multi,perlekar2012droplet,komrakova2015numerical,mukherjee_2019,rosti_ge_jain_dodd_brandt_2019}
the artificial large-scale forcing term in FHIT could contaminate the two-way coupling interactions,~\cite{2019DNS,rosti_ge_jain_dodd_brandt_2019} especially when the energy transfer across scales is analyzed. Therefore, in this study we choose DHIT as the background flow field.
To our knowledge, only Dodd \& Ferrante (2016)~\cite{dodd2016interaction} and Dodd \& Jofre (2019)~\cite{PhysRevFluids.4.064303} studied droplets evolution in DHIT.
To explain the basic mechanisms of droplet-turbulence interaction, Dodd \& Ferrante (2016)~\cite{dodd2016interaction} simulated motion of droplets of the Taylor length scale size in DHIT based on VOF, and studied the energy transfer and viscous dissipation during the flow evolution. They showed that the coalescence process would increase the turbulence kinetic energy (TKE) and the breakup process would decrease the TKE. Dodd \& Jofre (2019)~\cite{PhysRevFluids.4.064303} further showed that, increasing the droplet Weber number would decrease the interfacial shear stress and viscous length scale at the droplet surface, while increasing the density and viscosity ratio would increase the interfacial shear stress.
The breakup process of a droplet in a DHIT field has not been carefully studied with DNS.

In this paper, we explore the capabilities of a mesoscopic method based on the Boltzmann equation, coupled with the phase-field model, to simulate the breakup of a large droplet in a complex background flow field.
A relatively new gas kinetic scheme known as the discrete unified gas kinetic scheme (DUGKS)~\cite{Guo2013,Guo2015} is adopted here. 
DUGKS combines the advantages of the LBM~\cite{chen1998lattice,aidun2010lattice} and unified gas kinetic scheme (UGKS).~\cite{XU20107747} In DUGKS, a model Boltzmann equation is solved using an accurate finite-volume formulation coupling tightly the kinetic particle transport and particle collisions. 
Compared to LBM, DUGKS can more easily incorporate irregular meshes and different kinetic particle velocity models. The scheme has been applied to simulate single-phase HIT~\cite{WangPeng2016} and wall bounded turbulent flows.~\cite{Bo2017}  
The first objective of this paper is to incorporate the phase-field transport equation with the DUGKS framework, in order to simulate three-dimensional immiscible two-phase turbulent flows. This extends the capabilities of DUGKS. Specifically, as a first step we consider the breakup process of a spherical droplet in a turbulent background flow. 

Another objective of this study is to carefully investigate the evolution of droplet interfaces 
in DHIT. Droplet breakup in a turbulent flow is a complex physical phenomenon mainly controlled by the competition between the kinetic energy and the interface free energy.	
The kinetic energy makes the droplet deform and break up, increasing the overall interface area.
The free energy tends to drive droplet coalescence and make the droplets more spherical. Three distinct stages of evolution will be shown:
(1) the deformation stage, where the initially spherical droplet evolves into an irregular geometric pattern with complex structures, 
(2) the breakup stage, during which many small droplets are generated from the large droplet, 
(3) the restoration stage, during which all the droplets return to spherical and may coalesce with other nearby droplets, 
eventually reaching a quasi-stationary state.

The rest of the paper is organized as follows. In Section \ref{sec: Simu}, the simulation methodology is introduced, 
including both the mesoscopic and macroscopic descriptions of the system, and 
DUGKS.  
In Section \ref{sec: stationary},  
a stationary droplet is simulated first to test our code.
In Section \ref{sec: deform}, a freely-deforming droplet in decaying turbulent flow is simulated, and is compared to a single-phase flow with the same initial velocity field. In Section \ref{sec: breakup}, the breakup process of a large spherical droplet in DHIT is displayed in detail, and the three evolution stages, {\it i.e.}, the deformation stage, breakup stage, restoration stage, are revealed and analyzed. Main conclusions are presented in Section \ref{sec: Con}.   The appendices cover the inverse design of the Boltzmann equations for the Cahn-Hilliard-Navier-Stokes (CHNS) system, an algorithm for isolating and computing the volume of individual droplets, and
theoretical results concerning the kinetic energy of two-phase flow in the spectra space.

\section{Methodology} \label{sec: Simu}


\subsection{Phase-field model and macroscopic equations}

The phase-field model is a well-known diffuse-interface (DI)
method for solving interfacial flow
problems.~\cite{Anderson1998DIFFUSEaa,zhang2018discrete,chen2019simulation,liang2014phase,church2019high,zhang2019interface} In the phase-field model, the free energy 
of two-phase flow 
is~\cite{2002Molecular,swift1996lattice,liu2003phase,jacqmin1996energy,yue2004diffuse}
\begin{equation}
F(\phi)=\int_{V}\left[\psi(\phi)+\frac{\kappa}{2}|\nabla \phi|^{2}\right] d V,\label{Fphi}
\end{equation}
with a double-well form for $\psi(\phi)$,
\begin{equation}
\psi(\phi)=\beta\left(\phi-\phi_{A}\right)^{2}\left(\phi-\phi_{B}\right)^{2}.
\end{equation}
Here $ \phi $ is an order parameter to distinguish different phases, $\phi_{A}$ and $\phi_{B}$ are phase parameters corresponding to the two phases. In this paper, $\phi_{A}=1$ and $\phi_{B}=0$. $V$ is the volume of the system.
$\psi(\phi)$ is the bulk free-energy density acting as a {\it phobic} effect, which represents separation of the two phases into the bulk region.
${\kappa}|\nabla \phi|^{2}/2$ is the interfacial free-energy density acting as a {\it philic} effect, which prefers mixing with each other.
$\nabla$ is the gradient operator. 
The parameters $ \kappa $ and $\beta$ are  
related to the interfacial surface tension $\sigma$ and the interfacial thickness parameter $W$
as
\begin{equation}
\sigma=\frac{\left|\phi_{A}-\phi_{B}\right|^{3}}{6} \sqrt{2 \kappa \beta},\label{sigmaPF}\quad
W=\frac{1}{\phi_{A}-\phi_{B}} \sqrt{\frac{8 \kappa}{\beta}}.
\end{equation}

It is clear that $W\sim \sqrt{\kappa/\beta}$. Therefore, if $\kappa$ becomes larger (interfacial free-energy is larger) or $\beta$ becomes smaller (bulk free-energy is smaller), then $W$ becomes larger, as expected.
The variation of the free energy with respect to $\phi$ yields the chemical potential $\mu_{\phi}$, {\it i.e.},
\begin{equation}
\begin{aligned}
&\mu_{\phi}=\frac{\delta F}{\delta \phi}\\
=& 4 \beta\left(\phi-\phi_{A}\right)\left(\phi-\phi_{B}\right)\left(\phi-\frac{\phi_{A}+\phi_{B}}{2}\right)-\kappa \nabla^{2} \phi.\label{muphi}
\end{aligned}
\end{equation}
For a flat surface at equilibrium, $\mu_{\phi}=0$, leading to~\cite{chen2019simulation,jacqmin1996energy}
\begin{equation}
\phi(x)=\frac{\phi_{A}+\phi_{B}}{2}+\frac{\phi_{A}-\phi_{B}}{2} \tanh \left(\frac{2 x}{W}\right),\label{flatphi}
\end{equation}
where $x$ is the signed distance normal to the interface.

The macroscopic governing equations contain three parts, {\it i.e.,} the continuity equation, the momentum equation, and 
the Cahn-Hilliard (CH) equation,~\cite{cahn1958free,cahn1959free} which can maintain the local mass conservation and energy decay properties in theory.~\cite{2016A,2019Hessian} 
\begin{subequations}\label{macroEq}     	
	\begin{equation}\label{EqMa}
	\frac{1}{\rho RT}\frac{\partial p}{\partial t}+ \nabla \cdot \boldsymbol{u}=- \gamma \nabla \cdot\left(M_{C H} \nabla \mu_{\phi}\right),
	\end{equation}
	\begin{equation}\label{EqMo}
	\frac{\partial(\rho \boldsymbol{u})}{\partial t}+\nabla \cdot(\rho \boldsymbol{u} \boldsymbol{u})=-\nabla p+\nabla \cdot\left[\mu\left(\nabla \boldsymbol{u}+ \boldsymbol{u}\nabla\right)\right]+\boldsymbol{F},
	\end{equation}  
	\begin{equation}\label{EqCH}
	\frac{\partial \phi}{\partial t}+\nabla \cdot(\phi \boldsymbol{u})=\nabla\cdot \left( M_{CH}\nabla \mu_{\phi}\right) ,
	\end{equation}
\end{subequations}  
\begin{subequations} 	
	with fluid density $\rho$ and dynamic viscosity $\mu$ given by linear models~\cite{HE1999642} 
	as
	\begin{equation}\label{rholinear}
	\rho=\frac{\phi-\phi_{B}}{\phi_{A}-\phi_{B}} \rho_{A}+\frac{\phi-\phi_{A}}{\phi_{B}-\phi_{A}} \rho_{B},
	\end{equation}
	\begin{equation}
	{\mu}= \frac{\phi-\phi_{B}}{\phi_{A}-\phi_{B}}{\mu_A}+ \frac{\phi-\phi_{A}}{\phi_{B}-\phi_{A}}{\mu_B},    
	\end{equation}  
\end{subequations}  
where $R$ is the gas constant, $T$ is the reference temperature, $\boldsymbol{u}$ is the fluid velocity, $p$ is the pressure, $t$ is the time.
$\gamma=\left( \rho_{A}-\rho_{B}\right) /\left( \phi_{A} \rho_{B}-\phi_{B}\rho_{A}\right) $ is a constant related to the two-phase order parameters and densities.
The mobility $M_{CH}$ 
is assumed to be a constant.
$\boldsymbol{F}=-\phi\nabla\mu_{\phi}$ is the interfacial force.  
$\rho_{A}$, $\rho_{B}$ and $\mu_{A}$, $\mu_{B}$ are
the densities and
the dynamic viscosities of the two phases. The above macroscopic system is known as the Cahn-Hilliard-Navier-Stokes (CHNS) system.

It is noted that in the incompressible formulation,  
the density field $\rho$ is
solely determined by the phase field $\phi$, which is independent of the pressure $p$.
The usual divergence-free velocity ($\nabla \cdot \boldsymbol{u}=0$) is first modified by adding 
a term $- \gamma \nabla \cdot\left(M_{C H} \nabla \mu_{\phi}\right)$ to the right hand side to ensure that the local mass
conservation is satisfied even in the interfacial regions.~\cite{zhang2018discrete}
Furthermore, the first term on the left hand side of Eq.~(\ref{EqMa}) is added to facilitate the design of the model LB
equation (see Appendix~\ref{sec:inverse_design}), which essentially amounts to the artificial compressibility solver.~\cite{2017General}

\par 

\subsection{The mesoscopic model}
The double-distribution function model is used to recover the hydrodynamic equations and CH equation.  
The following two model Boltzmann equations with 
Bhatnager-Gross-Krook (BGK) collision model~\cite{PhysRev.94.511} are employed~\cite{zhang2018discrete,chen2019simulation}
\begin{subequations}\label{Boltzmann}
	\begin{equation}\label{Boltzmannf}
	\frac{\partial f_{\alpha}}{\partial t}+\boldsymbol{\xi}_{\alpha} \cdot \nabla f_{\alpha}=-\frac{f_{\alpha}-f_{\alpha}^{e q}}{\tau_{f}}+S_{\alpha}^{f},
	\end{equation}
	\begin{equation}\label{Boltzmanng}
	\frac{\partial g_{\alpha}}{\partial t}+\boldsymbol{\xi}_{\alpha} \cdot \nabla g_{\alpha}=-\frac{g_{\alpha}-g_{\alpha}^{e q}}{\tau_{g}}+S_{\alpha}^{g},
	\end{equation}
\end{subequations}
where the pressure/velocity distribution function $f_{\alpha} = f_{\alpha}\left(\boldsymbol{x}, t\right)$ and the order-parameter distribution function $g_{\alpha} = g_{\alpha}\left(\boldsymbol{x}, t\right)$   corresponding to a discrete particle velocity $\xi_\alpha$ 
are functions of position $\boldsymbol{x}$ and time $t$.  
$\tau_{f}$ and $\tau_{g}$ are the relaxation times.  
The key to recover the macroscopic governing equations is to properly design the two equilibrium distribution functions ($f_{\alpha}^{eq}$, $g_{\alpha}^{eq}$) and the two source terms ($S_{\alpha}^{f}$, $S_{\alpha}^{g}$). 
The macroscopic variables are obtained through the moments of the distribution functions,
\begin{subequations}
	\begin{equation}
	\phi\left(\boldsymbol{x}, t\right)=\sum_{\alpha=0}^{Q-1} {g}_{\alpha}\left(\boldsymbol{x}, t\right),
	\end{equation}
	\begin{equation}
	\boldsymbol{u}\left(\boldsymbol{x}, t\right)=\frac{1}{\rho\left(\boldsymbol{x}, t\right)RT}\sum_{\alpha=0}^{Q-1}  {f}_{\alpha}\left(\boldsymbol{x}, t\right) \boldsymbol{\xi}_{\alpha}
	,
	\end{equation}
	\begin{equation}
	\begin{aligned}
	&p\left(\boldsymbol{x}, t\right)=&
	\sum_{\alpha=0}^{Q-1} {f}_{\alpha}\left(\boldsymbol{x}, t\right),
	\end{aligned}
	\end{equation}
\end{subequations}
where $Q$ is the number of the discrete velocities 
in the particle velocity model.
In this study, the D3Q19 discrete-velocity model ($Q=19$) is used, with
\begin{equation}
\frac{\boldsymbol{\xi}_{\alpha}}{c}=\left\{\begin{array}{cc}{(0,0,0),}  & {\alpha=0,} \\ {(\pm 1,0,0),(0,\pm 1,0),(0,0,\pm 1),} & {\alpha=1-6,} \\ {(0,\pm 1,\pm 1),(\pm 1,0,\pm 1),(\pm 1,\pm 1,0),}  & {\alpha=7-18,}\end{array}\right.
\end{equation}
where $c=\sqrt{3RT}=1$ in the lattice units.
The weighting coefficients are $\omega_{0}=1/3$, $\omega_{1-6}=1/18$, $\omega_{7-18}=1/36$. 
Following Zhang {\it et al.}~(2018)~\cite{zhang2018discrete}'s work,
the equilibrium distribution functions are 
\begin{subequations}\label{feqgeq}
	\begin{equation}\label{feq}
	f_{\alpha}^{e q}=\omega_{\alpha} p+s_{\alpha}RT \rho, 	
	\end{equation}
	\begin{equation}\label{geq}
	g_{\alpha}^{e q} =\left\{\begin{array}{ll}{\left(s_{0}+1\right)\phi +\left(\omega_{0}-1\right) \eta \mu_{\phi},} & {\alpha=0,} \\ {s_{\alpha}\phi +\omega_{\alpha} \eta \mu_{\phi}}, & {\alpha \neq 0,}\end{array}\right.		
	\end{equation}
\end{subequations}
where 
\begin{equation}
s_{\alpha}=\omega_{\alpha}\left[\frac{\boldsymbol{\xi}_{\alpha} \cdot \boldsymbol{u}}{R T}+\frac{\left(\boldsymbol{\xi}_{\alpha} \cdot \boldsymbol{u}\right)^{2}}{2 (R T)^{2}}-\frac{u^{2}}{2 R T}\right].
\end{equation}
$\eta$ is an adjustable constant used to improve numerical stability. $\eta=1$ in our simulations.

The source terms are designed to take the following forms~\cite{zhang2018discrete}
\begin{subequations}\label{SfSg}
	\begin{equation}\label{Sf}
	\begin{aligned}
	S_{\alpha}^{f}=&\left(\boldsymbol{\xi}_{\alpha}-\boldsymbol{u}\right) \cdot\left[\left(\omega_{\alpha}+s_{\alpha}\right)\boldsymbol{F}+s_{\alpha} RT \nabla \rho\right]\\& -\omega_{\alpha}\gamma RT \rho  \nabla\cdot \left( M_{CH}\nabla \mu_{\phi}\right), 
	\end{aligned}
	\end{equation}
	\begin{equation}\label{Sg}
	S_{\alpha}^{g}=\left(\omega_{\alpha}+s_{\alpha}\right)\frac{\phi}{RT \rho}(\boldsymbol{\xi}_\alpha-\boldsymbol{u}) \cdot(\boldsymbol{F}-\nabla p) .
	\end{equation}
\end{subequations}
Here the two relaxation times are
\begin{equation}
\tau_{f}=\frac{\mu}{\rho RT},\label{tauf}\quad
\tau_{g}=\frac{M_{CH}}{\eta RT}.
\end{equation}
This can be shown by
applying the Chapman-Enskog analysis.~\cite{Chapman1970,zhang2018discrete}
It is noted that the design of equilibrium distribution functions and source terms is not unique.
They are applied to recover the macroscopic governing equations, 
Eqs.~(\ref{macroEq}a-c). 
Essentially, they are constrained by a few moment-integral conditions, which are derived systematically  
 in Appendix~\ref{sec:inverse_design}.

When the density is constant in the whole flow field ({\it i.e.}, $\rho_A = \rho_B$), the source term $S_{\alpha}^{f}$ is then simplified to
\begin{equation}
S_{\alpha}^{f}=\left(\boldsymbol{\xi}_{\alpha}-\boldsymbol{u}\right) \cdot\left(\omega_{\alpha}+s_{\alpha}\right)\boldsymbol{F}.
\end{equation}

\subsection{The numerical algorithm: discrete unified gas kinetic scheme (DUGKS)}

The DUGKS approach is briefly summarized here. More details can be found in Guo {\it et al.}.~\cite{Guo2013,Guo2015} 
Eqs.~(\ref{Boltzmann}) can be written in an unified form,
\begin{equation}\label{Boltzmannvarphi}
\frac{\partial \varphi_{\alpha}}{\partial t}+\boldsymbol{\xi}_{\alpha} \cdot \nabla \varphi_{\alpha}=\Omega_{\alpha}^{\varphi}+S_{\alpha}^{\varphi},
\end{equation} 
where $\varphi=f$ or $g$ represents the distribution function, and $\Omega_{\alpha}^{\varphi}=-\left(\varphi_{\alpha}-\varphi_{\alpha}^{e q}\right) / \tau_{\varphi}$ is the corresponding collision term.

DUGKS is a finite-volume scheme. The computational domain is divided into a set of control volumes. Integrating~Eq.~(\ref{Boltzmannvarphi}) over a control volume $V_j$ centered at $\boldsymbol{x}_j$ from $t_n$ to $t_{n+1}$, with the midpoint rule for the advection term and trapezoidal rule for $\Omega$ and $S$, yields
\begin{equation}
\begin{aligned}
&\varphi_{\alpha}^{n+1}-\varphi_{\alpha}^{n}+\frac{\delta t}{\left|V_{j}\right|} J_{\alpha}^{n+1 / 2}\\
=& \frac{\delta t}{2}\left(\Omega_{\alpha}^{\varphi, n+1}+\Omega_{\alpha}^{\varphi, n}\right)+\frac{\delta t}{2}\left(S_{\alpha}^{\varphi, n+1}+S_{\alpha}^{\varphi, n}\right),\label{intvarphi}
\end{aligned}
\end{equation}
with
\begin{subequations}
	\begin{equation}
	\varphi_{\alpha}^{n} = \frac{1}{\left|V_{j}\right|} \int_{V_{j}} \varphi_{\alpha}\left(\boldsymbol{x}, t_{n}\right) d V,
	\end{equation}
	\begin{equation}
	J_{\alpha}^{n+1 / 2}=\int_{\partial V_{j}}(\boldsymbol{\xi}_{\alpha} \cdot \boldsymbol{n}) \varphi_{\alpha}\left(\boldsymbol{x}, t_{n+1 / 2}\right) d A,\label{flux}
	\end{equation}
	\begin{equation}
	\Omega_{\alpha}^{\varphi, n} = \frac{1}{|V_{j}|} \int_{V_{j}} \Omega_{\alpha}^{\varphi}\left(\boldsymbol{x}_{j}, t_{n}\right) d V,
	\end{equation}
	\begin{equation}
	S_{\alpha}^{\varphi, n} = \frac{1}{|V_{j}|} \int_{V_{j}} S_{\alpha}^{\varphi}\left(\boldsymbol{x}_{j}, t_{n}\right) d V,
	\end{equation}
\end{subequations}
where $|V_{j}|$ and $\partial V_{j}$ are the volume and surface of the grid cell $V_{j}$.  
Introducing linear transformations
\begin{subequations}
	\begin{equation}
	\begin{aligned}
	\widetilde{\varphi}_{\alpha}=&\varphi_{\alpha}-\frac{\delta t}{2}\left(\Omega_{\alpha}^{\varphi}+S_{\alpha}^{\varphi}\right) \\
	=&\frac{2 \tau_{\varphi}+\delta t}{2 \tau_{\varphi}} \varphi_{\alpha}-\frac{\delta t}{2 \tau_{\varphi}} \varphi_{\alpha}^{e q}-\frac{\delta t}{2} S_{\alpha}^{\varphi} ,
	\end{aligned}
	\end{equation}
	\begin{equation}
	\begin{aligned}
	\widetilde{\varphi}_{\alpha}^{+} =&\varphi_{\alpha}+\frac{\delta t}{2}\left(\Omega_{\alpha}^{\varphi}+S_{\alpha}^{\varphi}\right) \\
	=&\frac{2 \tau_{\varphi}-\delta t}{2 \tau_{\varphi}+\delta t} \widetilde{\varphi}_{\alpha}+\frac{2 \delta t}{2 \tau_{\varphi}+\delta t} \varphi_{\alpha}^{e q}+\frac{2 \tau_{\varphi} \delta t}{2 \tau_{\varphi}+\delta t} S_{\alpha}^{\varphi},
	\end{aligned}
	\end{equation}
\end{subequations}
then
Eq.~(\ref{intvarphi}) can be written as
\begin{equation}
\widetilde{\varphi}_{\alpha}^{n+1}=\widetilde{\varphi}_{\alpha}^{+, n}-\frac{\delta t}{\left|V_{j}\right|} J_{\alpha}^{n+1 / 2}, \label{plusone}
\end{equation}
namely, the implicity in the collision term and source term are removed.

To update $\widetilde{\varphi}_{\alpha}^{n+1}$, the key now is the evaluation of the flux across the cell interface at half time step $t_{n+1/2}$. 
Integrating Eq.~(\ref{Boltzmannvarphi}) for a half time step $h=\delta t / 2$ along the characteristic line, yields
\begin{equation}
\begin{aligned}
&\varphi_{\alpha}\left(\boldsymbol{x}_{b}, t_{n}+h\right)-\varphi_{\alpha}\left(\boldsymbol{x}_{b}-\boldsymbol{\xi}_{\alpha} h, t_{n}\right)
\\=&\frac{h}{2}\left[\Omega_{\alpha}^{\varphi}\left(\boldsymbol{x}_{b}, t_{n}+h\right)+\Omega_{\alpha}^{\varphi}\left(\boldsymbol{x}_{b}-\boldsymbol{\xi}_{\alpha} h, t_{n}\right)\right]
\\&+\frac{h}{2}\left[S_{\alpha}^{\varphi}\left(\boldsymbol{x}_{b}, t_{n}+h\right)+S_{\alpha}^{\varphi}\left(\boldsymbol{x}_{b}-\boldsymbol{\xi}_{\alpha} h, t_{n}\right)\right] 
,
\end{aligned}\label{characteristicline}
\end{equation}
where the interface location $\boldsymbol{x}_{b}=\left(\boldsymbol{x}_{j}+\boldsymbol{x}_{j+1}\right) / 2$ for the uniform grid.
Similarly, by introducing the linear transformations
\begin{subequations}
	\begin{equation}
	\overline{\varphi}_{\alpha}=\frac{2 \tau_{\varphi}+h}{2 \tau_{\varphi}} \varphi_{\alpha}-\frac{h}{2 \tau_{\varphi}} \varphi_{\alpha}^{e q}-\frac{h}{2} S_{\alpha}^{\varphi}, \label{transform3}
	\end{equation}
	\begin{equation}
	\overline{\varphi}_{\alpha}^{+}=\frac{2 \tau_{\varphi}-h}{2 \tau_{\varphi}+h} \overline{\varphi}_{\alpha}+\frac{2 h}{2 \tau_{\varphi}+h} \varphi_{\alpha}^{e q}+\frac{2 \tau_{\varphi} h}{2 \tau_{\varphi}+h} S_{\alpha}^{\varphi},
	\end{equation}
\end{subequations}
Eq.~(\ref{characteristicline}) can be reduced in an explicit form as
\begin{equation}
\overline{\varphi}_{\alpha}\left(\boldsymbol{x}_{b}, t_{n}+h\right)=\overline{\varphi}_{\alpha}^{+}\left(\boldsymbol{x}_{b}-\boldsymbol{\xi}_{\alpha} h, t_{n}\right).
\label{varphibar}
\end{equation}
The right hand side of Eq.~(\ref{varphibar}) can be approximated using the first-order Taylor expansion,
\begin{equation}
\overline{\varphi}_{\alpha}^{+}\left(\boldsymbol{x}_{b}-\boldsymbol{\xi} h, t_{n}\right) \approx \overline{\varphi}_{\alpha}^{+}\left(\boldsymbol{x}_{b}, t_{n}\right)-\boldsymbol{\xi}_{\alpha} h \cdot \boldsymbol{\sigma}_{b},
\end{equation}                                                             
where $\boldsymbol{\sigma}_{b} = \nabla \overline{\varphi}_{\alpha}^{+}\left(\boldsymbol{x}_{b}, t_{n}\right)$.

Inverting the transformation, Eq.~(\ref{transform3}), the original distribution $\varphi_{\alpha}\left(\boldsymbol{x}_{b}, t_{n}+h\right)$ is
\begin{equation}
\varphi_{\alpha}=\frac{2 \tau_{\varphi}}{2 \tau_{\varphi}+h} \overline{\varphi}_{\alpha}+\frac{h}{2 \tau_{\varphi}+h} \varphi_{\alpha}^{e q}+\frac{\tau_{\varphi} h}{2 \tau_{\varphi}+h}S_{\alpha}^{\varphi},
\end{equation}
which can now be used to evaluate $\boldsymbol{J}_{\alpha}^{n+1 / 2}$ in Eq.~(\ref{flux}).

All the transformations are linear and only two of the distributions are independent. It is straightforward to show that
\begin{equation}
\widetilde{\varphi}_{\alpha}^{+}=\frac{4}{3} \overline{\varphi}_{\alpha}^{+}-\frac{1}{3} \widetilde{\varphi}_{\alpha},
\end{equation}
\begin{equation}
\overline{\varphi}_{\alpha}^{+}=\frac{2 \tau_{\varphi}-h}{2 \tau_{\varphi}+\delta t} \widetilde{\varphi}_{\alpha}+\frac{3 h}{2 \tau_{\varphi}+\delta t} \varphi_{\alpha}^{e q}+\frac{3 \tau_{\varphi} h}{2 \tau_{\varphi}+\delta t} S_{\alpha}^{\varphi}.
\end{equation}

Finally, $\widetilde{\varphi_{\alpha}}^{n+1}$ can be obtained from Eq.~(\ref{plusone}). 

In the code, we track $\widetilde{g}_{\alpha}$ and $\widetilde{f}_{\alpha}$, then the macroscopic variables are evaluated as~\cite{zhang2018discrete,chen2019simulation}
\begin{subequations}
	\begin{equation}
	\phi\left(\boldsymbol{x}_{j}, t_{n}+\delta t\right)=\sum_{\alpha=0}^{Q-1} \widetilde{g}_{\alpha},
	\end{equation}
	\begin{equation}
	\boldsymbol{u}\left(\boldsymbol{x}_{j}, t_{n}+\delta t\right)=\frac{1}{RT\rho}\left(\sum_{\alpha=0}^{Q-1} \boldsymbol{\xi}_{\alpha} \widetilde{f}_{\alpha} +\frac{\delta t}{2} \boldsymbol{F}\right)
	,
	\end{equation}
	\begin{equation}
	\begin{aligned}
	&p\left(\boldsymbol{x}_{j}, t_{n}+\delta t\right)=\\&
	\sum_{\alpha=0}^{Q-1} \widetilde{f}_{\alpha}+\frac{\delta t}{2}RT\left[  \boldsymbol{u} \cdot \nabla \rho-\gamma  \rho  \nabla \cdot\left(M_{C H} \nabla \mu_{\phi}\right)\right]   . \label{pressuremoment}
	\end{aligned}
	\end{equation}
\end{subequations}
For the special case of $\rho_A =\rho_B$, Eq.~(\ref{pressuremoment}) can be simplified to 
\begin{equation}
\begin{aligned}
&p\left(\boldsymbol{x}_{j}, t_{n}+\delta t\right)=&
\sum_{\alpha=0}^{Q-1} \widetilde{f}_{\alpha}.
\end{aligned}
\end{equation}


The time step size is determined 
by the Courant-Friedrichs-Lewy (CFL) condition~\cite{XU20107747,Guo2013,zhang2018discrete,chen2019simulation,2020Simulation}
\begin{equation}
\delta t= \CFL  \frac{\delta x_{\min }} { {c + \left| \boldsymbol{u}\right| _{\max} } },
\end{equation}
where $\delta x_{\min}$ and $\left| \boldsymbol{u}\right| _{\max}$ represent the minimal grid spacing and maximum hydrodynamic velocity magnitude, respectively, $\CFL$ denotes the CFL number.

Another detail is that there are first-order and second-order spatial derivatives involved in the phase-field DUGKS (PF-DUGKS) approach.
These are evaluated by second-order central finite difference schemes.

\section{A stationary droplet}\label{sec: stationary}

We first simulate a stationary droplet to validate our code based on the PF-DUGKS approach.
All the parameters are in lattice units in this paper unless otherwise stated. Initially, a spherical droplet with a radius $R_0=32.0$ is placed at the center of the computational domain of size $128^3$. 
The periodic boundary conditions are applied to
all spatial directions. 
The density ratio and viscosity ratio are both one. 
As reported by Zu \& He~(2013)~\cite{zu2013phase} and Chen {\it et al.}~(2019),~\cite{chen2019simulation} a small mobility may cause numerical instability, 
while a larger mobility would increase the parasitic currents.
As a compromise,
$M_{CH}=0.01$ is chosen. $W=3.0$, implies the Cahn number {\it Ch}$\hspace{4pt}=W/R_0=0.0938$.
The other parameters are $\tau_{f}=\tau_{g}=0.5$, $\sigma=8.0\times 10^{-4}$,  {\it CFL}$\hspace{4pt}=0.25$. 
The order parameter is initialized as~\cite{zhang2018discrete,zu2013phase,chen2019simulation,ZHANG20191128} 
\begin{equation}
\phi\left(r_D\right)=\frac{\phi_{A}+\phi_{B}}{2}+\frac{\phi_{A}-\phi_{B}}{2} \tanh \left(2 \frac{R_0-r_D}{W}\right),\label{phini}
\end{equation}
where $D=2,3$ for 2D and 3D cases, respectively.
The distance between any point ($(x,y)$ or $(x,y,z)$) and the droplet center ($(x_{c},y_{c})$ or $(x_{c},y_{c},z_{c})$) are $r_2=\sqrt{\left(x-x_{c}\right)^{2}+\left(y-y_{c}\right)^{2}}$, $r_3=\sqrt{\left(x-x_{c}\right)^{2}+\left(y-y_{c}\right)^{2}+\left(z-z_{c}\right)^{2}}$
for 2D and 3D.
The initial velocity and hydrodynamic pressure are both set to zero in the whole domain. In this case, the dimensionless time is defined as $t^*=\frac{t}{R_0}\sqrt{\frac{\sigma}{\rho R_0}}$.

\begin{figure}[t!]
	\centering    
	\begin{minipage}[t]{0.125\linewidth}
		\centering
		\includegraphics[width=0.8\columnwidth,trim={2cm 4cm 2cm 2cm},clip]{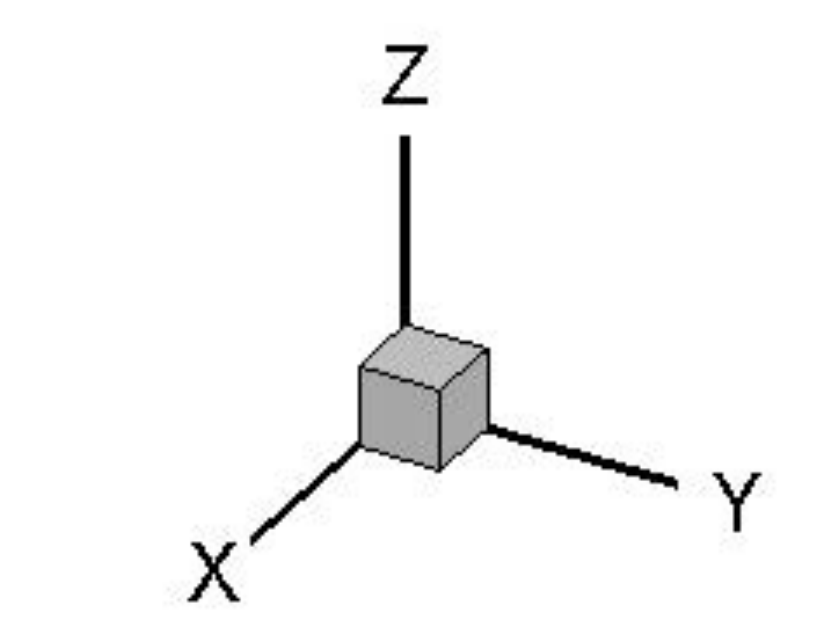}
	\end{minipage}
	\\  	
	\subfigure[$t^*=0$ (LHS) and $t^*=39.1$ (RHS)]{
		\begin{minipage}[t]{0.25\linewidth}
			\centering
			\includegraphics[width=1.\columnwidth,trim={3cm 0cm 3cm 0cm},clip]{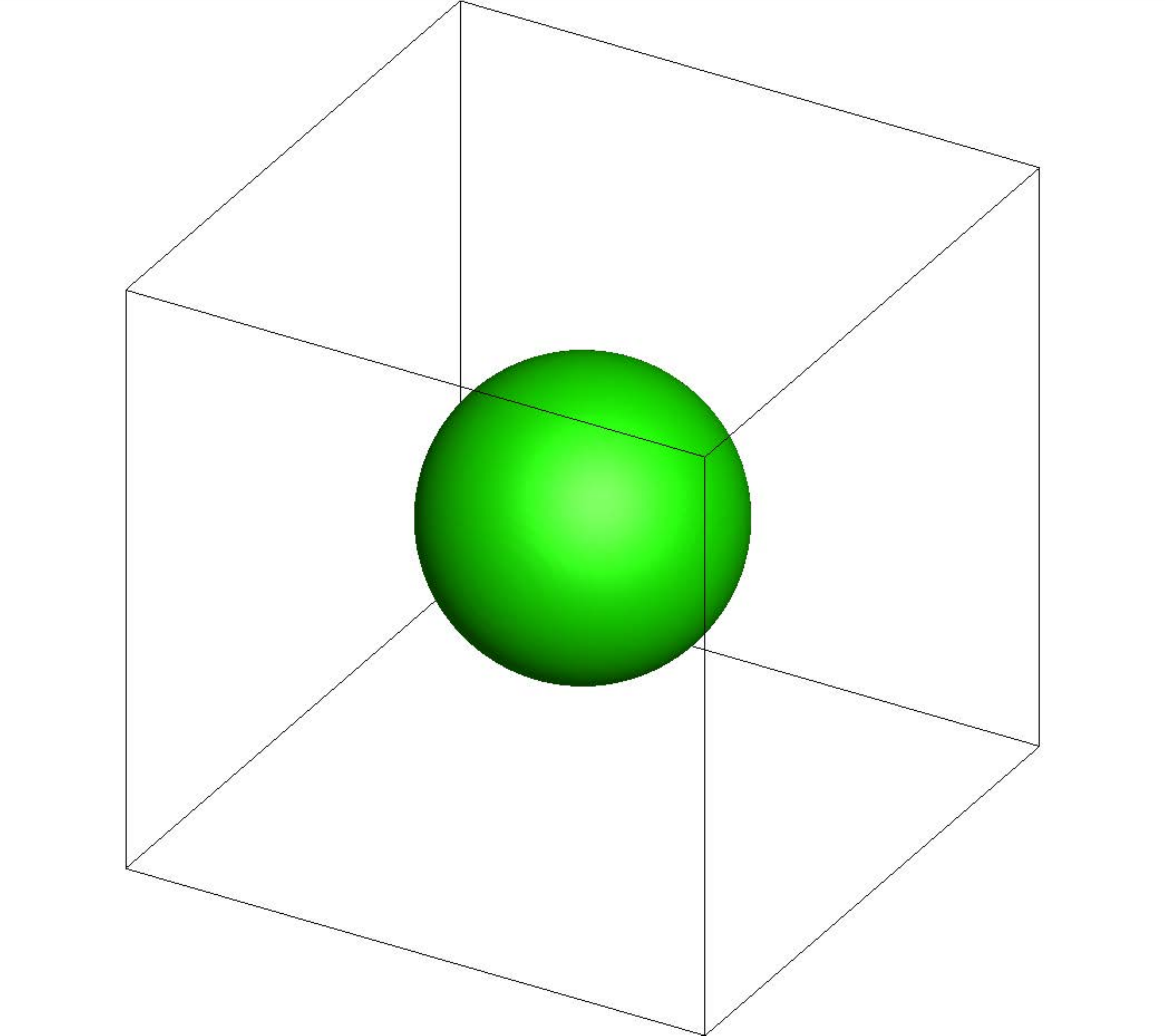}
			\label{phi1}
		\end{minipage}	
		\begin{minipage}[t]{0.25\linewidth}
			\centering
			\includegraphics[width=1.\columnwidth,trim={3cm 0cm 3cm 0cm},clip]{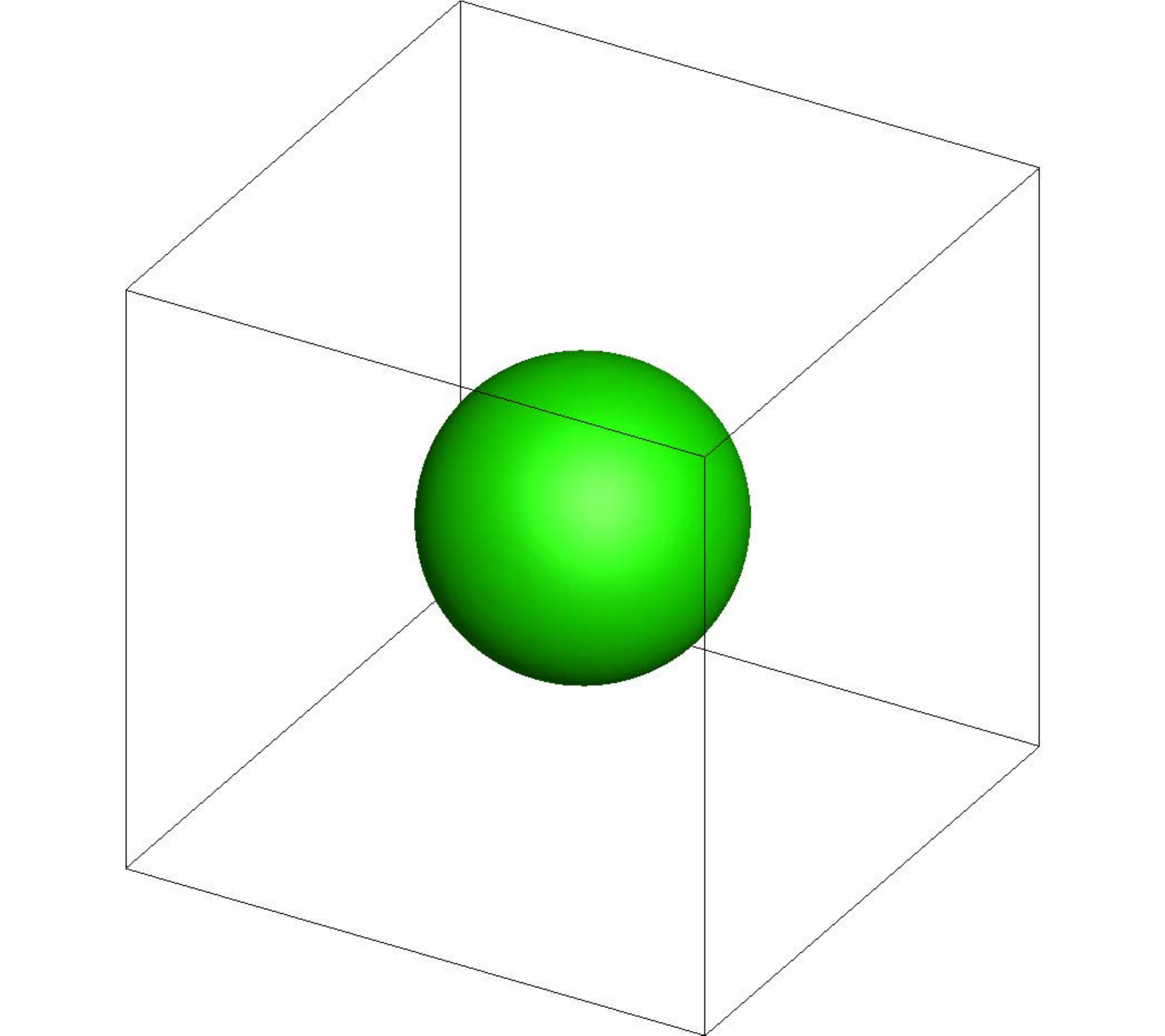}
			\label{phi2}
		\end{minipage}
	}\\
	\subfigure[$\phi$ in the center line of $x$ direction]{
		\begin{minipage}[t]{0.45\linewidth}
			\centering
			\includegraphics[width=1.\columnwidth,trim={0cm 0cm 0cm 0cm},clip]{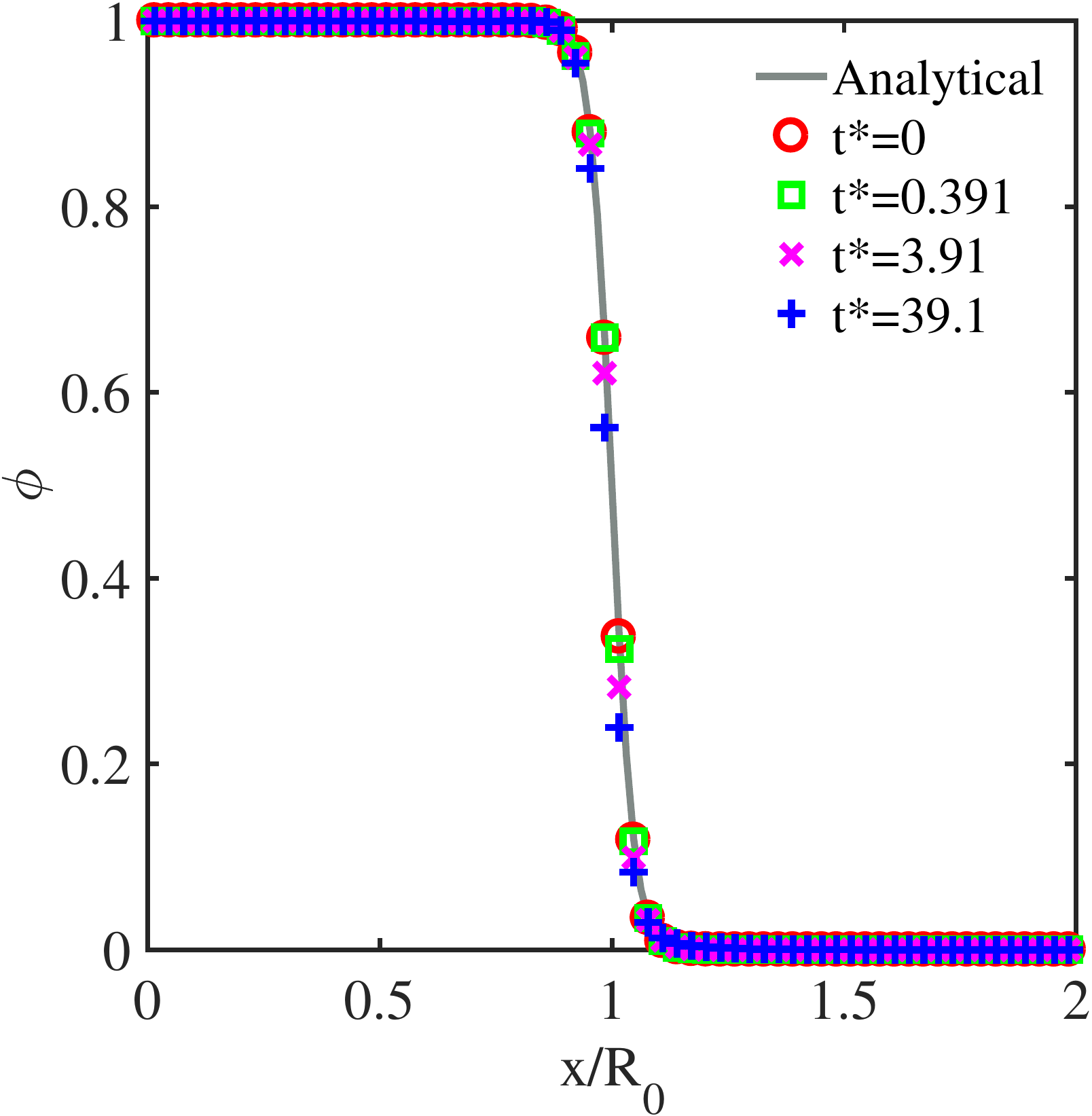}
			\label{philine}
		\end{minipage}
	}	
	\centering
	\caption{The profiles of the order parameter at different times. The  curve marked ``Analytical" represents Eq.~(\ref{phini}).
	}
	\label{phist}
\end{figure}

The order parameter  as a function of the distance from the droplet center 
at different times are showed in Fig.~\ref{phist}. We observe that the distribution of the order parameter $\phi$ is nearly
stable for a long time and the interface between the two phases remains unaltered. For a fixed position near the interface, $\phi$ decreases slightly, which indicates that the droplet becomes slightly smaller over time because of the minor mass loss.~\cite{yue2007spontaneous,PhysRevE.89.033302,2016A,PhysRevE.100.061302}
At $t^*=39.1$, {\it i.e.}, after $1\times 10^6$ time steps, the volume $V(t^*=39.1)$ is found to be $0.990V(t^*=0)$, while it is $0.995V(t^*=0)$ for the 2D case with the same simulation parameters. In terms of the radius ratio, we have $R(t^*=39.1)/R(t^*=0)=0.997$ and $0.998$ for the 3D and 2D cases, respectively, which means that the droplet size in 3D case decreases slightly faster than that in the 2D case. The observed values of $R(t^*)/R(t^*=0)$ are similar as that reported by Lee \& Fischer~(2006)~\cite{PhysRevE.74.046709} and Zu \& He~(2013).~\cite{zu2013phase}

\begin{figure}[htbp]
	\centering    
	\includegraphics[width=0.45\columnwidth,trim={0cm 0cm 0cm 0cm},clip]{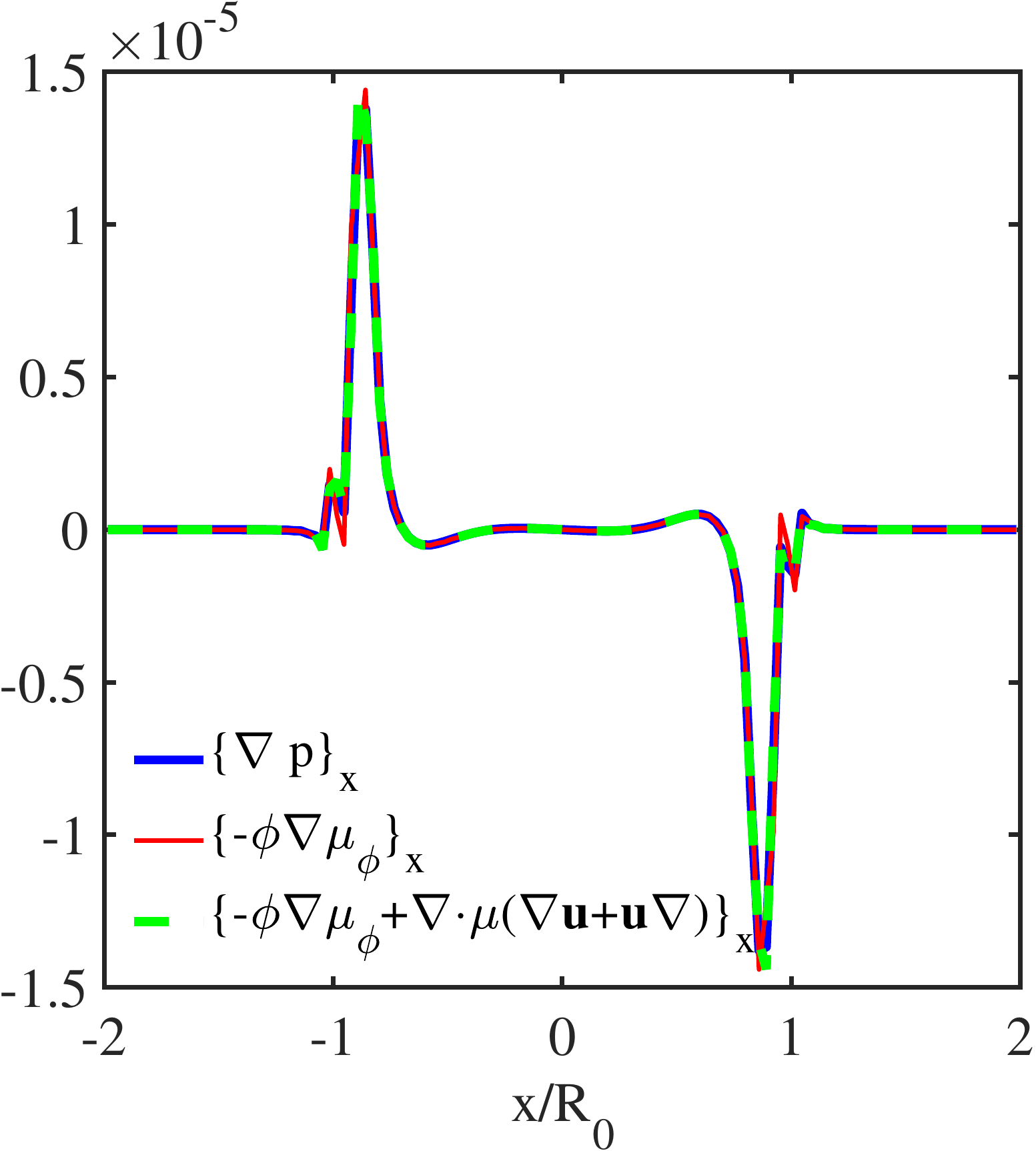}
	\centering
	\caption{Momentum balance along the center line in $x$ direction at $t^*=39.1$. }
	\label{Momentum}
\end{figure}

Then we examine the momentum equation, Eq.~(\ref{EqMo}), along the same centerline in the $x$ direction at $t^*=39.1$. 
We calculate all the terms in the $x$ direction of Eq.~(\ref{EqMo}), the maximum and minimum values 
are presented in Table \ref{TabMomen}. 
From this table, we observe that the sum of maximum and minimum is always zero for each of these six terms, because of the symmetry of the problem with respect to the center of the droplet, or equivalently asymmetry in Cartesian coordinates. Furthermore, the time derivative term and the advection term can be neglected. The viscous term, which is separated into two parts in the table, is also small compared to the pressure gradient term and the interfacial force term. In Fig.~\ref{Momentum}, we can observe that the pressure gradient term is well balanced by the interfacial force term in the bulk region. In the interface region, they also balance each other but not precisely. If we add the viscous term, the balance in the interface region is closely met.  
In summary, the pressure gradient term and the interfacial force term roughly balance each other, and the viscous term makes a noticeable contribution near the interface region only. Very similar results were shown in Chen {\it et al.}~(2019).~\cite{chen2019simulation}

\begin{table}[]
	\centering
	\caption{Maximum and minimum of each term in $x$ direction of momentum equation}
	\label{TabMomen}
	\begin{tabular}{ccccccc}
		\toprule
		\multirow{2}{*}{} & \multicolumn{1}{c}{$\left\lbrace
			\frac{\partial(\rho \boldsymbol{u})}{\partial t}
			\right\rbrace _x$} & \multicolumn{1}{c}{$\left\lbrace
			\nabla \cdot(\rho \boldsymbol{u} \boldsymbol{u})
			\right\rbrace _x$} & \multicolumn{1}{c}{$\left\lbrace
			-\nabla p
			\right\rbrace _x$} & \multicolumn{1}{c}{$\left\lbrace
			\nabla \cdot\left(\mu\nabla \boldsymbol{u}\right)
			\right\rbrace _x$}& \multicolumn{1}{c}{$\left\lbrace
			\nabla \cdot\left(\mu \boldsymbol{u}\nabla\right)
			\right\rbrace _x$} & \multicolumn{1}{c}{$\left\lbrace
			-\phi\nabla\mu_{\phi}
			\right\rbrace _x$}\\
		\midrule
		max             &4.5727e-12                          & 4.5970e-11                    & 1.3882e-05                   & 3.1267e-06           & 8.2250e-07           & 1.4410e-05                               \\
		min             &-4.5727e-12                          & -4.5970e-11                    & -1.3882e-05                   & -3.1267e-06           & -8.2250e-07           & -1.4410e-05                   \\
		\bottomrule
	\end{tabular}\\
\end{table}

\begin{figure}[htbp]
	\centering    
	\subfigure[Kinetic energy]{
		\begin{minipage}[t]{0.47\linewidth}
			\centering
			\includegraphics[width=1.\columnwidth,trim={0cm 0cm 0cm 0cm},clip]{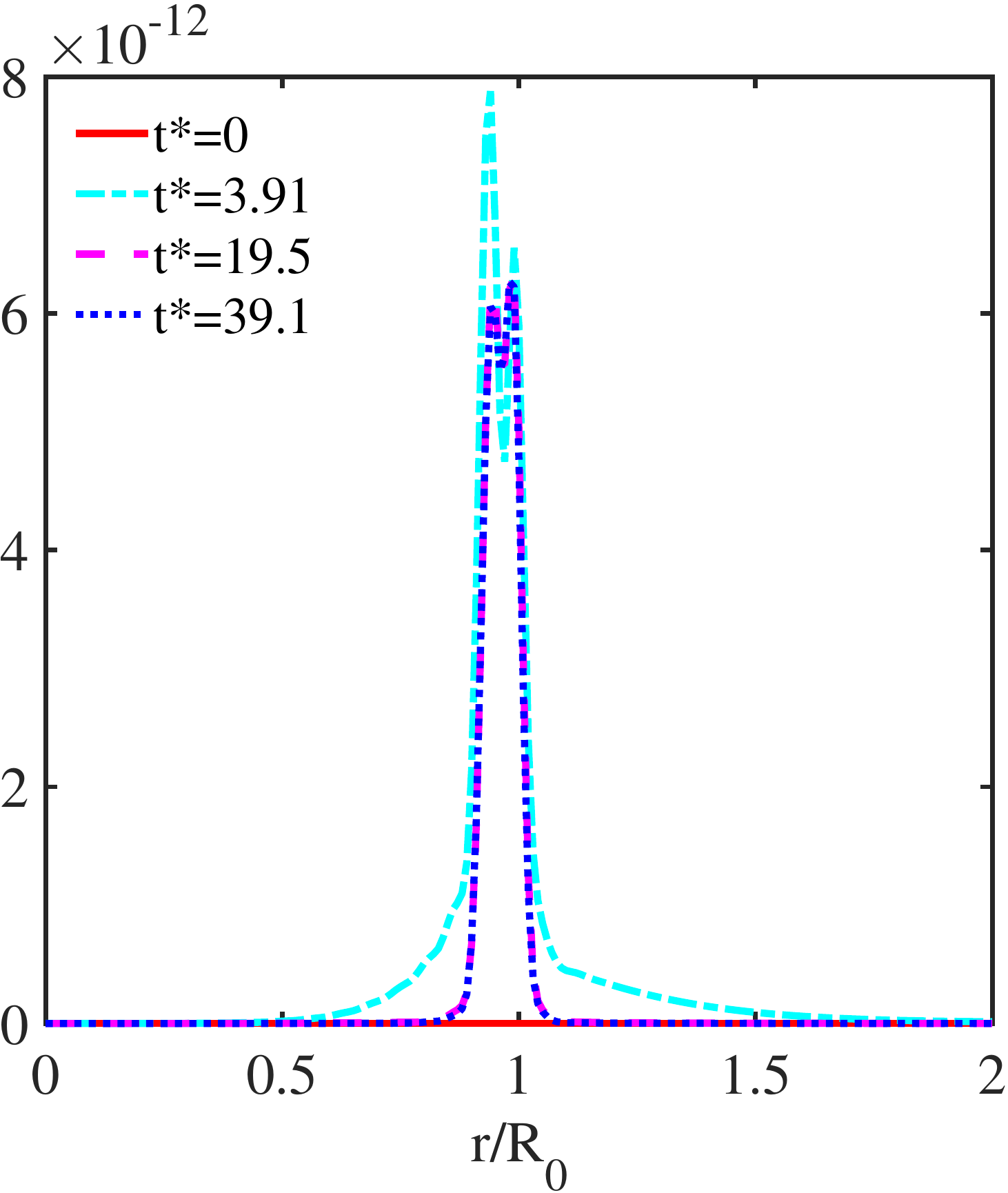}
			\label{StationaryKE201less}
		\end{minipage}
	}
	\subfigure[Free energy]{
		\begin{minipage}[t]{0.47\linewidth}
			\centering
			\includegraphics[width=1.\columnwidth,trim={0cm 0cm 0cm 0cm},clip]{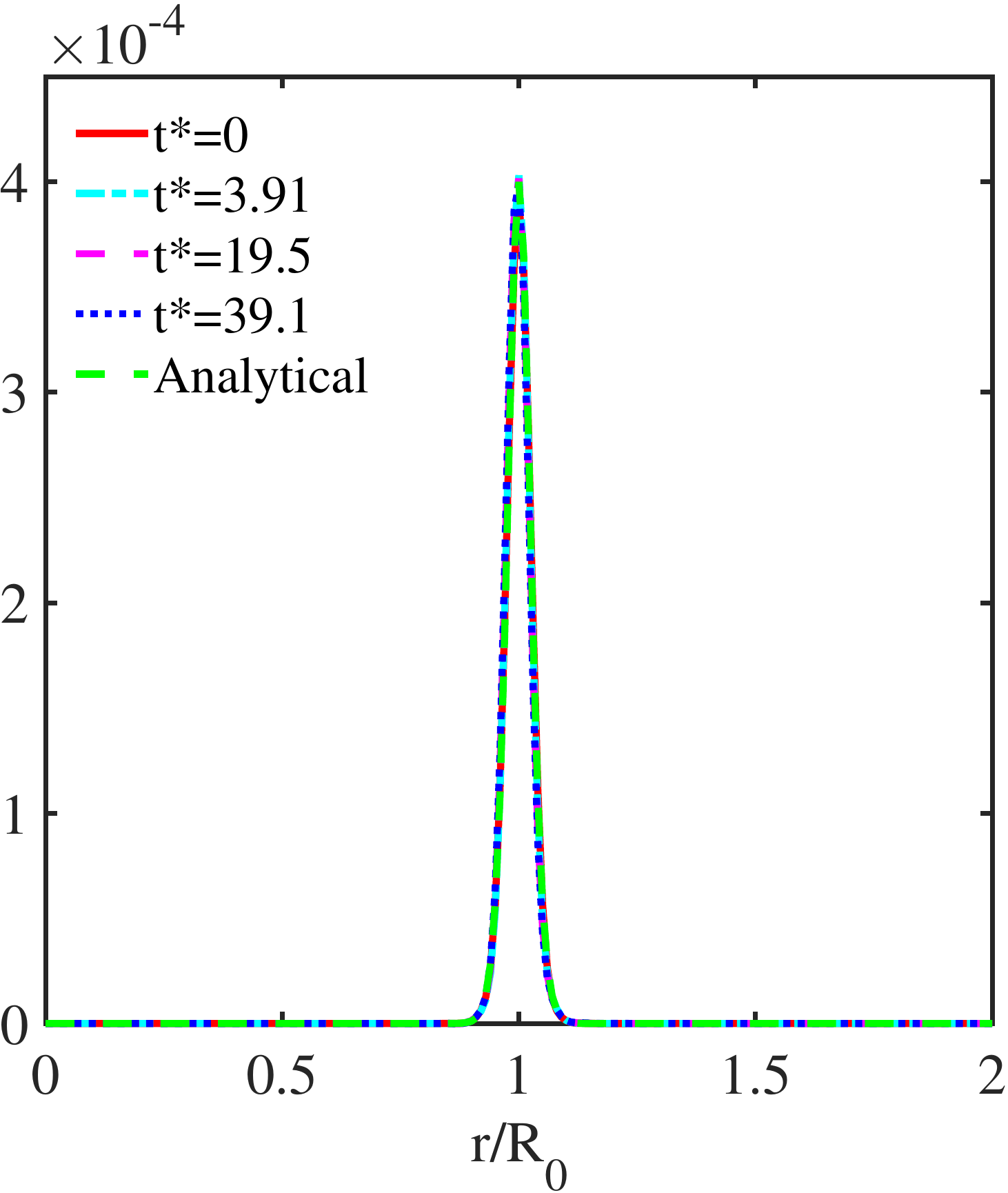}
			\label{StationaryFE201spline}
		\end{minipage}
	}
	\centering
	\caption{Spherically averaged energies as a function of radial distance from  the center of the initial droplet, at different times.   The "Analytical" result in (b) is based on Eq.~(\ref{phini}).}
	\label{Es}
\end{figure}

Finally, we compare the values of  kinetic energy and free energy in Fig.~\ref{Es}, with each averaged over
spherical surfaces, as a function of distance from the droplet center. For the stationary droplet,  physically the kinetic energy should be zero. However, a low level of spurious currents usually exists near the fluid-fluid interface in simulations.~\cite{PhysRevE.74.046709,ZHENG2006353,pooley2008eliminating,van2008emulsion,PhysRevE.83.036707,zu2013phase,dodd2016interaction,PENG2019104257,ZHANG20191128,mukherjee_2019,HU2020103432} 
Indeed, a small non-zero kinetic energy is found to exist in the stationary droplet simulation near the interface. Fig.~\ref{StationaryKE201less} shows that the kinetic energy remains
stable after about $t^*=19.5$. Compared to Fig.~\ref{StationaryFE201spline}, we conclude that the energy from 
spurious currents is much smaller than the free energy, by a factor of about $2\times 10^{-8}$ in this case, and
thus the former can be safely neglected. Fig.~\ref{StationaryFE201spline} also shows that the free energy distribution is almost
independent of time and close to the approximate analytical result during the evolution process of the stationary droplet.

In summary, the simulation results for a stationary droplet in 2D and 3D indicate that the PF-DUGKS approach yields physically accurate results, with a stable spherical drop being maintained over time, a good mass conservation and momentum balance, 
and negligible spurious currents. This provides a basis for us to apply the approach to turbulent immiscible two-phase flows, to
be discussed next.

\section{A freely-deforming droplet in decaying turbulence} \label{sec: deform}

In this section, we apply the PF-DUGKS code to perform DNS of a freely-deforming droplet in decaying turbulence to (1) test our code in a turbulent background flow, (2) demonstrate that the PF-DUGKS model can also reproduce a single-phase flow in the limit of negligible surface tension, (3) demonstrate the application of a spherical harmonic spectral analysis, 
and (4) show the energy evolution in such a decaying background flow field. 

A few main statistics of the flow are defined here, which are also used in the next section. The kinetic energy and dissipation rate per unit mass are computed from the energy spectrum $E({k},t)$, namely, 
$
K\left( t\right) =  (\boldsymbol{u}' )^2 /2 =\int E\left( {k},t\right)dk$, $\epsilon\left( t\right)=\int 2\nu k^2E\left( {k},t\right)dk
$,
where $\nu=\mu/\rho$ is the kinematic viscosity. The Taylor microscale Reynolds number is defined as $Re_\lambda = u'\lambda/\nu$, where the root-mean-square (r.m.s.) turbulent fluctuating velocity $u^{\prime}$ and the Taylor miscroscale $\lambda$ are computed from
$u^{\prime}=\sqrt{\langle\boldsymbol{u}^{\prime} \cdot \boldsymbol{u}^{\prime}\rangle_V/3}$, $\lambda=\sqrt{15 \nu/\epsilon} u^{\prime}$,
here $\left\langle \cdot\right\rangle_V $ represents the volume average over the whole computational domain.   

This droplet-deformation simulation case follows the work of Chen {\it et al.}~(2019),~\cite{chen2019simulation} where
they showed that the PF-DUGKS results are in good agreement with those based on solving the macroscopic
governing equations such as the ARCHER code,~\cite{chen2019simulation} implying that the PF-DUGKS approach can generate reliable results for a two-phase decaying turbulence.  
The spatial grid resolution is $128^3$  with periodic boundary conditions in all three directions. Wang~{\it et al.}~(2016)~\cite{WangPeng2016} showed that DUGKS has a superior numerical stability particularly for high Reynolds number flows and can adequately resolve the turbulent flow when $k_{\max}\eta > 3$, where $k_{\max}$ is the maximum resolved wave number and $\eta=\left(\nu^{3}/\epsilon\right)^{1 / 4}$ is the Kolmogorov length scale. In our simulation, the initial $k_{max}\eta = 7.0$. The density ratio and the viscosity ratio of two-phase flow are both set to 1. The CFL number is 0.25. 
The initial maximum velocity magnitude is 0.034, satisfying the model assumption of an incompressible flow.

We consider two different settings. 
The first setting is referred to as the two-phase setting.
Under  the two-phase setting, we expect that the two-phase flow resembles the single-phase flow when
the  surface tension  is set to a very small value, say  $\sigma = 1.0\times 10^{-20}$, yielding very large
Weber number and Capillary number (see below). 
Alternatively, in a second setting, referred to as the single-phase setting, $\phi$ is simply 
set  to $\phi = \phi_B = 0$ in the whole computational domain, 
thus the chemical potential $\mu_\phi$ and the interfacial force ${\bf F}$ are both zero everywhere at all times.
Then the CH equation (Eq.~(\ref{EqCH})) and ${\bf F}$ in the momentum equation (Eq.(\ref{EqMo})) can be neglected in theory. The governing equations then become identical to those in a  single-phase flow. 
It follows that in the single-phase setting, the value of $\sigma$ plays no dynamic role and we can set
this to any value; we set $\sigma = 1.0\times 10^{-3}$ in the single-phase setting.

\begin{figure}[t!]
	\centering    
	\begin{minipage}[t]{0.6\linewidth}
		\centering
		\includegraphics[width=0.6\columnwidth,trim={0cm 0cm 0cm 0cm},clip]{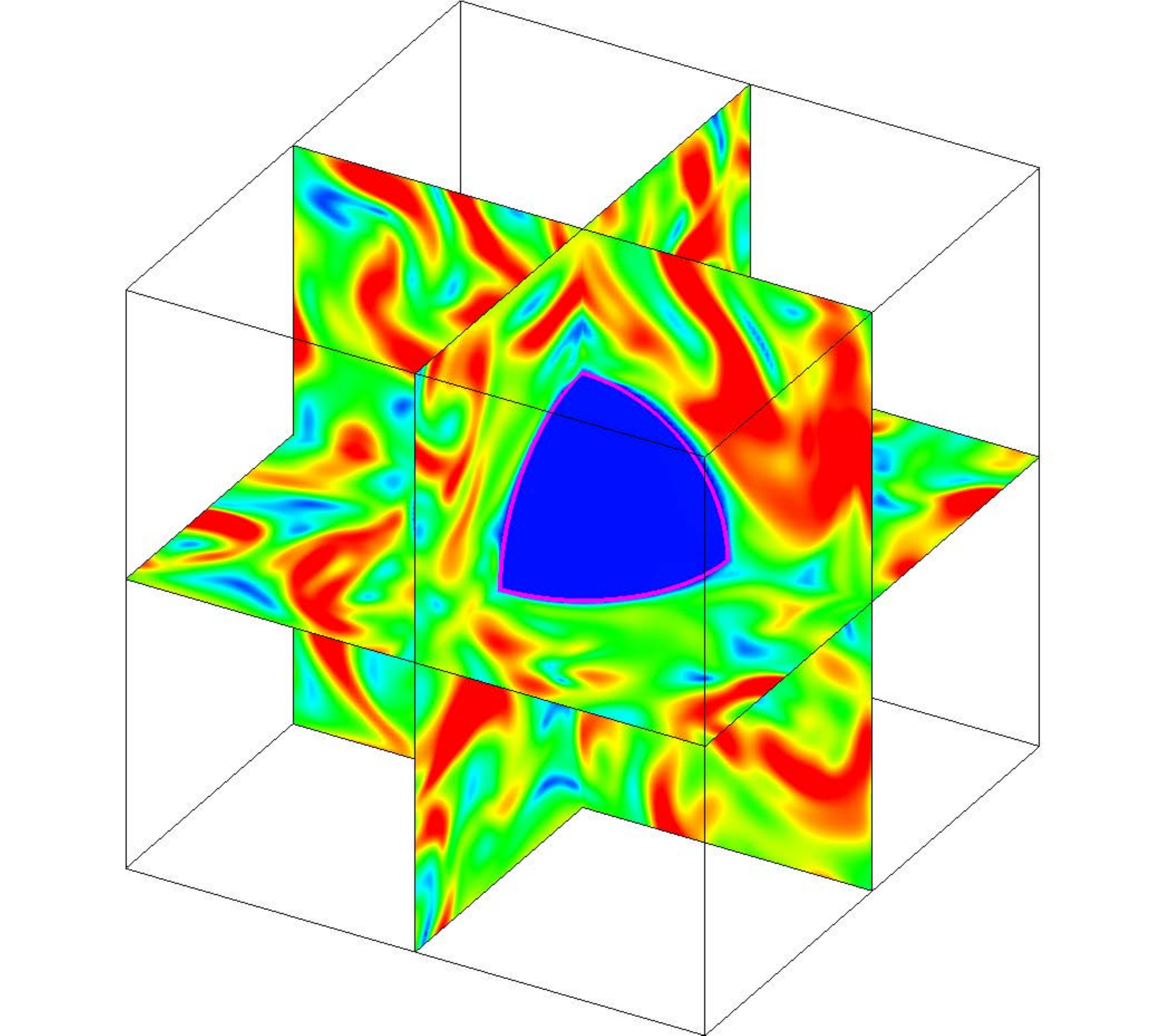}
	\end{minipage}
	\\
	\begin{minipage}[t]{0.12\linewidth}
		\centering
		\includegraphics[width=1\columnwidth,trim={2cm 4cm 2cm 2cm},clip]{pngpdf/xyz1.pdf}
	\end{minipage}\\
	\begin{minipage}[t]{0.48\linewidth}
		\centering
		\includegraphics[width=1.0\columnwidth,trim={0cm 1.5cm 0cm 1.5cm},clip]{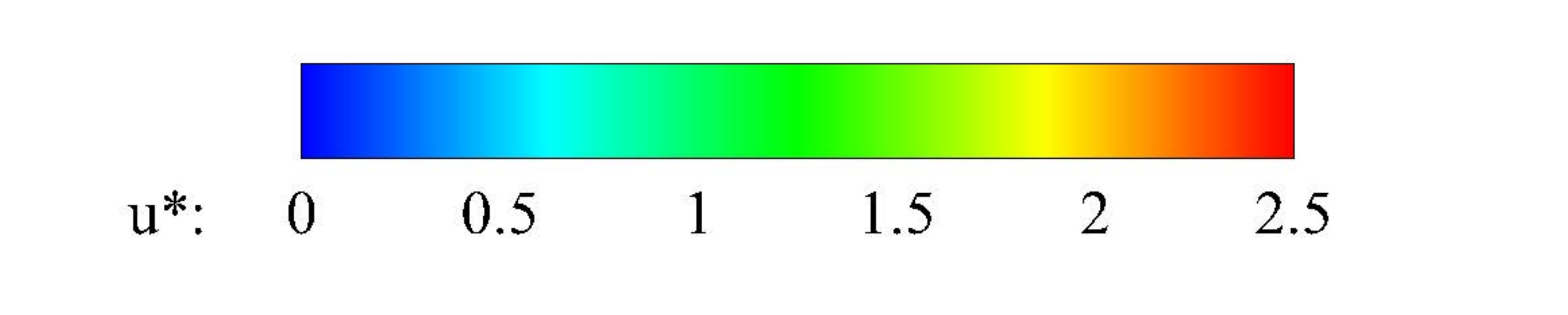}
	\end{minipage}
	\centering
	\caption{Dimensionless velocity magnitude $u^*=\left| \boldsymbol{u}\right|/\left[ u^{\prime}(t^*=0)\right] $ 
		on the center planes at the initial state. The purple lines mark the intersection between droplet surface $\phi=0.5$ and the central planes for the two-phase setting.}
	\label{veldi}
\end{figure}

In order to set up a physical initial velocity field across the droplet interface,  we follow the initialization procedure in Chen {\it et al.}~(2019),~\cite{chen2019simulation} where a turbulent flow containing a solid particle is 
first simulated as follows. First, single-phase FHIT is generated using a linear forcing scheme.~\cite{rosales2005linear,duret2012dns} Second, a solid particle is introduced and the flow containing the solid particle continues to evolve without the large-scale forcing, for  several large-eddy turnover times. This step allows 
the flow near the solid particle surface to settle properly. The above flow simulations were performed using
a finite-volume approach combined with the immersed boundary method.~\cite{menard2007coupling} This 
flow field was then used to initialize our PF-DUGKS code by replacing the solid particle with a droplet of the same 
volume. The initial values of $\tilde{f}$ and $\tilde{g}$ are constructed based on the given macroscopic flow field,
namely, with the non-equilibrium distributions provided by 
the Chapmann-Enskog analysis.
The initialization is now done and the time is denoted as $t=0$. 
Since the initialization is spherically symmetric and the magnitudes of the flow velocity inside and outside
the droplet are quite different, 
a spectral analysis based on the spherical harmonics (See Appendix~\ref{subsec: Energyspherical}) is more appropriate when analyzing the droplet
deformation after its initial release.  
For the two-phase setting, the initial radius of the droplet is $R_0=31.74$ lattice units, ensuring a good resolution of the flow inside the droplet. Therefore, the initial Weber number for two-phase flow is $We = \rho u^{\prime 2}R_0/\sigma = 2.48\times 10^{17}$
and the initial Capillary number $Ca = \mu u'/\sigma = 1.64\times 10^{16}$, which means that the surface tension
plays no dynamic role in this case. The volume fraction is $V_{drop}/V_{box}= 0.064$. The initial interfacial thickness parameter $W$ = 3.0, giving a small $\Ch$ number $\Ch=W/R_0=0.095$. 
The dimensionless time is defined as $t^*=t\epsilon\left( 0\right) /K\left( 0\right)$, where $K\left( 0\right)$ and 
$\epsilon\left( 0\right)$ are the initial kinetic energy and initial dissipation rate, respectively. 
The initial velocity magnitude field is shown in Fig.~\ref{veldi}.

\begin{figure*}
	\centering    
	\subfigure[$t^*=0$]{
		\begin{minipage}[t]{0.22\linewidth}
			\centering
			\includegraphics[width=1.04\columnwidth,trim={3cm 0cm 3cm 0cm},clip]{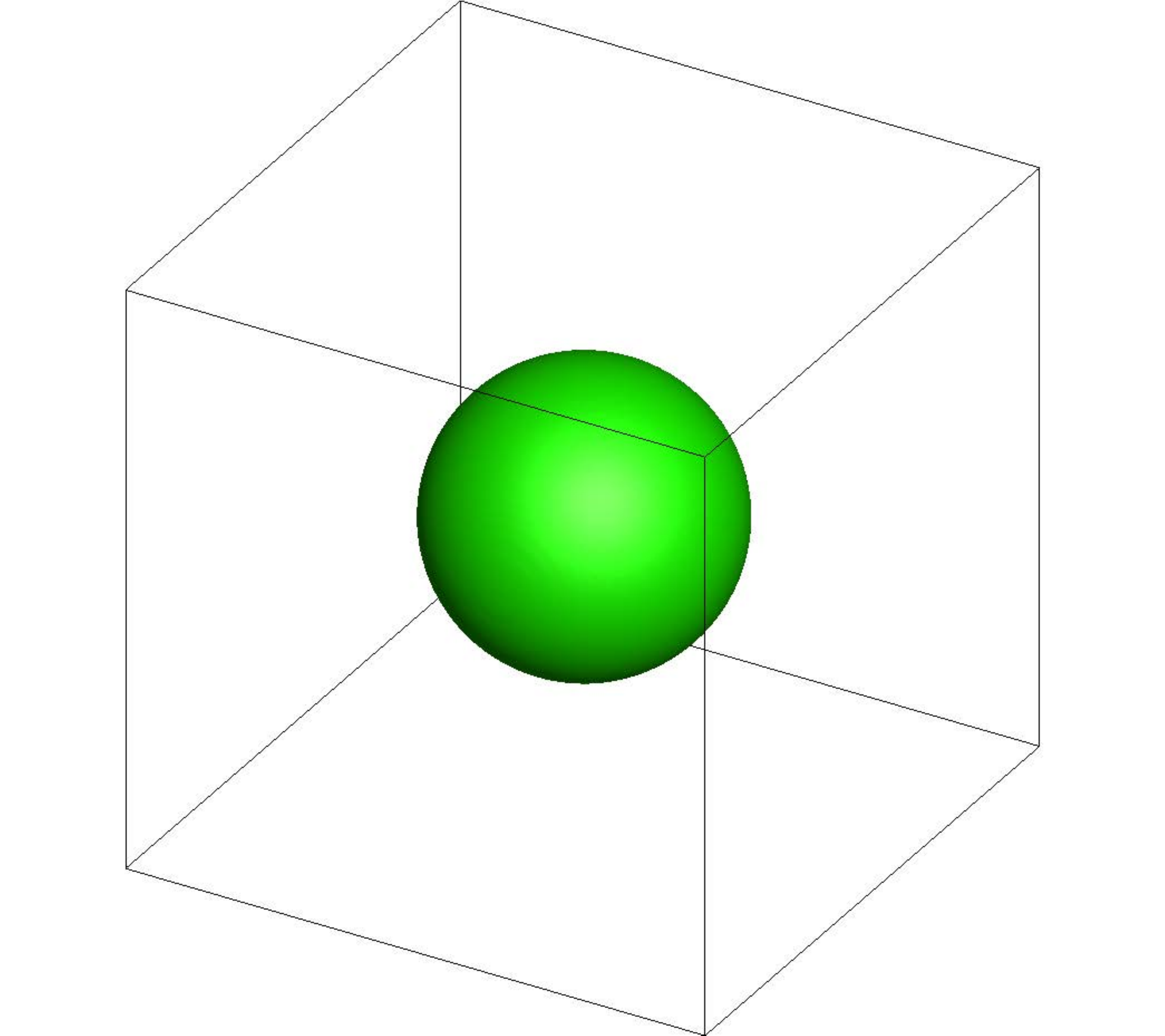}
			\label{multi0phi}
		\end{minipage}
	}
	\subfigure[$t^*=0.98$]{
		\begin{minipage}[t]{0.22\linewidth}
			\centering
			\includegraphics[width=1.04\columnwidth,trim={3cm 0cm 3cm 0cm},clip]{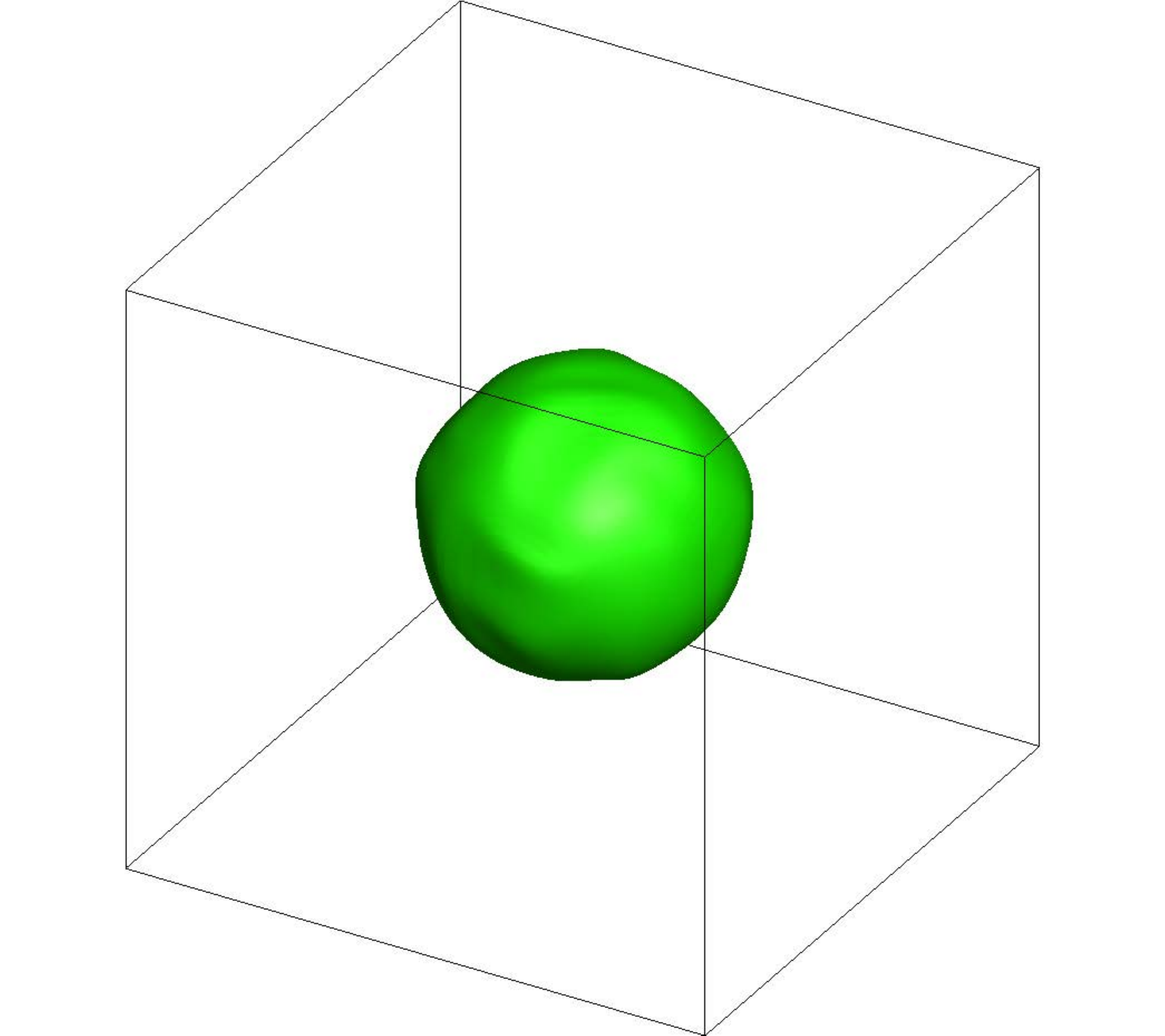}
			\label{multi2500phi}
		\end{minipage}
	}
	\subfigure[$t^*=3.91$]{
		\begin{minipage}[t]{0.22\linewidth}
			\centering
			\includegraphics[width=1.04\columnwidth,trim={3cm 0cm 3cm 0cm},clip]{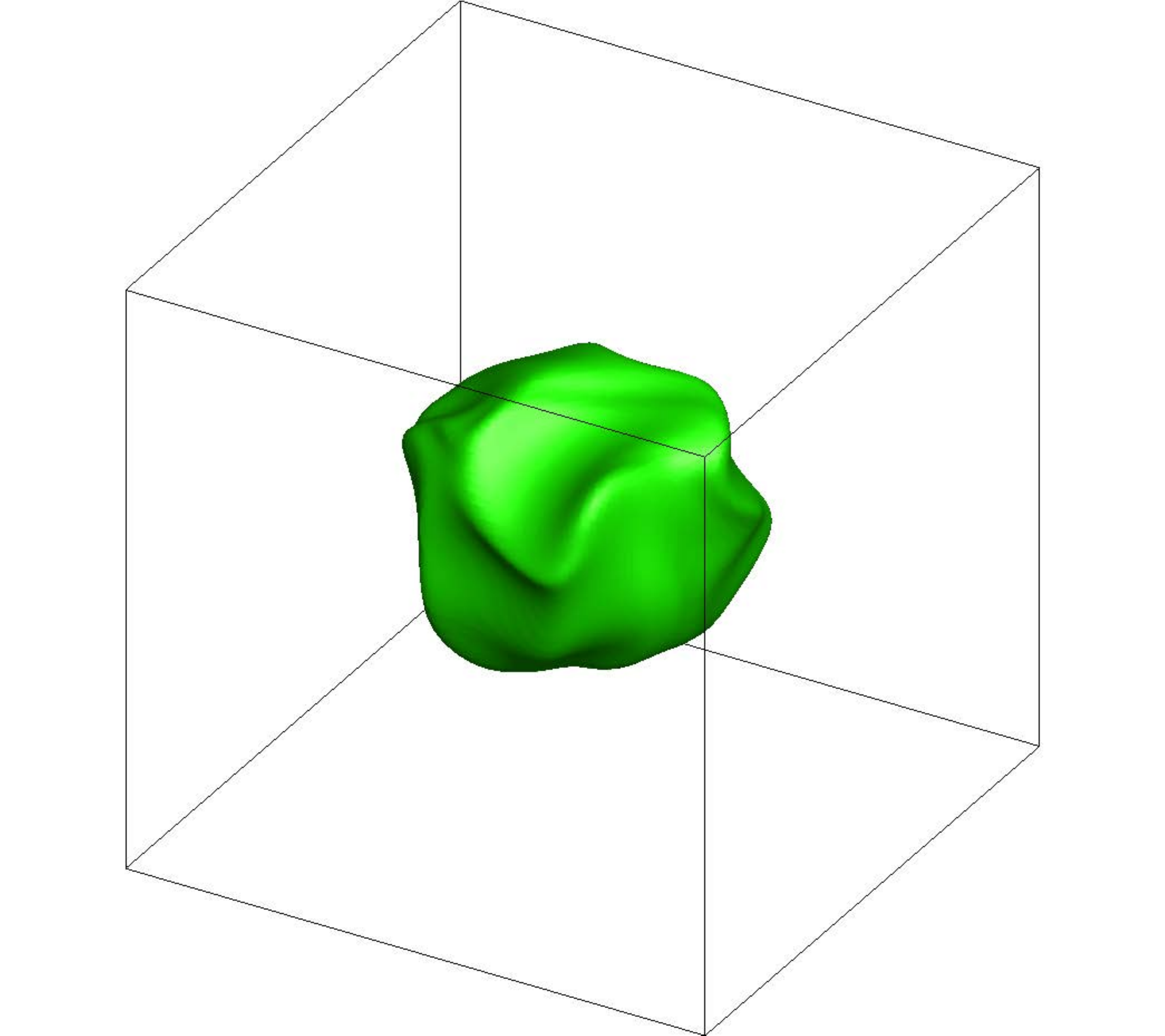}
			\label{multi10000phi}
		\end{minipage}
	}
	\subfigure[$t^*=7.83$]{
		\begin{minipage}[t]{0.22\linewidth}
			\centering
			\includegraphics[width=1.04\columnwidth,trim={3cm 0cm 3cm 0cm},clip]{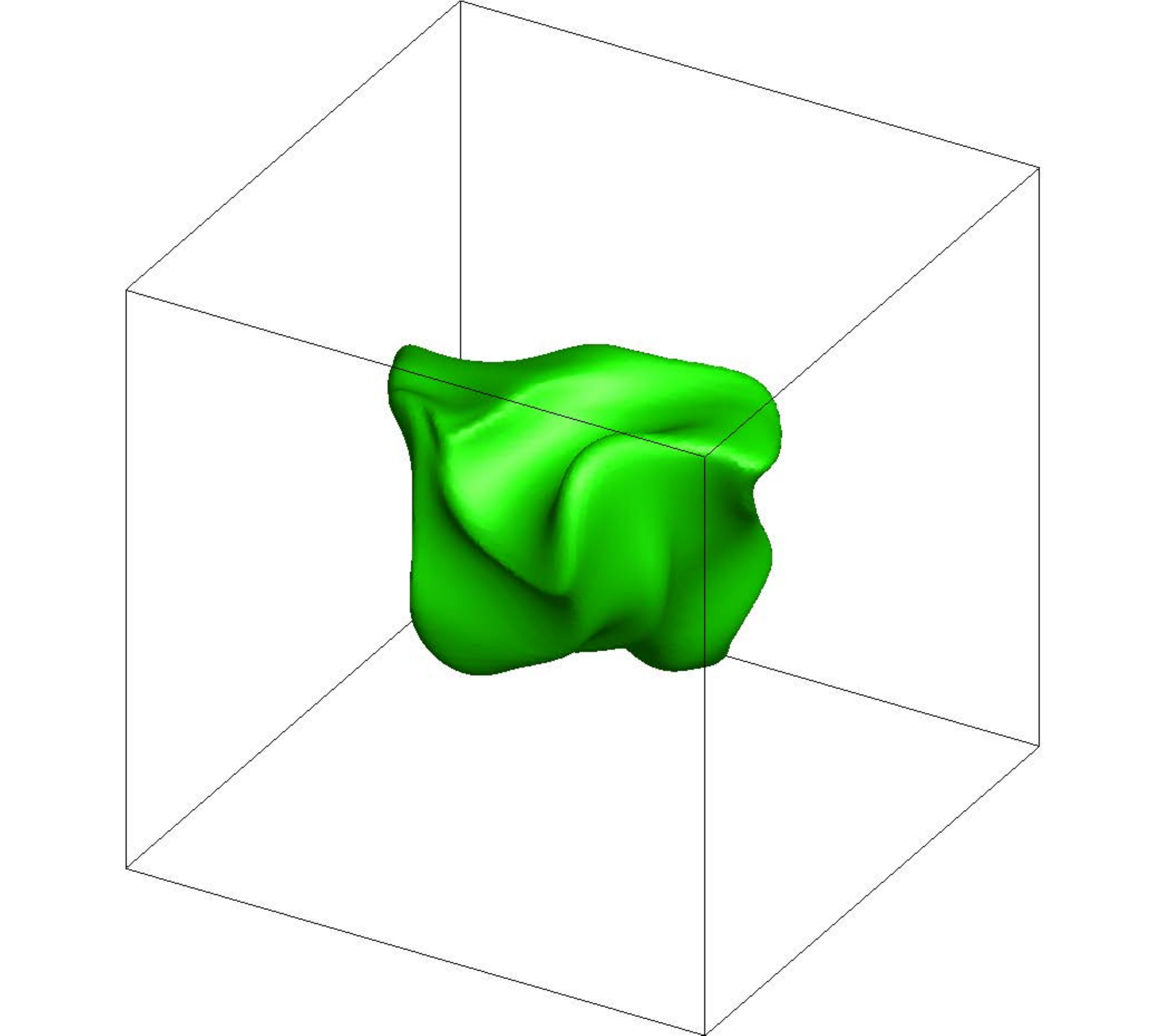}
			\label{multi20000phi}
		\end{minipage}
	}
	\begin{minipage}[t]{0.15\linewidth}
		\centering
		\includegraphics[width=0.6\columnwidth,trim={2cm 4cm 2cm 2cm},clip]{pngpdf/xyz1.pdf}
	\end{minipage}
	\centering
	\caption{3D visualization of a droplet deformation in a decaying turbulent flow at different dimensionless times. The surface is given by $\phi$ = 0.5.}
	\label{phid}
\end{figure*}

Fig.~\ref{phid} shows the evolution of fluid-fluid interface in the decaying turbulent flow under the two-phase settings, illustrating
the complex distortion of the
passive 
droplet interface as a function of time. 
At the early time $t^*=0.98$, the droplet is only slightly deformed. 
At $t^*=3.91$, the interface becomes much more complicated, with distinct depressions and bulges.

\begin{figure*}[t!]
	\centering    
	\subfigure[$t^*=0$]{
		\begin{minipage}[t]{0.45\linewidth}
			\centering
			\includegraphics[width=1.\columnwidth,trim={0cm 0cm 0cm 0cm},clip]{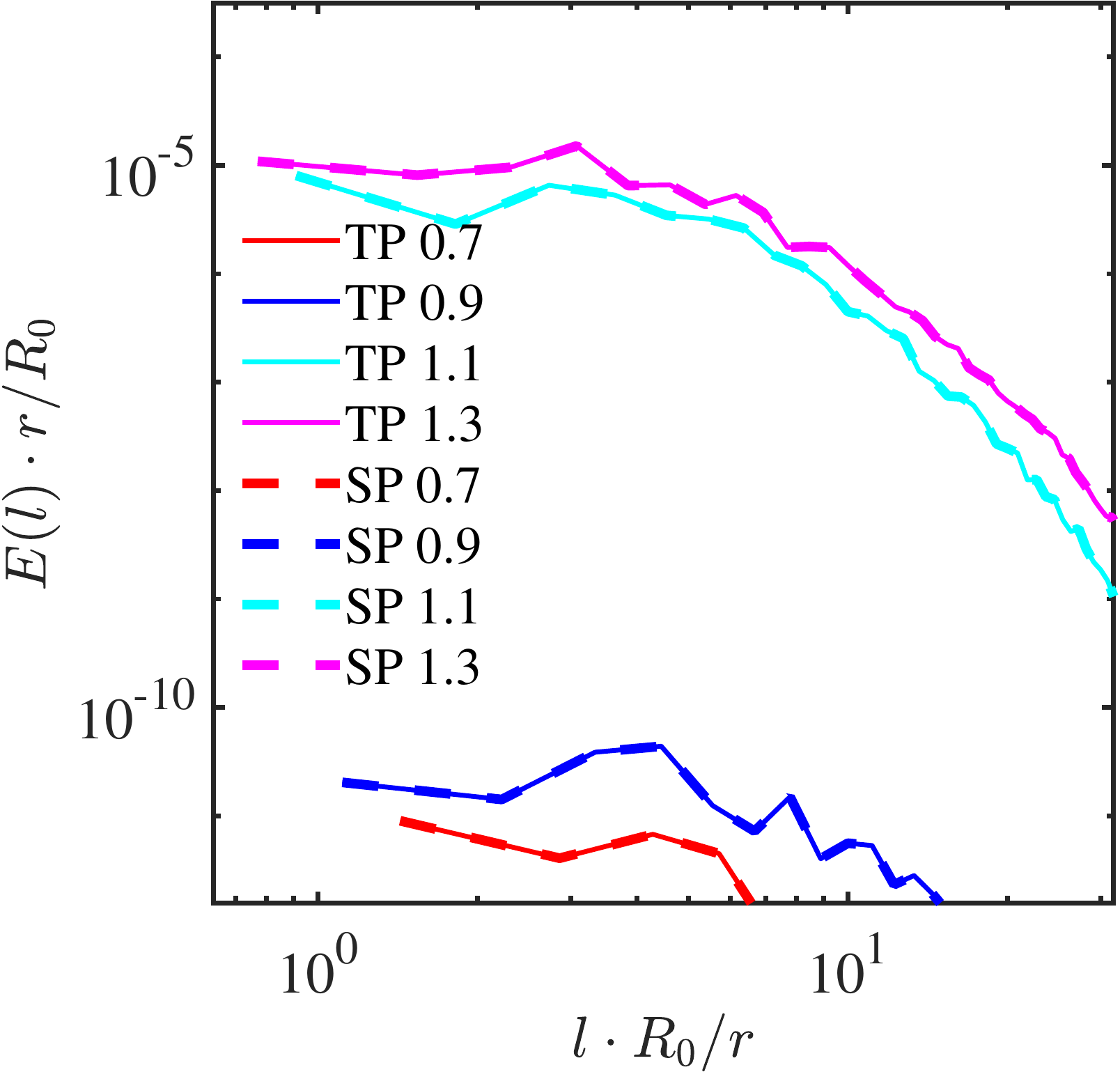}
			\label{El0rr}
		\end{minipage}
	}
	\subfigure[$t^*=0.98$]{
		\begin{minipage}[t]{0.45\linewidth}
			\centering
			\includegraphics[width=1.\columnwidth,trim={0cm 0cm 0cm 0cm},clip]{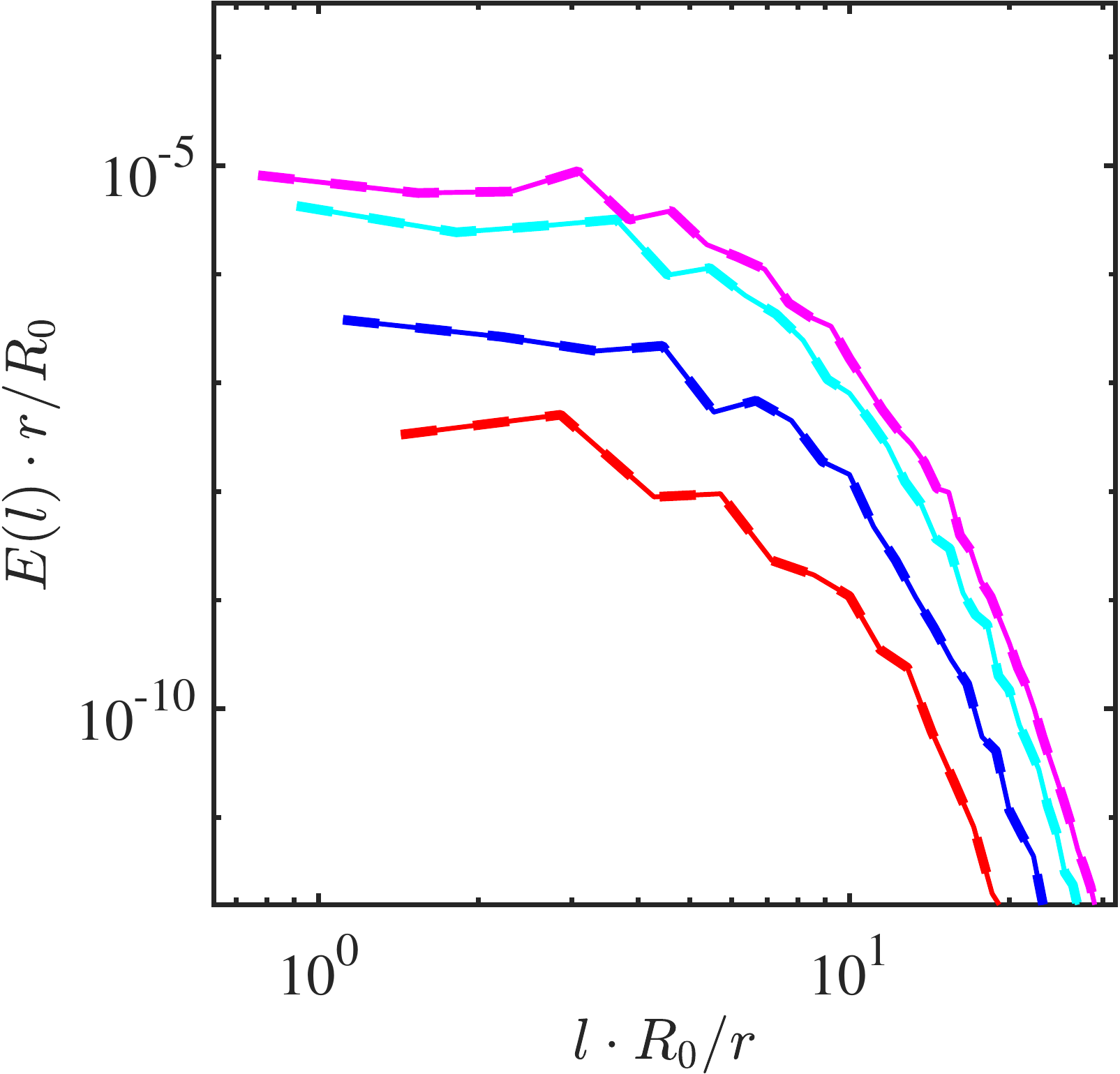}
			\label{El2500rrno}
		\end{minipage}
	}
	\\
	\subfigure[$t^*=3.91$]{
		\begin{minipage}[t]{0.45\linewidth}
			\centering
			\includegraphics[width=1.\columnwidth,trim={0cm 0cm 0cm 0cm},clip]{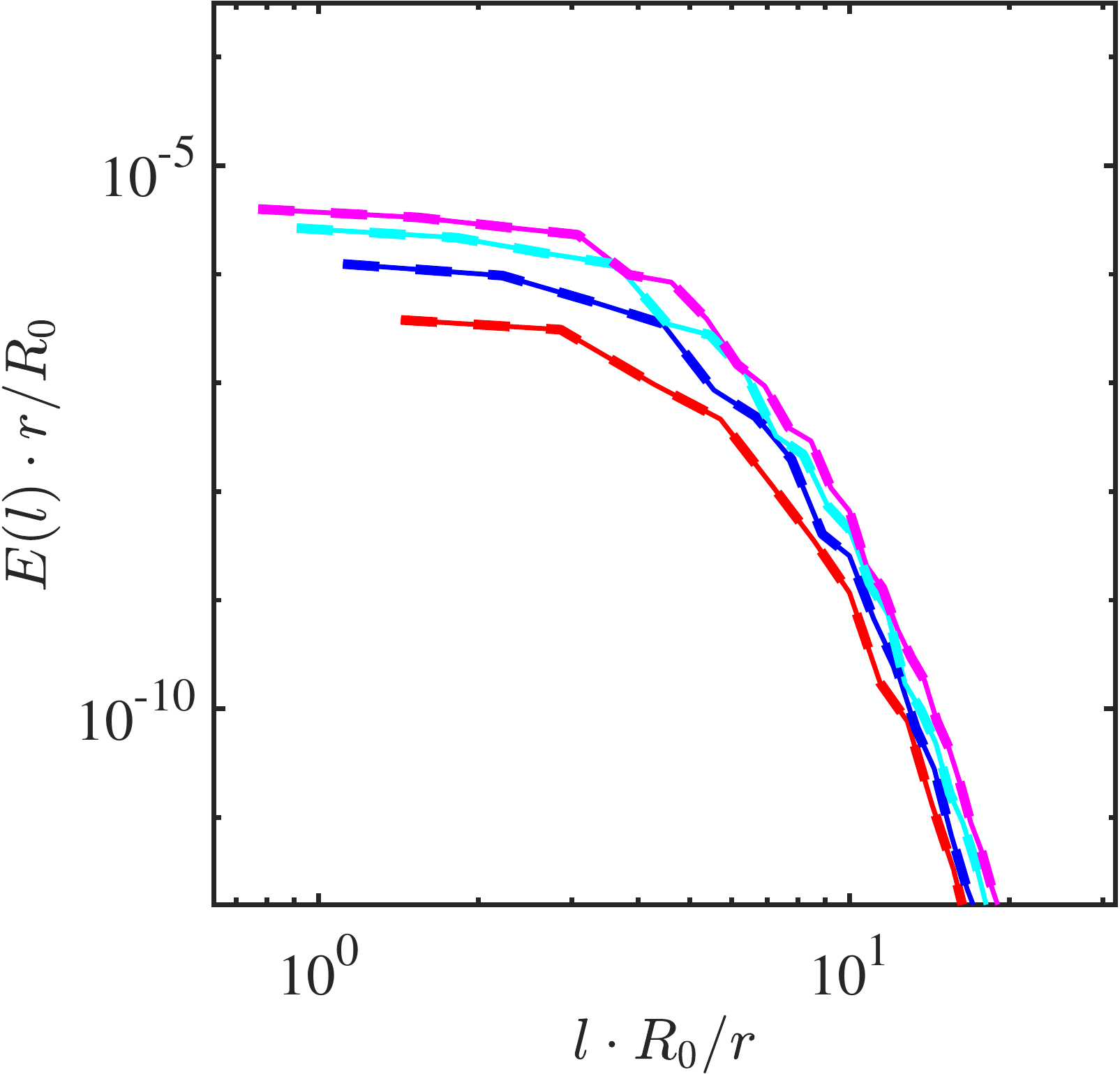}
			\label{El10000rrno}
		\end{minipage}
	}
	\subfigure[$t^*=7.83$]{
		\begin{minipage}[t]{0.45\linewidth}
			\centering
			\includegraphics[width=1.\columnwidth,trim={0cm 0cm 0cm 0cm},clip]{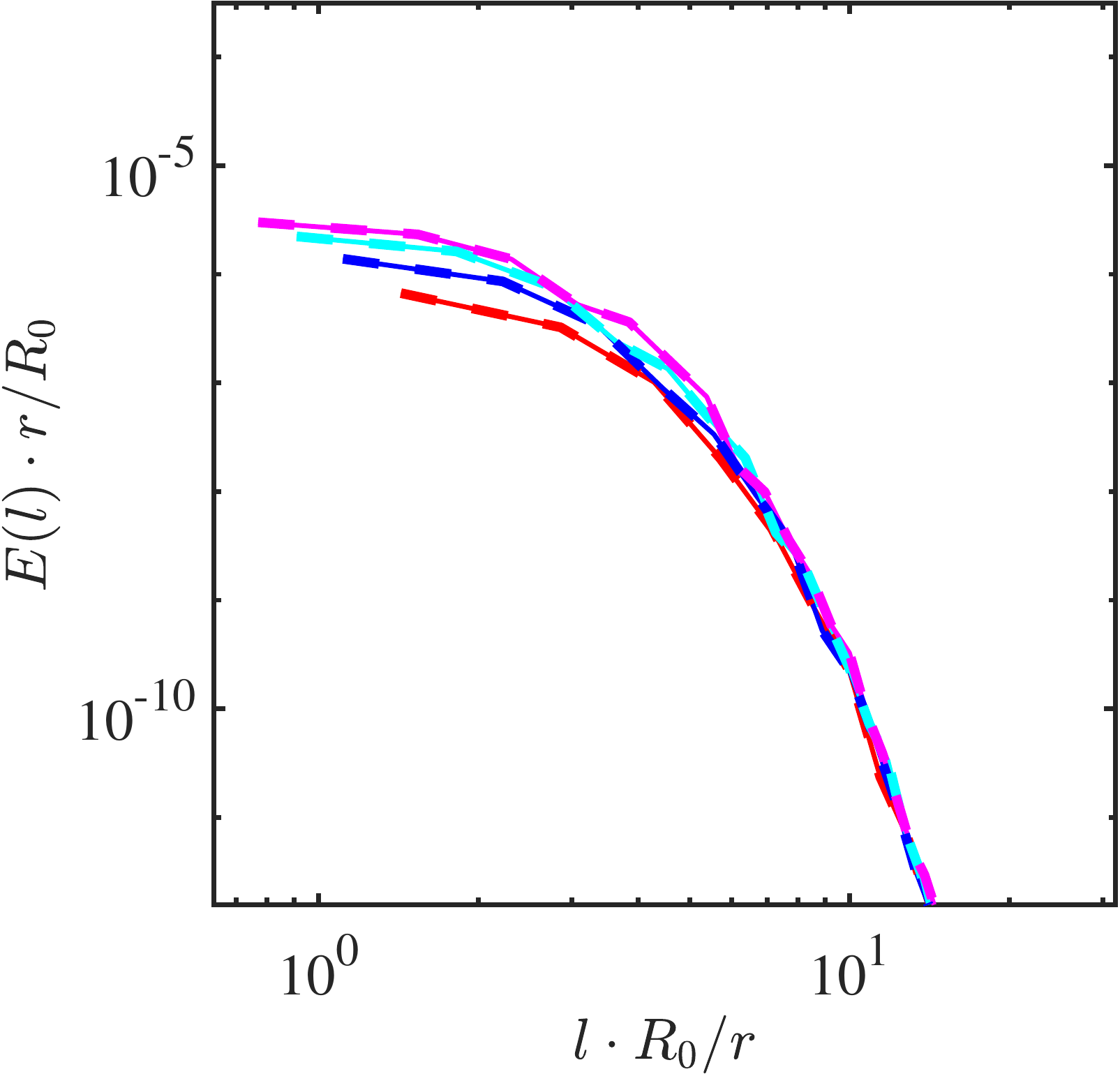}
			\label{El20000rrno}
		\end{minipage}
	}
	\centering
	\caption{ 
Energy spectra on the spherical surfaces at different times for the two-phase (TP) flow ($We= 2.48\times 10^{17}$) and
  the single-phase (SP) flow. The surface location is indicated by the normalized radii, $r/R_0=0.7,0.9,1.1,1.3$.
  The initial radius of the droplet is   $R_0=31.74$.}
	\label{sphericaldeform}
\end{figure*}

The total kinetic energy decreases rapidly between $t^*=0$ and $t^*=0.98$ due to the specific initialization used, 
followed by
a relatively slow decay of the total kinetic energy from $t^*=0.98$ to $t^*=3.91$.
The total kinetic energy at $t^*=0, 0.98, 3.91$ are 152.80, 83.34, 32.74, respectively. 
Fig.~\ref{sphericaldeform} compares the energy spectra based on the spherical harmonics,
showing an excellent agreement between the two-phase setting and single-phase setting of the PF-DUGKS code, for all distances from the initial droplet center and all times. 
Here $l$ represents the wavenumber on a spherical surface of a given radius $r$, meaning that $E(l)$ 
represents the kinetic energy associated with velocity fluctuations at the length scale of $r/l$. To properly compare the energy
level at a same length scale, we therefore set the horizontal axis in Fig.~\ref{sphericaldeform} to $l R_0/r$.
Since both the single-phase and two-phase settings are designed to eliminate the 
surface tension effect, the identical results obtained here validate the PF-DUGKS code.
Furthermore, Fig.~\ref{sphericaldeform} indicates that the kinetic energy outside the initial droplet region will decay due to  two reasons: first, the viscous dissipation takes away kinetic energy as no external forcing is
applied. Second, the kinetic energy in the outer region is transported to the region near and inside the droplet
where the initial kinetic energy is low,
causing a fast decay of energy outside and a quick increase of kinetic energy inside, during the initial evolution stage.
Eventually, the spectrum shapes are similar inside and outside the droplet, for high wavenumbers of wavelengths much less than
the radius $r$, as confirmed in Fig.~\ref{sphericaldeform}(d).  For lower wave numbers of wavelengths comparable to $r$, the spherical harmonic modes are
different for different $r$, yielding different energy spectra for different $r$.  Therefore, the spherical harmonic energy spectra
provide a way to represent energy transport between different $r$ as well as different scales at a given $r$.

\begin{figure}[t!]
	\centering    
	\subfigure[Kinetic energy]{
		\begin{minipage}[t]{0.47\linewidth}
			\centering
			\includegraphics[width=1.\columnwidth,trim={0cm 0cm 0cm 0cm},clip]{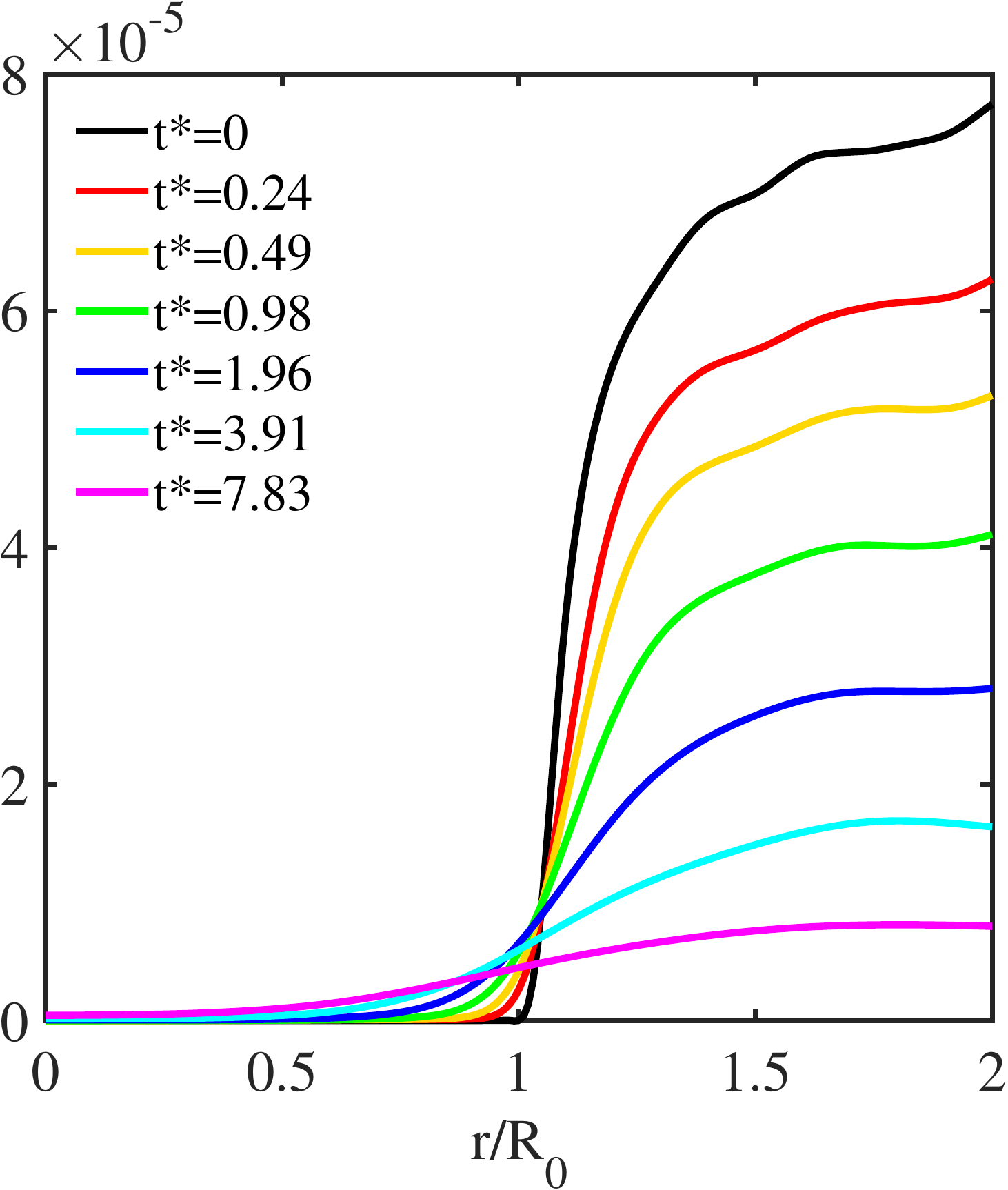}
			\label{Kr}
		\end{minipage}
	}
	\subfigure[Free energy]{
		\begin{minipage}[t]{0.47\linewidth}
			\centering
			\includegraphics[width=1.\columnwidth,trim={0cm 0cm 0cm 0cm},clip]{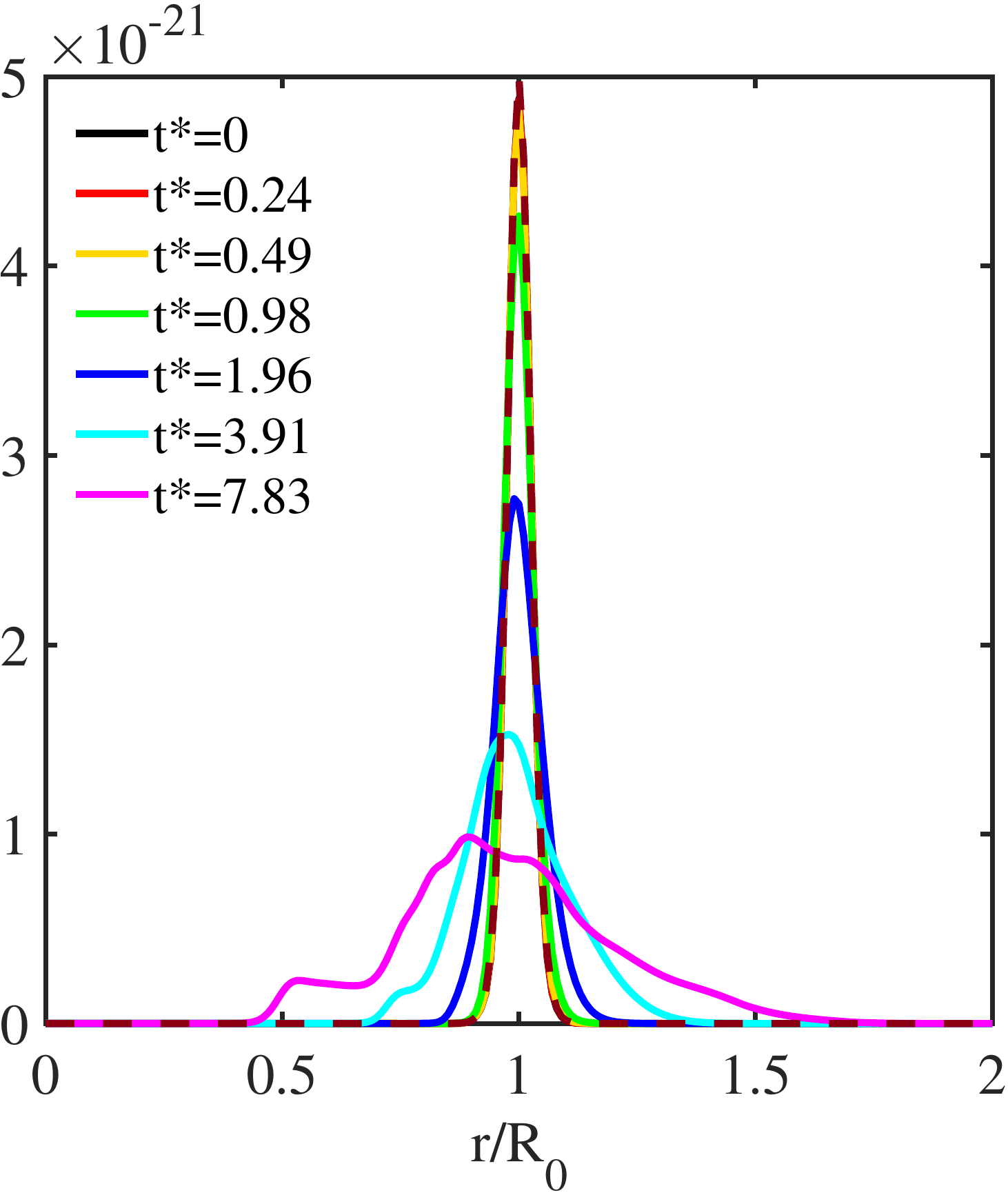}
			\label{FEr}
		\end{minipage}
	}
	\centering
	\caption{Spherical-surface averaged energies as a function of radial distance, at different times for the two-phase setting. 
 The dark red dashed line in (b) is the initial analytical result based on Eq.~(\ref{phini}).}
	\label{Ed}
\end{figure}

Fig.~\ref{Ed} shows the kinetic energy and free surface energy averaged over a spherical surface as a function of the distance from the 
initial droplet center. Fig.~\ref{Kr} clearly demonstrates the fast decay of kinetic energy outside the droplet and transport of kinetic energy into the droplet, as noted already in Fig.~\ref{sphericaldeform}. 
Fig.~\ref{FEr} shows the evolution of free energy from the two-phase setting. Initially, the profiles of free energy $F\left( \phi\right) $ are localized near the interface $r/R_0=1$. As time evolves, the interface expands to the regions inside and outside the initial spherical droplet. Therefore, the profiles of free energy also expand, being wider with smaller peak amplitudes, and then their shape becomes more and more irregular. The magnitude of the free energy is on the order of $10^{-21}$, 16 orders of magnitude smaller
than that of the kinetic energy, confirming that the free energy can be safely neglected here. Note that the relative difference in magnitude roughly mimics the initial Weber number ($2.48\times 10^{17}$) noted before, so the Weber number could be viewed as the ratio of 
kinetic energy to 
free energy.

\section{Breakup of a spherical droplet in decaying homogeneous isotropic turbulence} \label{sec: breakup}

\subsection{Problem setup and initialization}

In this section, we apply the PF-DUGKS method to simulate the breakup process of a spherical droplet in DHIT. 
The computational domain is a cubic box containing $256^3$ lattice nodes with periodic boundary conditions in all directions. The density ratio and the viscosity ratio are set to be unity. We first initialize a DHIT field using a single-phase setting. 
The initial order parameter $\phi$ and pressure field $p$ are both set to zero. The initial velocity field $\boldsymbol{u}$ in this single-phase setting is specified by a Gaussian field with a prescribed kinetic energy spectrum,~\cite{2010Comparison,WangPeng2016}
$
E(k)=A k^{4} e^{-0.14 k^{2}}
$ for $k_{min} \le k \le k_{max}$, 
where $k$ is the wavenumber, $k_{\min }=2$ and $k_{\max }=10$ are selected in our simulation. The constant parameter $A$ is set $5.0\times 10^{-6}$. The initial velocity field of single-phase flow in the physical space is obtained by inverse Fourier transform. The maximum velocity magnitude of this single-phase initialization is 0.0988, which ensures a sufficiently small Mach number that is consistent with our nearly incompressible DUGKS formulation. 

\begin{figure}[]
	\centering    
	\subfigure[PF-DUGKS method]{
		\begin{minipage}[t]{0.46\linewidth}
			\centering
			\includegraphics[width=1.1\columnwidth,trim={0cm 0cm 0cm 0cm},clip]{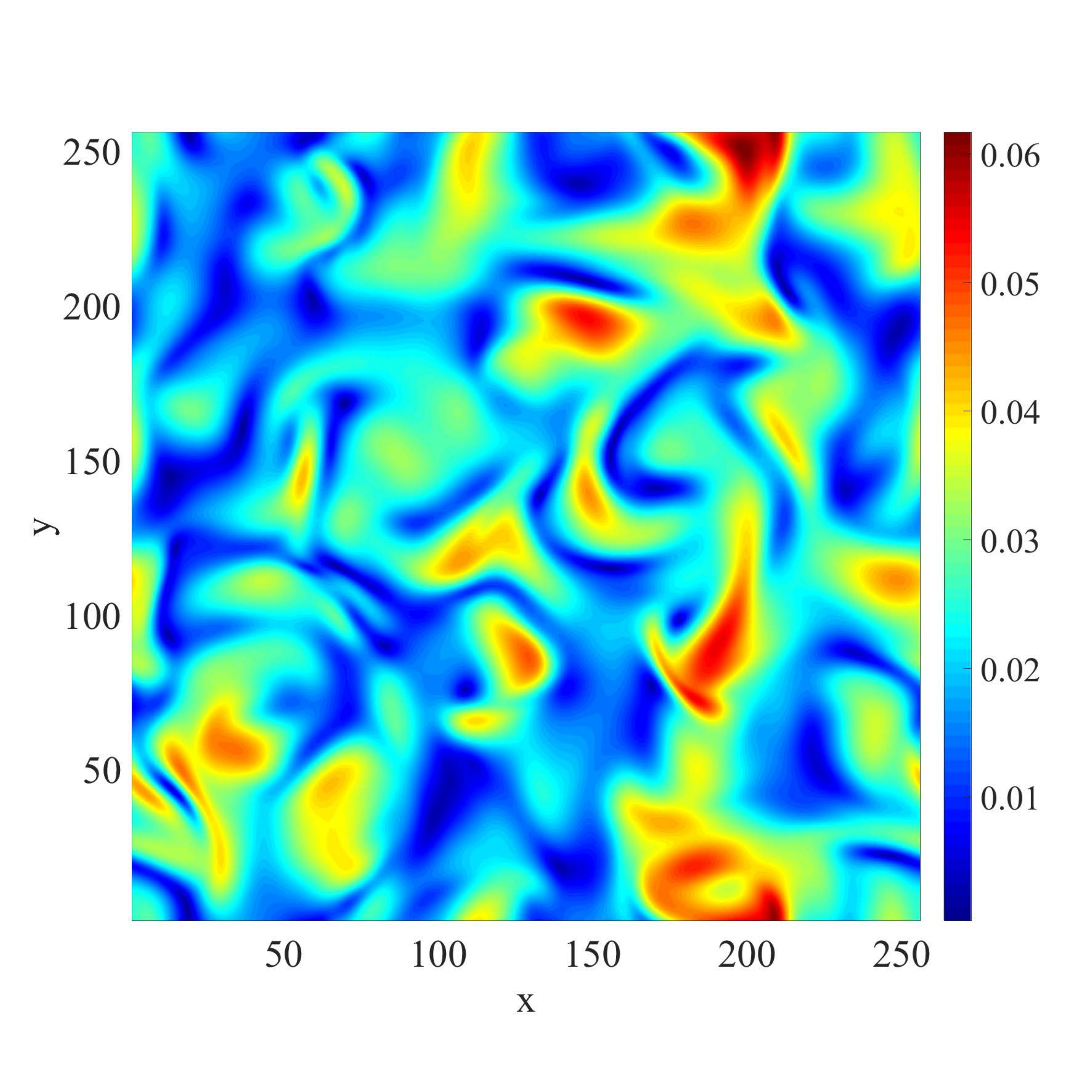}
			\label{DUGKSIni}
		\end{minipage}
	} \hspace{5pt}
	\subfigure[PS method]{
		\begin{minipage}[t]{0.46\linewidth}
			\centering
			\includegraphics[width=1.1\columnwidth,trim={0cm 0cm 0cm 0cm},clip]{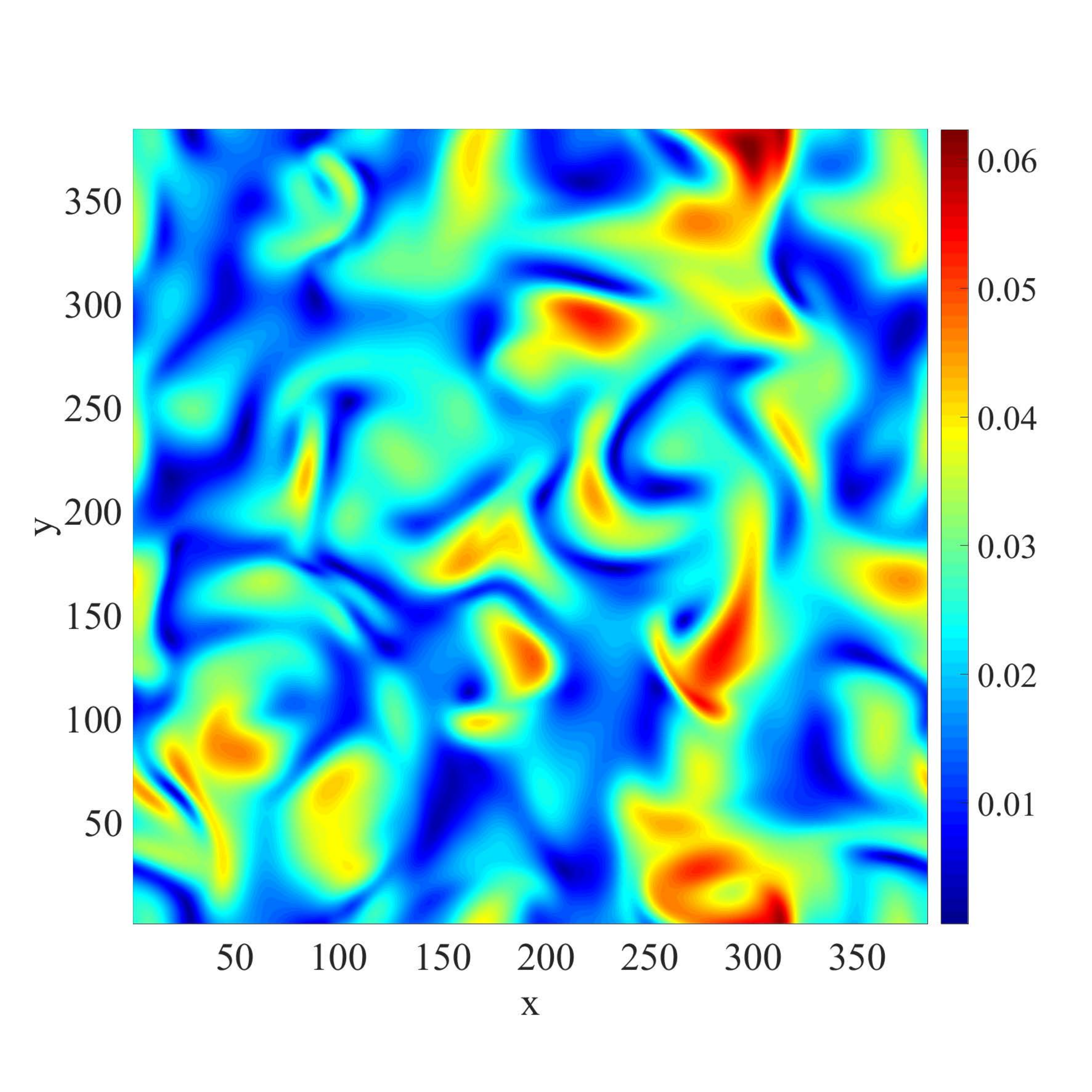}
			\label{SpecIni}
		\end{minipage}
	}
	\centering
	\caption{Contours of velocity magnitude on the center plane in the $z$ direction at $t^*=0$.}
	\label{IniXY}
\end{figure}

\begin{figure}[t!]
	\centering
	\includegraphics[width=0.85\columnwidth,trim={0cm 0cm 0cm 0cm},clip]{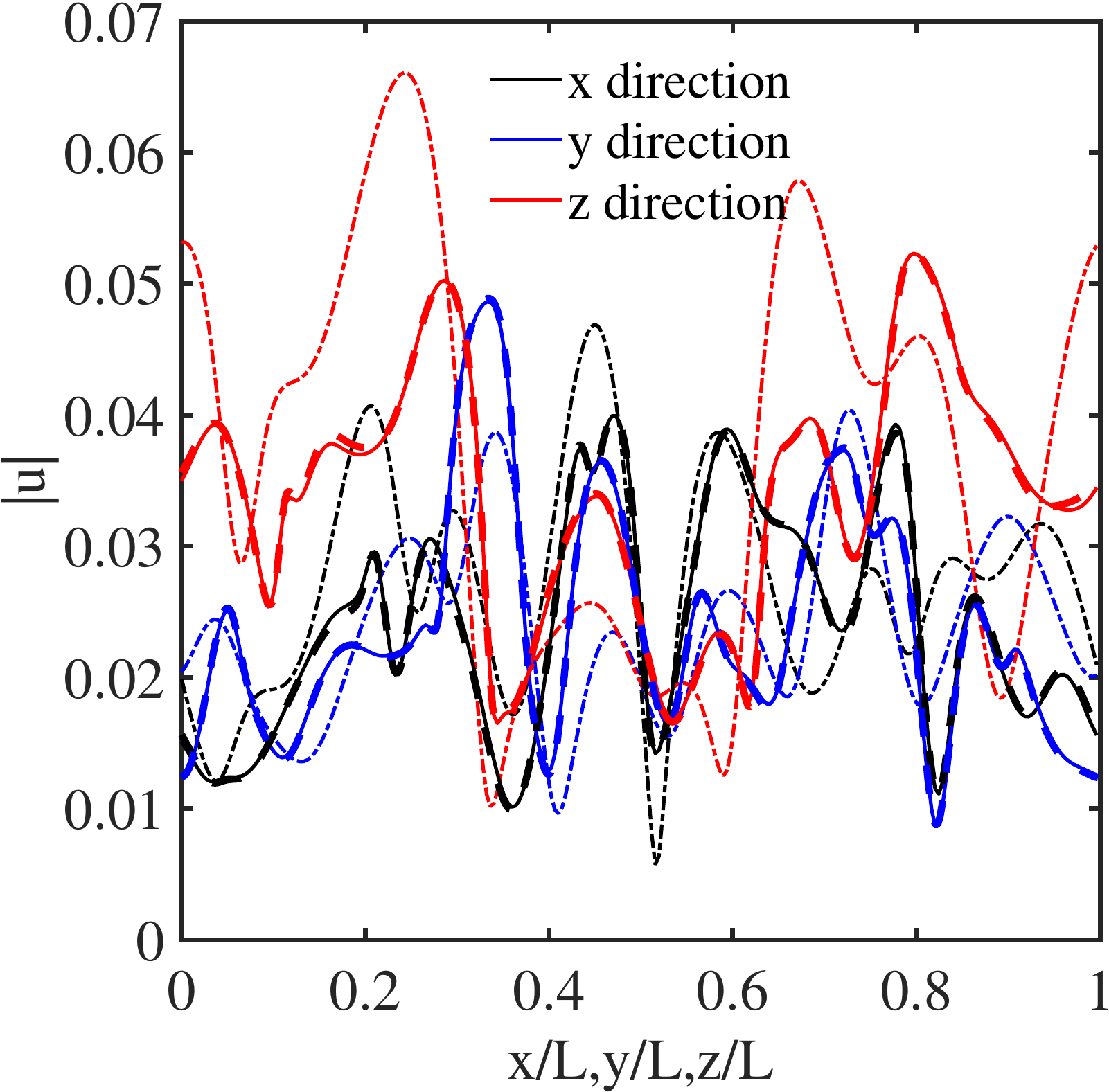}
	\centering
	\caption{Velocity magnitude profiles of the single-phase flows along the centerlines in the $x,y,z$ directions. $L$ is the computational box size. $L=256$ for the PF-DUGKS method, while $L=384$ for the PS method. The thin solid lines represent the velocity magnitude calculated by the PF-DUGKS method at $t^*=0$, and the thick dashed lines represent the velocity magnitude calculated by the PS method at $t^*=0$. The thin dash-dot lines represent the velocity profiles that are shared with the PS method, at $t^*=-0.1$.}
	\label{initxyzline}
\end{figure}

We first compare the single-phase velocity magnitude field computed by the PF-DUGKS method, under the
single-phase setting ({\it i.e.}, $\phi = 0$), with that computed by the pseudo-spectral (PS) method. 
We first run the PF-DUGKS code for a period of time till the pressure field is properly initialized to be consistent
with the velocity field.
At this point, 
$Re_{\lambda}=76$,
$k_{\max}\eta = 3.6$, 
and we extract the velocity field and share with the PS code. 
We use the $2/3-rule$ in the spectral method to remove the aliasing error,~\cite{2006Spectral,Potherat2015The} hence the computational domain is set to
a cubic box with $384^3$ nodes in the PS code. After about 0.1 large-eddy turnover times, we obtain the velocity fields from these two codes, and compare with each other. Fig.~\ref{IniXY} shows the velocity magnitude contours on a same center plane. Fig.~\ref{initxyzline} plots the velocity magnitude profiles on the same center lines in three directions. Both figures show almost the same results from the two approaches. Furthermore, we also calculate the correlation coefficient of three components of the velocity over the whole domain, {\it i.e.}, $u_x$, $u_y$, $u_z$, between these two approaches. The results for Corr($u_x$(PF-DUGKS), $u_x$(PS)), Corr($u_y$(PF-DUGKS), $u_y$(PS)), Corr($u_z$(PF-DUGKS), $u_z$(PS)) are 0.9999, 0.9999, 0.9998, respectively. 
Here the correlation coefficient Corr$(s_1,s_2)$ of arbitrary scalar fields $s_1$ and $s_2$ is
$
\text{Corr}\left(s_1,  s_2 \right)
= \frac{\left\langle\left( s_1-\left\langle s_1\right\rangle_V\right) \left( s_2-\left\langle s_2\right\rangle_V\right) \right\rangle_V}
{\sqrt{
		\left\langle\left( s_1-\left\langle s_1\right\rangle_V\right)^2\right\rangle_V
		\left\langle\left( s_2-\left\langle s_2\right\rangle_V\right)^2\right\rangle_V
}}.\label{Corr}
$
These comparison results again demonstrate that the PF-DUGKS code can accurately simulate the evolution of a single-phase turbulence with adequate resolution.~\cite{WangPeng2016}

Now the above velocity field calculated by PF-DUGKS method (Fig.~\ref{DUGKSIni}) is used as the initial velocity field for the two-phase flow setting. We add a spherical droplet in this flow field, {\it i.e.}, reset the order parameter $\phi$ with Eq.~(\ref{phini}). 
This moment is now viewed as $t=0$ for the discussions below.

The maximum velocity and vorticity magnitudes at the two-phase flow initialization are $\left| \boldsymbol{u}(t^*=0) \right|_{\max} = 0.0822$ and $\left| \boldsymbol{\omega}(t^*=0) \right|_{\max} = 0.0238$, respectively. The r.m.s. velocity and vorticity of two-phase initialization are $u^{\prime}(t^*=0) = 0.0165$ and $\omega^{\prime}(t^*=0) = 0.0023$, respectively, where $\omega^{\prime}= \sqrt{\langle\boldsymbol{\omega}^{\prime} \cdot \boldsymbol{\omega}^{\prime}\rangle_V/3}$. The surface tension is set to
$\sigma = 8.0\times 10^{-4}$, the density and kinematic viscosity are set to
1.0 and $ 4.5\times 10^{-3}$, respectively, for both phases.
The initial droplet radius $R_0$ is 64 (the initial droplet diameter is $D_0=128$), the interfacial thickness parameter is $W$ = 3. 
These settings imply the following initial dimensionless parameters: the Taylor microscale Reynolds number $Re_\lambda$ is 58, $\Ch$ = 0.0469, the dispersed phase volume fraction $\varphi$ is 6.54\%, the Weber number $\We$ is 21.66, and $k_{max}\eta$ is 3.28 (which is sufficient to resolve the smallest turbulence scales).~\cite{WangPeng2016} Since no external forcing is applied, both of the Weber number and Reynolds number decrease with time. 

\begin{figure}[t!]
	\centering
	\includegraphics[width=0.75\columnwidth,trim={0cm 0cm 0cm 0cm},clip]{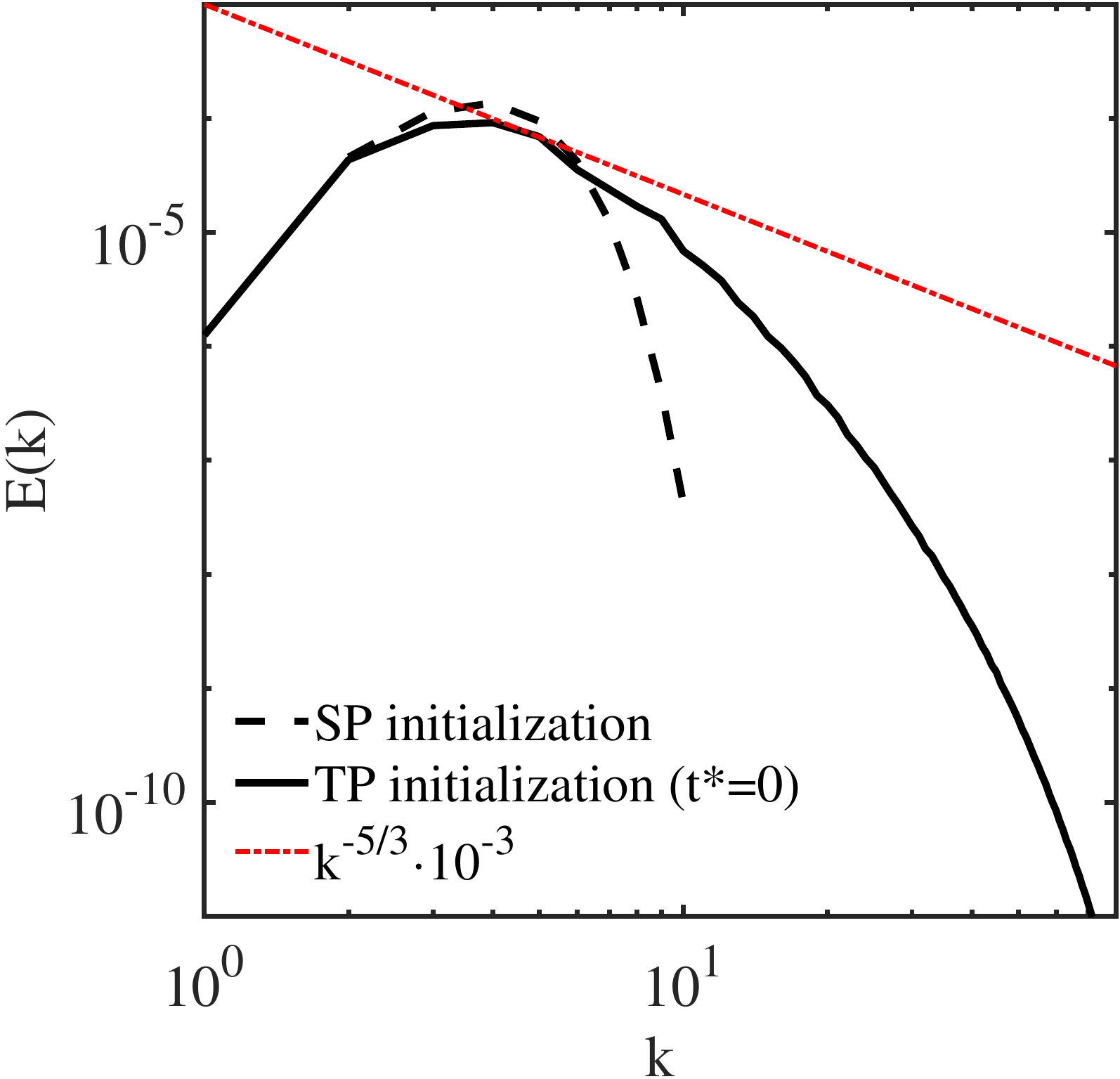}
	\centering
	\caption{The kinetic energy spectrum used to initialize the single-phase (SP) and two-phase (TP) flow simulation, respectively.}
	\label{fig:init}
\end{figure}

The energy spectra of single-phase initialization and two-phase initialization 
are showed in Fig.~\ref{fig:init}. The kinetic energy mainly contains large scales at the single-phase initialization. Then the
energy is transferred to the small scales over time when the two-phase flow simulation begins.
At this time, the velocity derivative skewness $S_u$ is between $-0.6$ and $-0.5$, which is reasonable for this flow field.~\cite{wang1993settling,antonia2015boundedness,WangPeng2016,2020Simulation} Here $S_{u}=\frac{\left\langle\left[\left(\partial u_{x} / \partial x\right)^{3}+\left(\partial u_{y} / \partial y\right)^{3}+\left(\partial u_{z} / \partial z\right)^{3}\right] / 3\right\rangle}{\left\langle\left[\left(\partial u_{x} / \partial x\right)^{2}+\left(\partial u_{y} / \partial y\right)^{2}+\left(\partial u_{z} / \partial z\right)^{2}\right] / 3\right\rangle^{3 / 2}}$.

Additionally, we also simulate another single-phase ({\it i.e.}, $\phi=0$) DHIT using the same code and the same velocity field and pressure field at $t^*=0$, in order to compare the results with those from the two-phase flow setting when necessary.
It is noted that, unlike Section~\ref{sec: deform}, here the two-phase setting is expected to yield
different results from the single-phase setting as the surface tension now can play a role. In fact, due to the turbulence decay,
the Weber number will decrease monotonically, from 21.66 to less than 0.01 (see later discussions in Section~\ref{threestages}), implying that the surface tension eventually plays a more important role in determining the interfacial dynamics, than the turbulent kinetic energy.

\subsection{Three stages of fluid-fluid interface evolution}
\label{threestages}

\begin{figure}[t!]
	\centering    
	\subfigure[$t^*=0$]{
		\begin{minipage}[t]{0.22\linewidth}
			\centering
			\includegraphics[width=1.04\columnwidth,trim={3cm 0cm 3cm 0cm},clip]{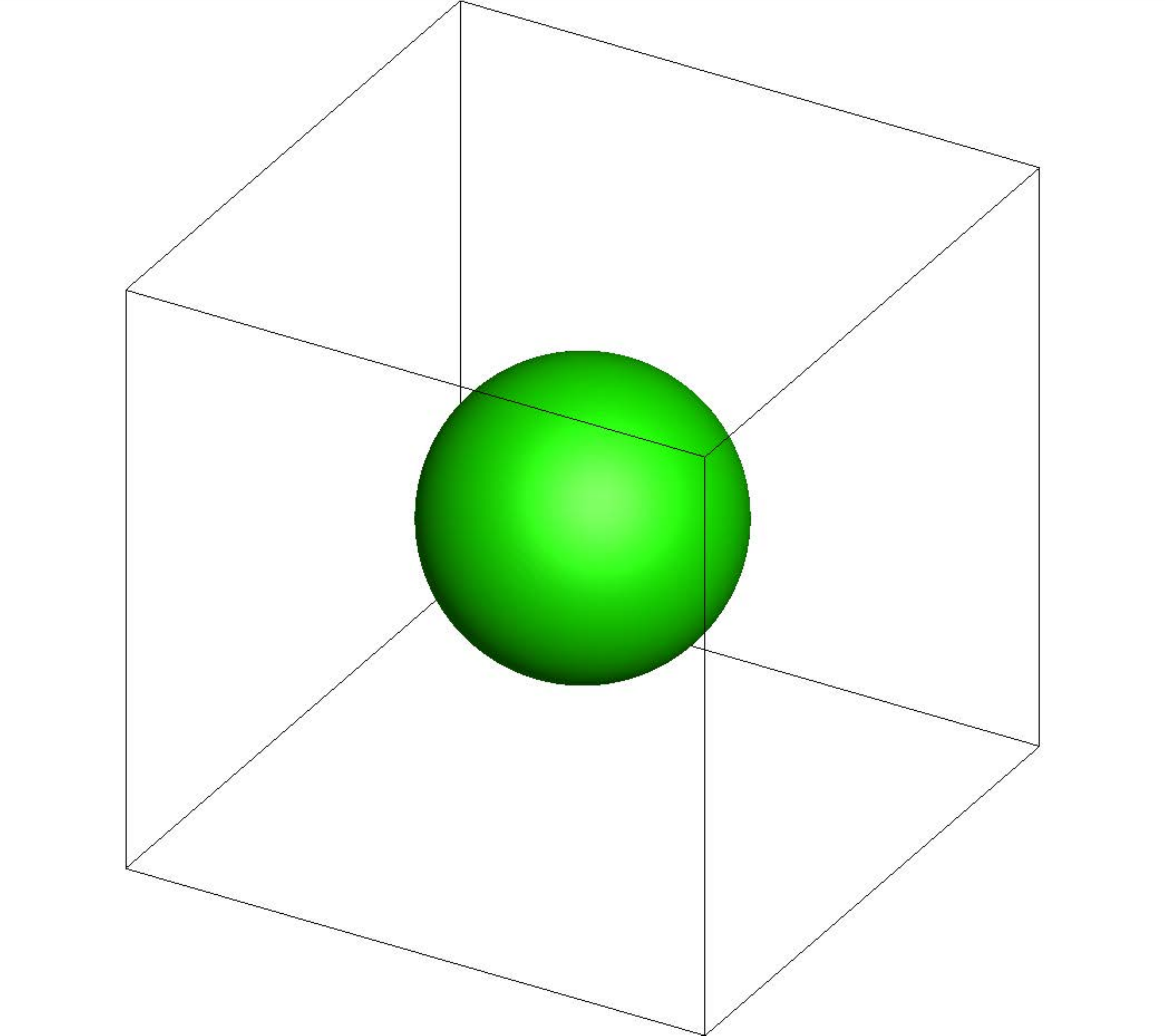}
			\label{3000}
		\end{minipage}
	}
	\subfigure[$t^*=0.0448$]{
		\begin{minipage}[t]{0.22\linewidth}
			\centering
			\includegraphics[width=1.04\columnwidth,trim={3cm 0cm 3cm 0cm},clip]{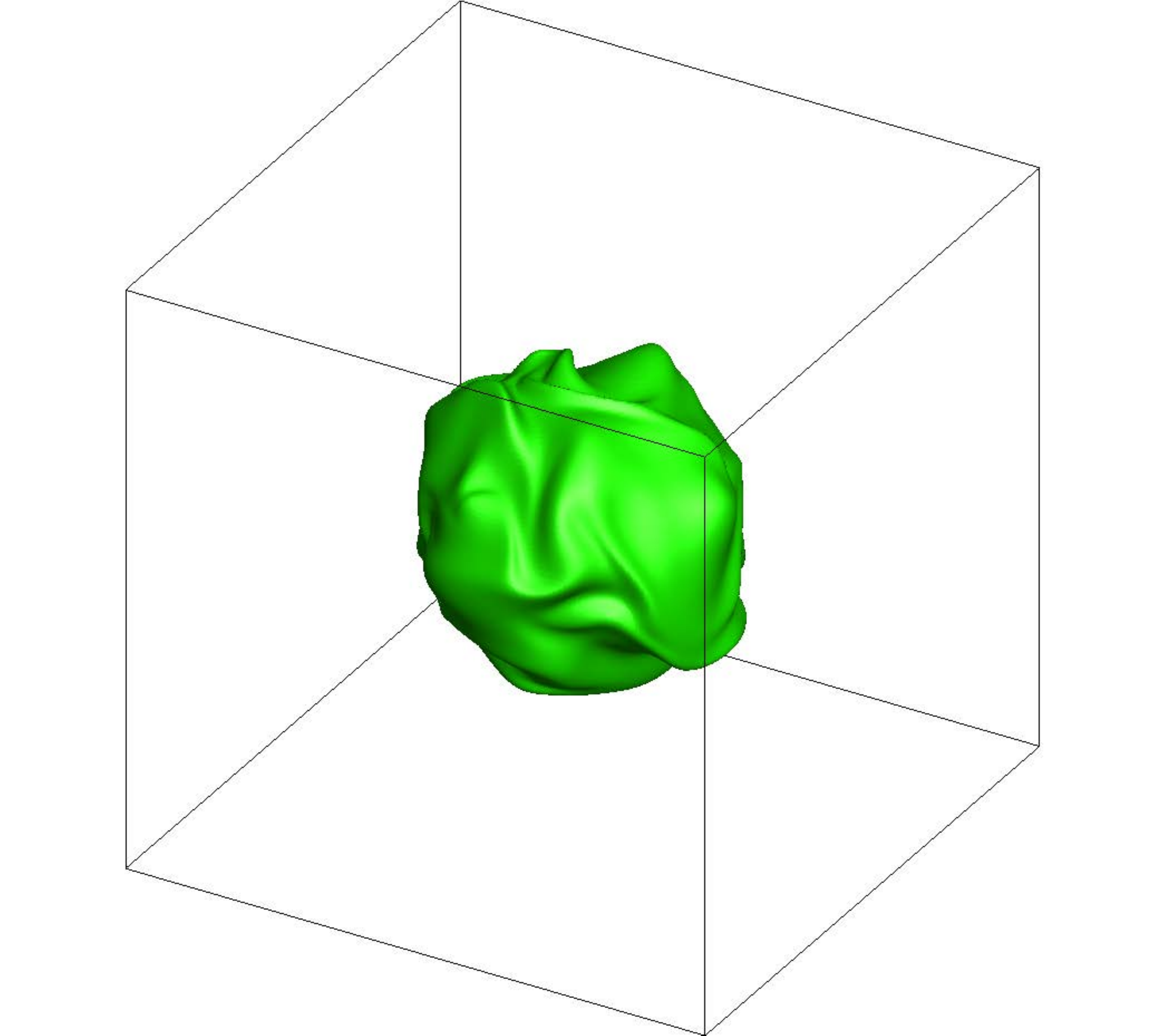}
			\label{4000}
		\end{minipage}
	}
	\subfigure[$t^*=0.179$]{
		\begin{minipage}[t]{0.22\linewidth}
			\centering
			\includegraphics[width=1.04\columnwidth,trim={3cm 0cm 3cm 0cm},clip]{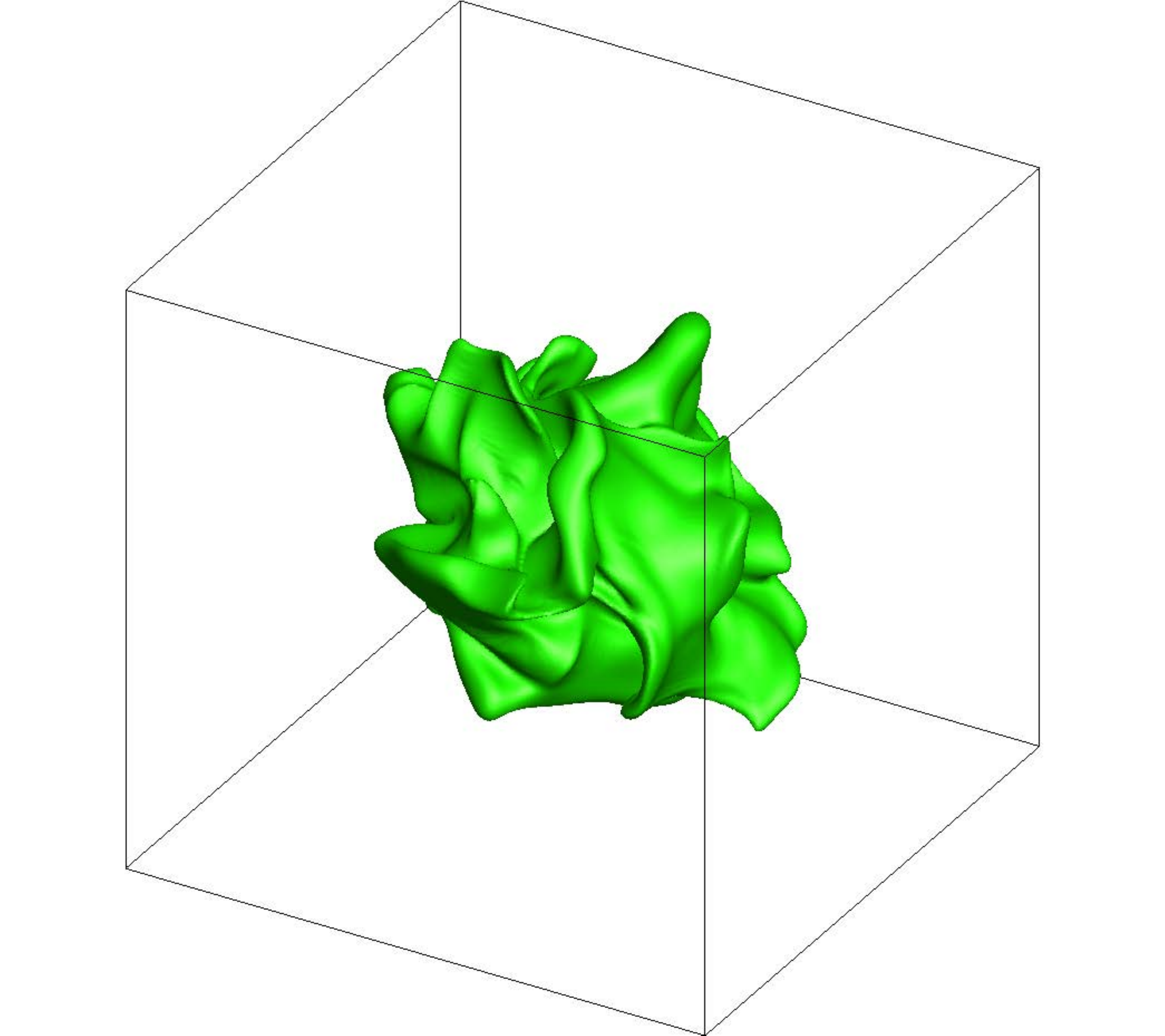}
			\label{7000}
		\end{minipage}
	}
	\subfigure[$t^*=0.403$]{
		\begin{minipage}[t]{0.22\linewidth}
			\centering
			\includegraphics[width=1.04\columnwidth,trim={3cm 0cm 3cm 0cm},clip]{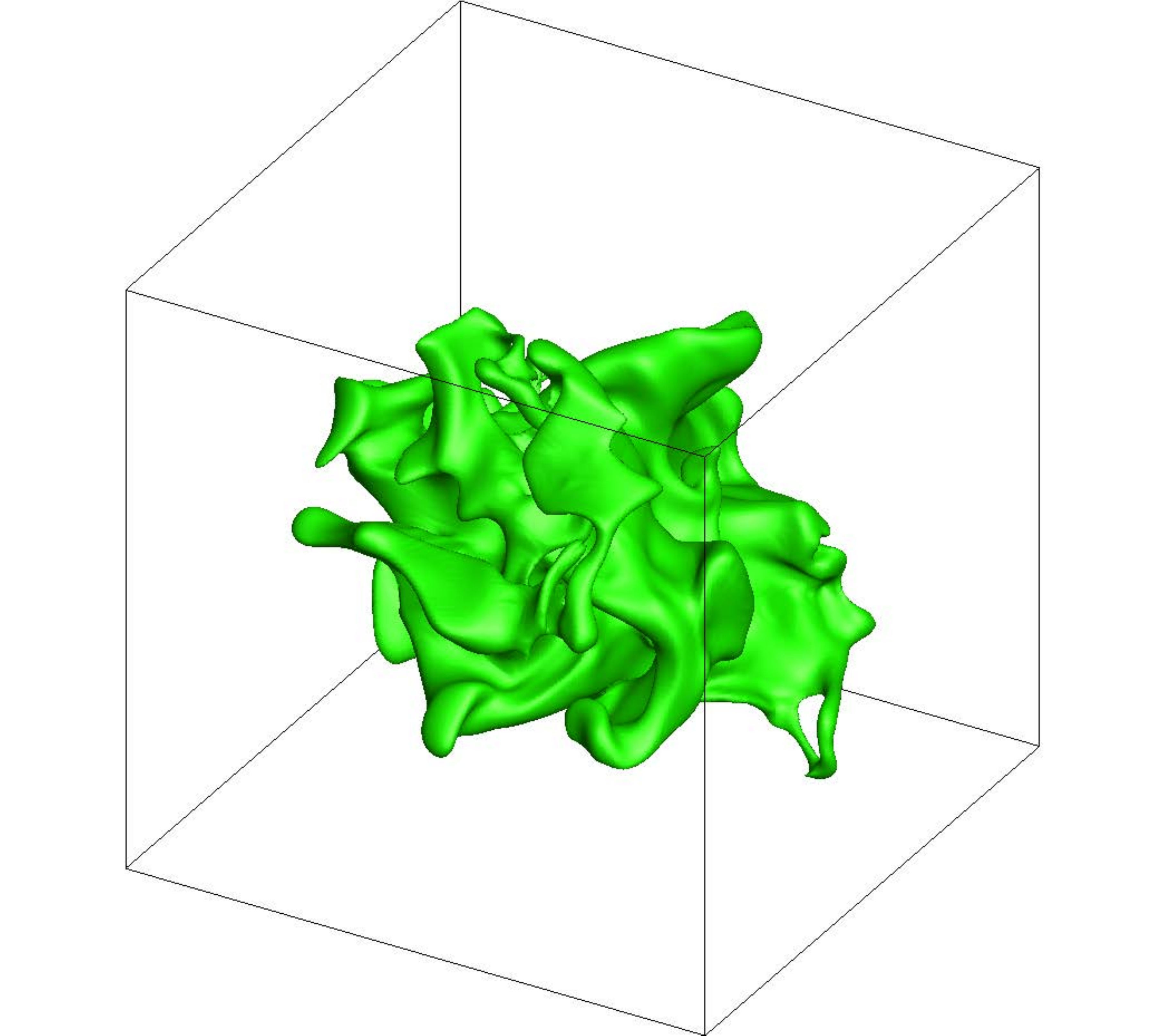}
			\label{12000}
		\end{minipage}
	}
	\\
	\subfigure[$t^*=0.896$]{
		\begin{minipage}[t]{0.22\linewidth}
			\centering
			\includegraphics[width=1.04\columnwidth,trim={3cm 0cm 3cm 0cm},clip]{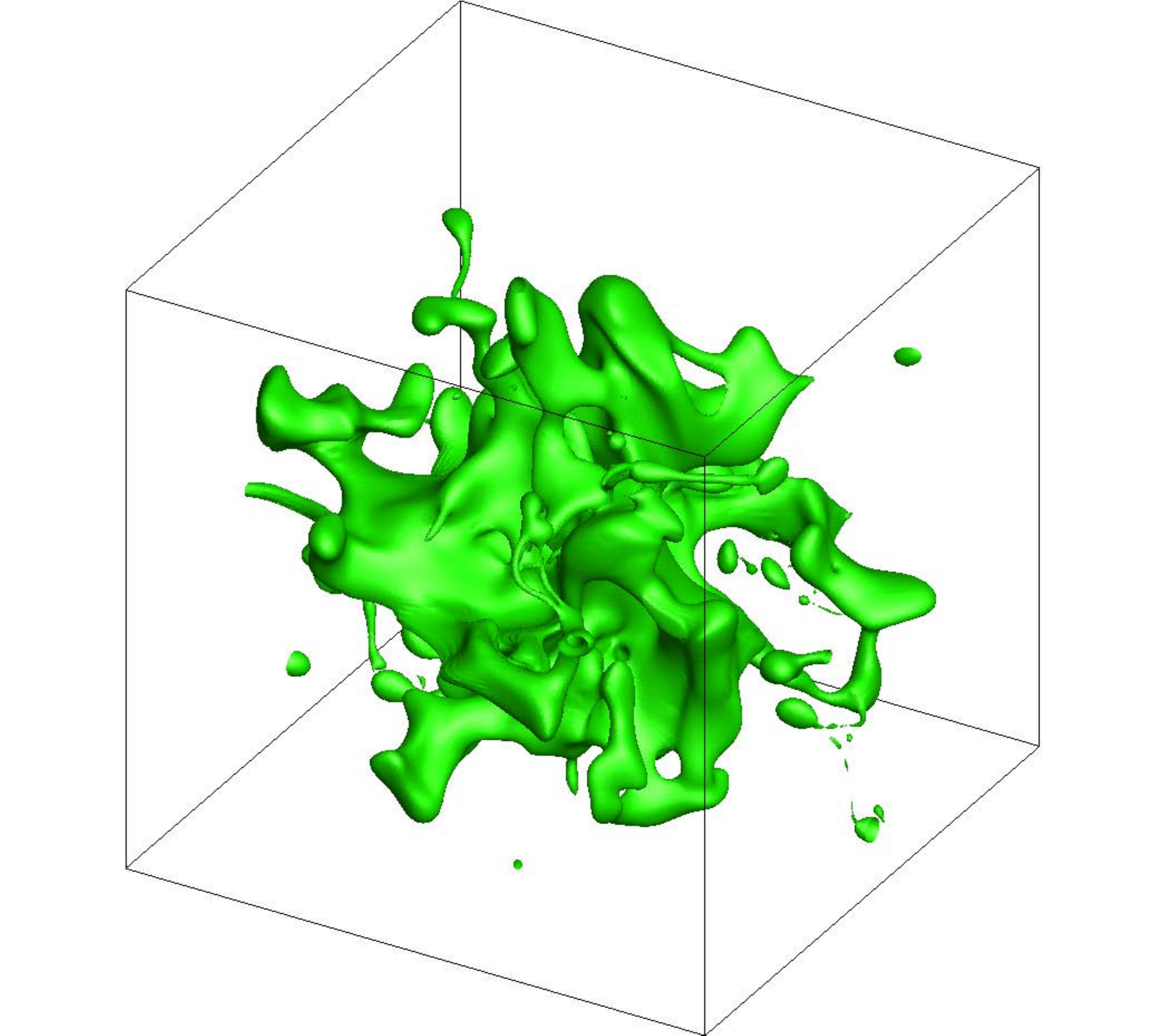}
			\label{23000}
		\end{minipage}
	}
	\subfigure[$t^*=1.21$]{
		\begin{minipage}[t]{0.22\linewidth}
			\centering
			\includegraphics[width=1.04\columnwidth,trim={3cm 0cm 3cm 0cm},clip]{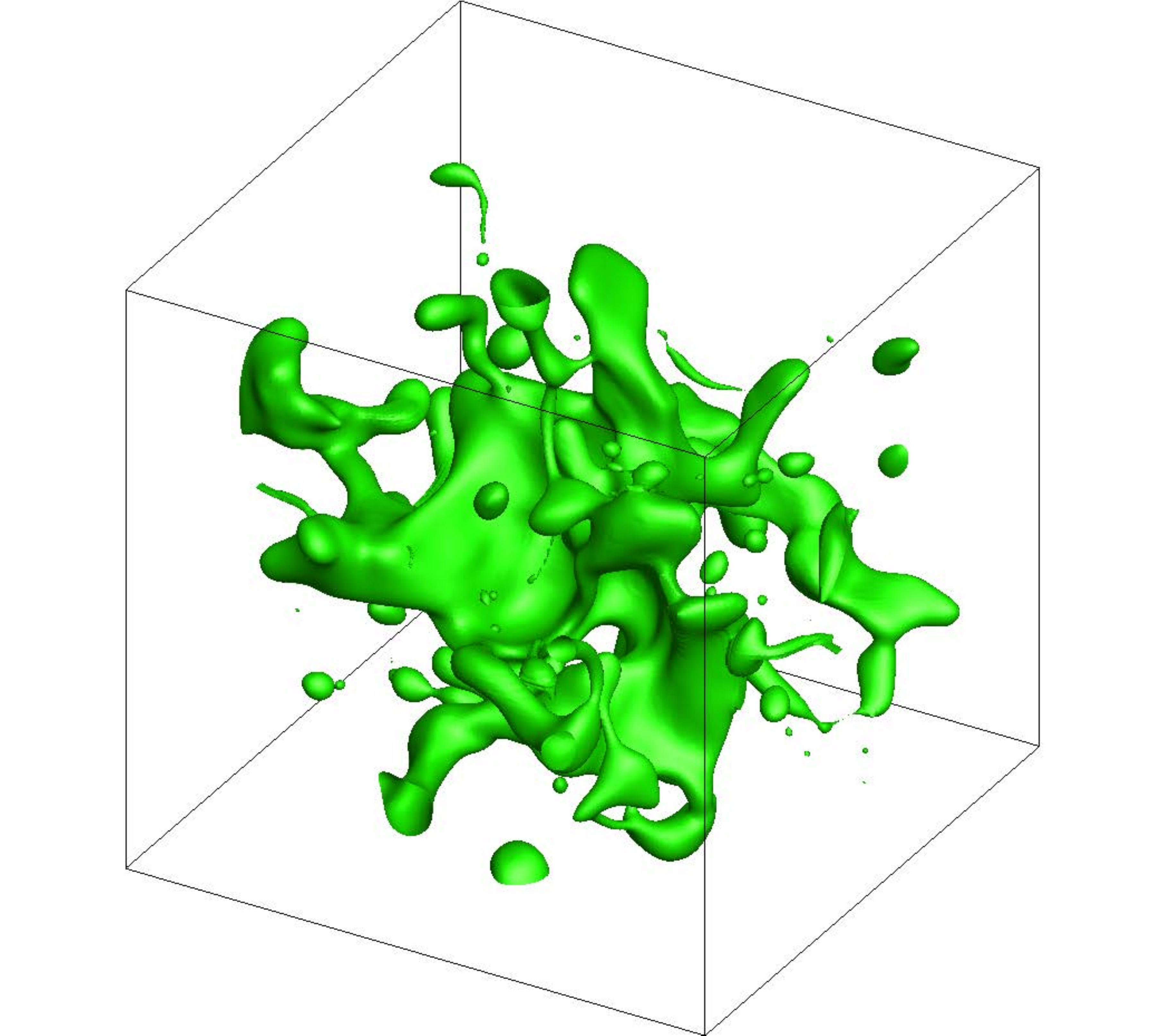}
			\label{30000}
		\end{minipage}
	}
	\subfigure[$t^*=1.57$]{
		\begin{minipage}[t]{0.22\linewidth}
			\centering
			\includegraphics[width=1.04\columnwidth,trim={3cm 0cm 3cm 0cm},clip]{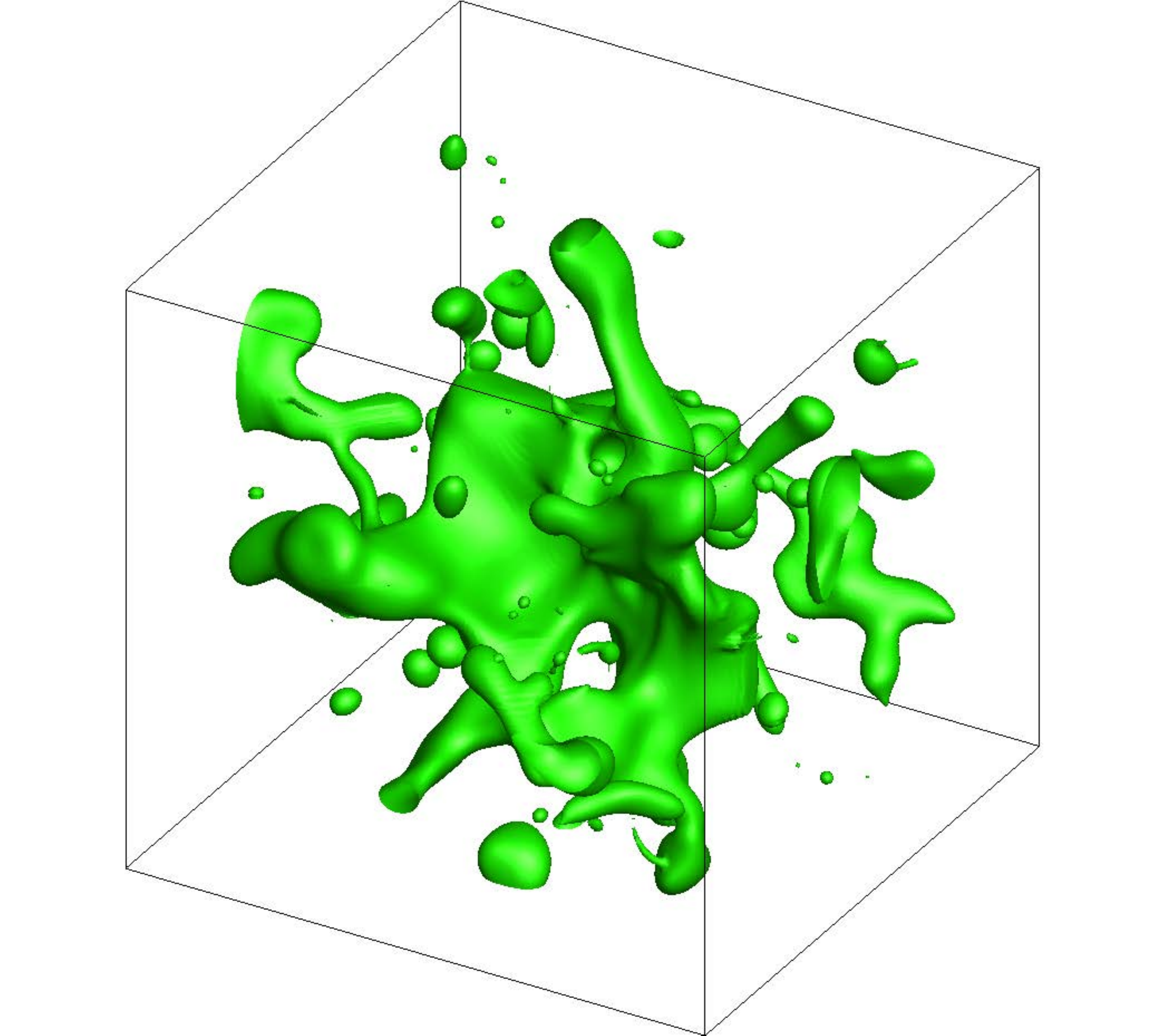}
			\label{38000}
		\end{minipage}
	}
	\subfigure[$t^*=5.15$]{
		\begin{minipage}[t]{0.22\linewidth}
			\centering
			\includegraphics[width=1.04\columnwidth,trim={3cm 0cm 3cm 0cm},clip]{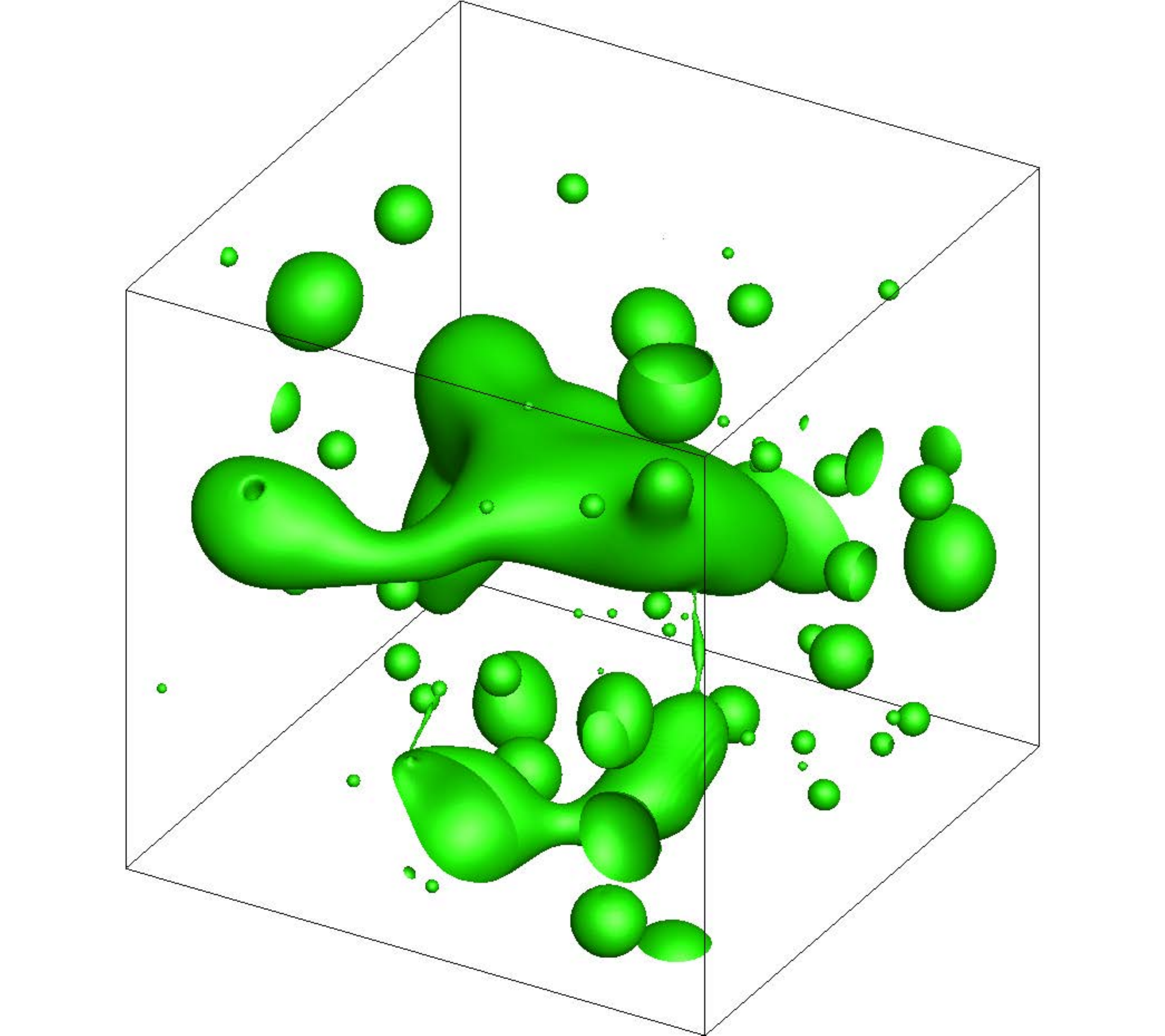}
			\label{118000}
		\end{minipage}
	}
	\\
	\subfigure[$t^*=7.84$]{
		\begin{minipage}[t]{0.22\linewidth}
			\centering
			\includegraphics[width=1.04\columnwidth,trim={3cm 0cm 3cm 0cm},clip]{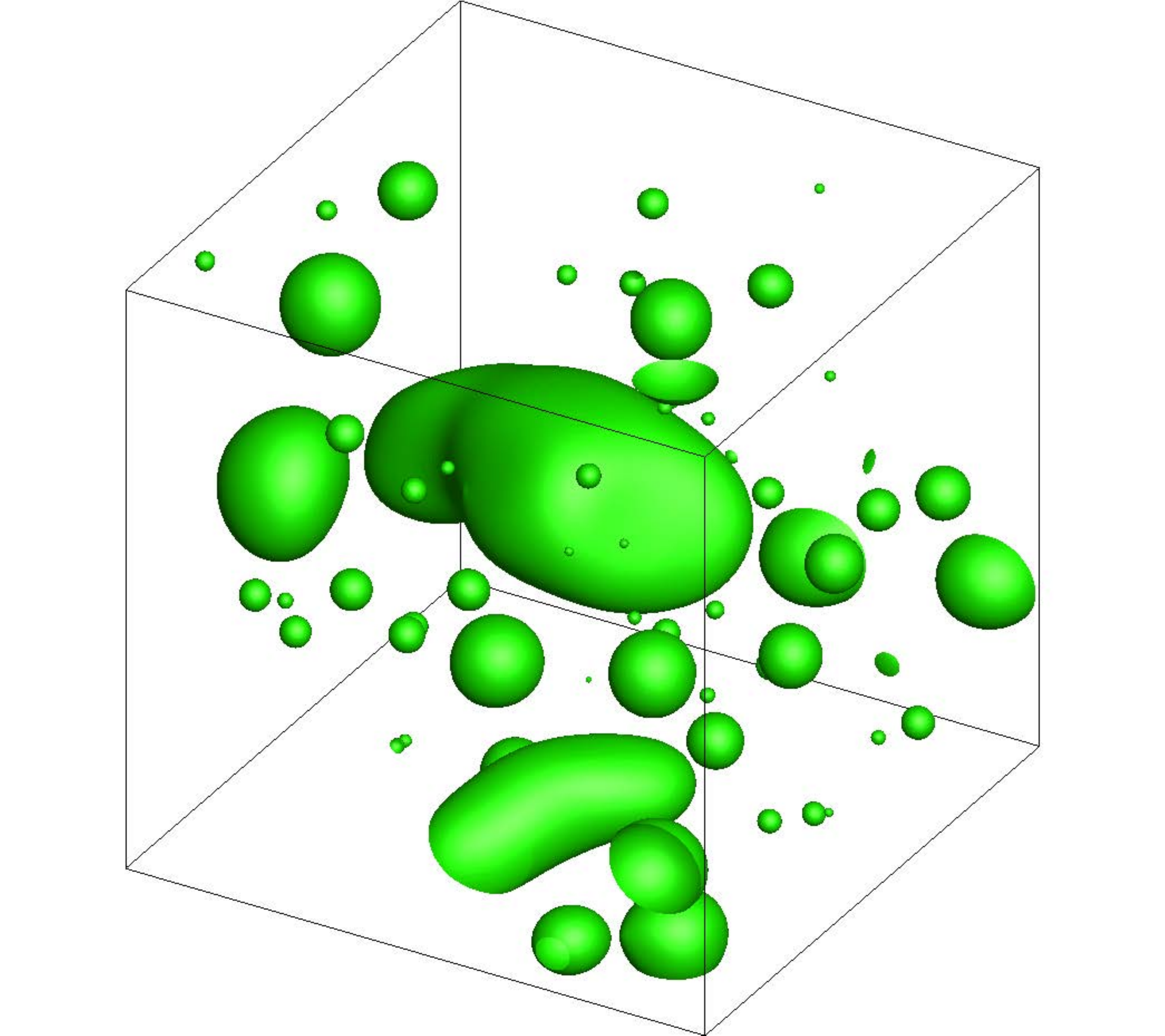}
			\label{178000}
		\end{minipage}
	}
	\subfigure[$t^*=11.5$]{
		\begin{minipage}[t]{0.22\linewidth}
			\centering
			\includegraphics[width=1.04\columnwidth,trim={3cm 0cm 3cm 0cm},clip]{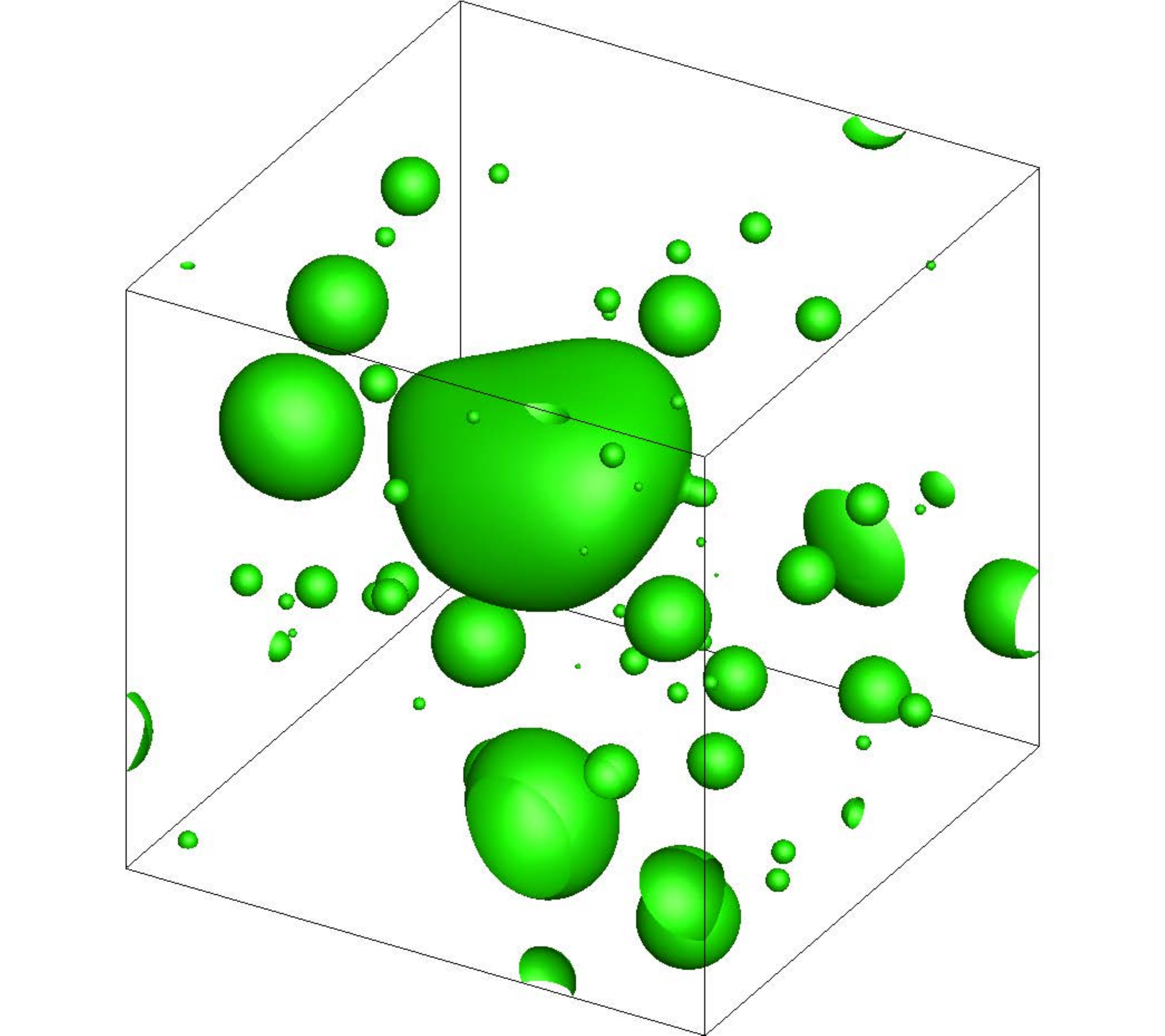}
			\label{259000}
		\end{minipage}
	}
	\subfigure[$t^*=14.6$]{
		\begin{minipage}[t]{0.22\linewidth}
			\centering
			\includegraphics[width=1.04\columnwidth,trim={3cm 0cm 3cm 0cm},clip]{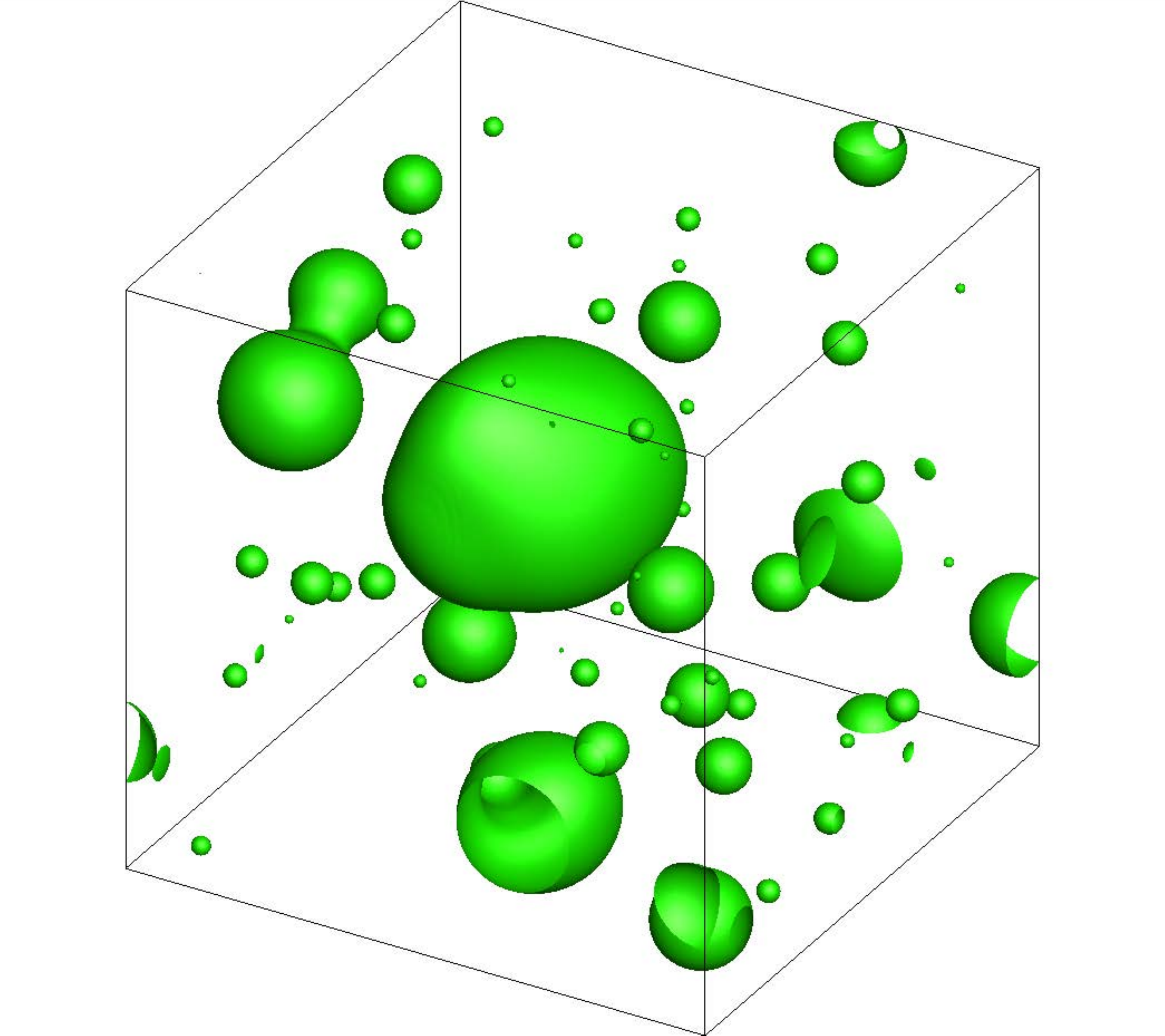}
			\label{330000}
		\end{minipage}
	}
	\subfigure[$t^*=22.4$]{
		\begin{minipage}[t]{0.22\linewidth}
			\centering
			\includegraphics[width=1.04\columnwidth,trim={3cm 0cm 3cm 0cm},clip]{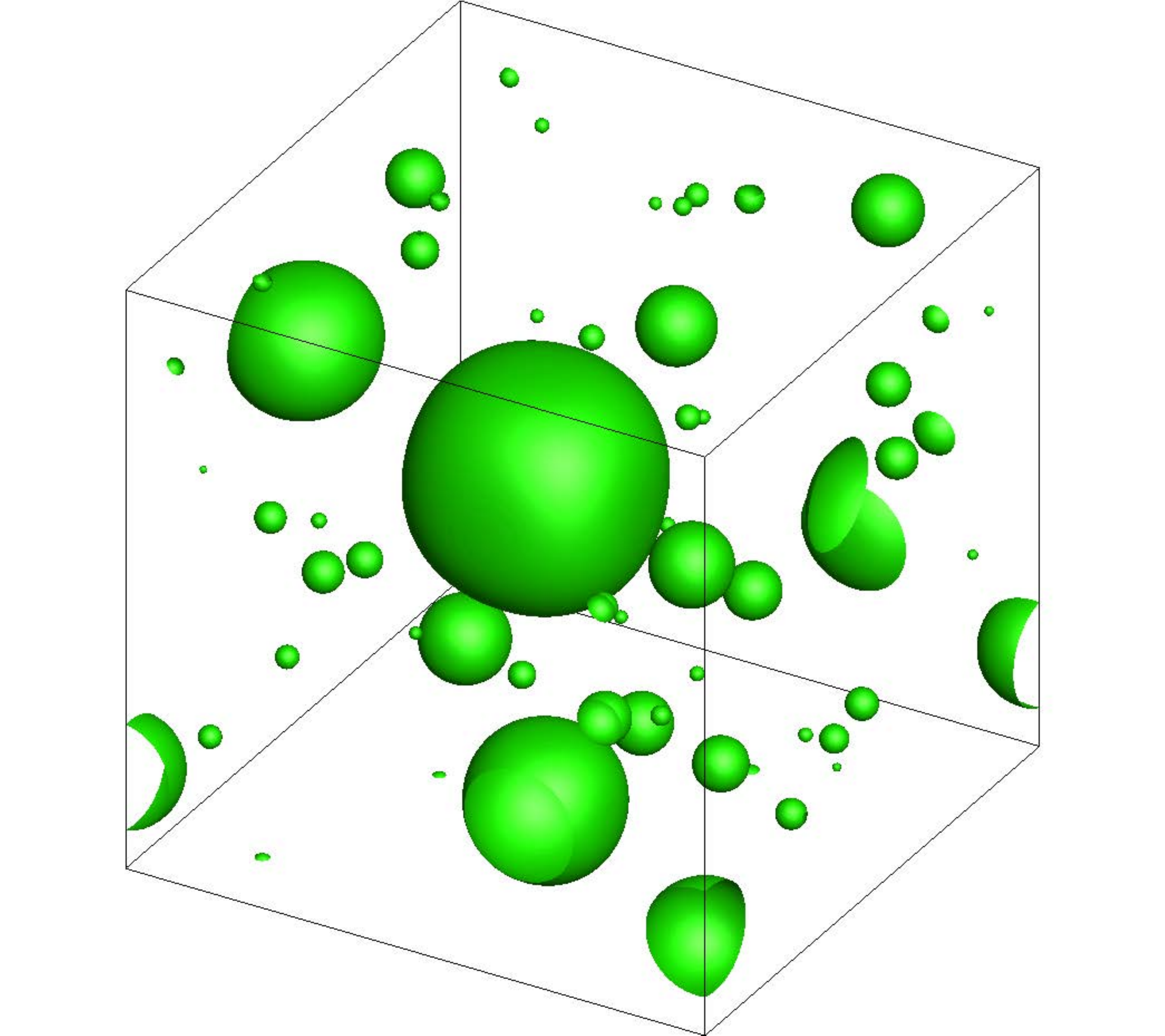}
			\label{503000}
		\end{minipage}
	}
	\\
	\begin{minipage}[t]{0.15\linewidth}
		\centering
		\includegraphics[width=0.6\columnwidth,trim={2cm 4cm 2cm 2cm},clip]{pngpdf/xyz1.pdf}
	\end{minipage}
	\centering
	\caption{3D visualization of a droplet breakup in a DHIT flow at different normalized times. The surface is given by $\phi = 0.5$.}
	\label{iso}
\end{figure}

\begin{figure}[t!]
	\centering    
	\subfigure[log-log plot]{
		\begin{minipage}[t]{0.47\linewidth}
			\centering
			\includegraphics[width=1.\columnwidth,trim={0cm 0cm 0cm 0cm},clip]{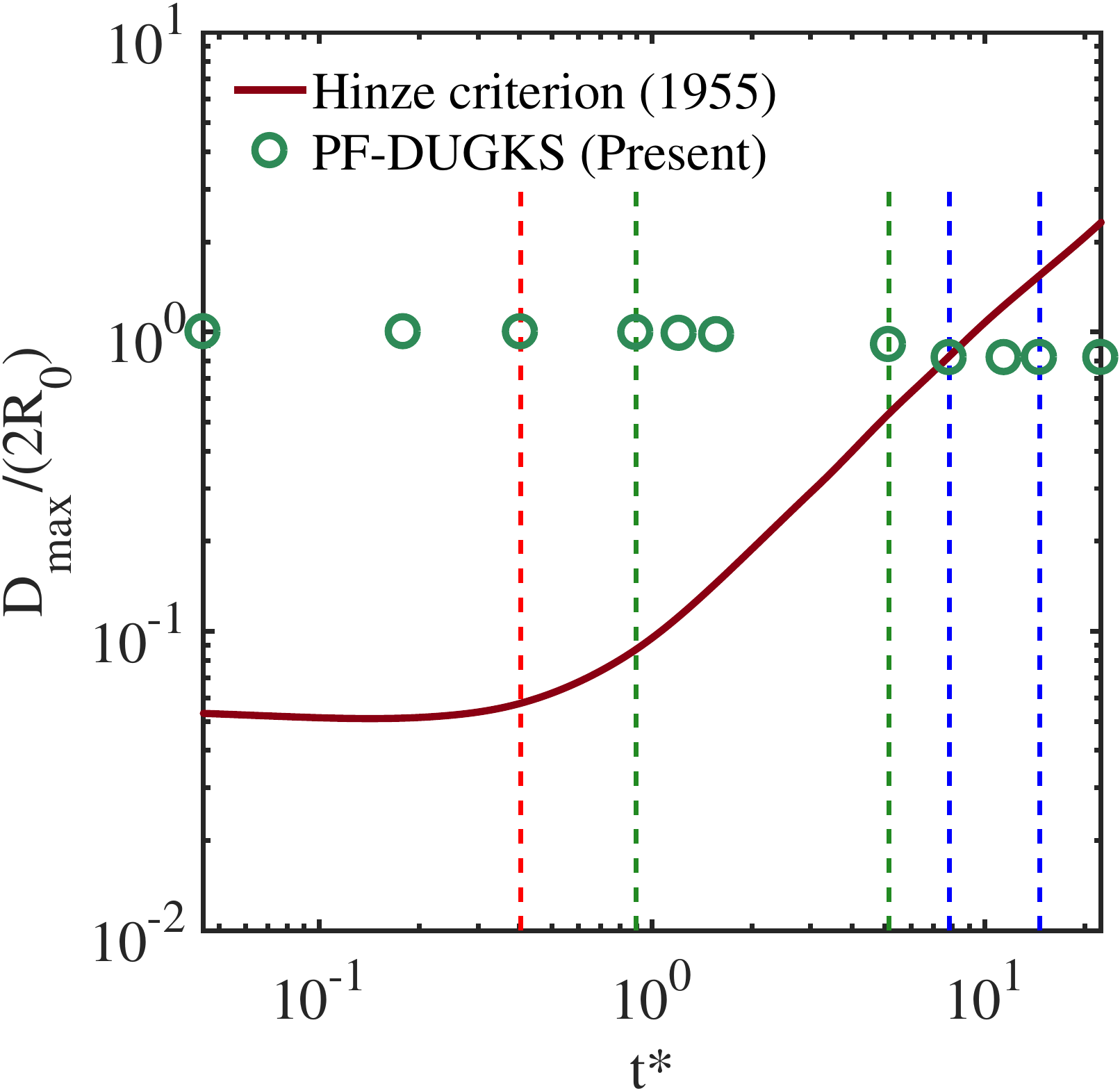}
			\label{Dlog}
		\end{minipage}
	}
	\subfigure[linear-linear plot]{
		\begin{minipage}[t]{0.47\linewidth}
			\centering
			\includegraphics[width=1.\columnwidth,trim={0cm 0cm 0cm 0cm},clip]{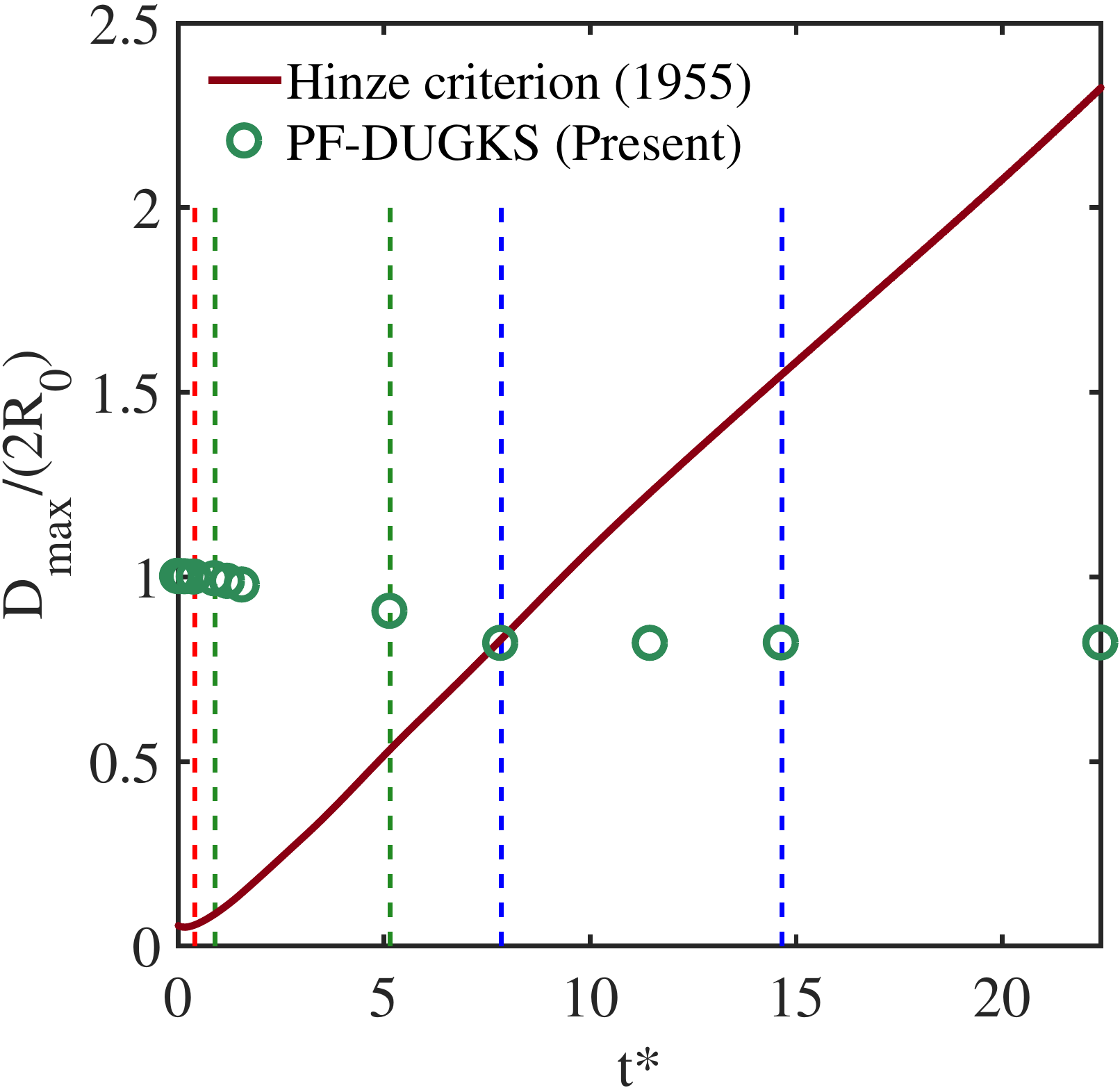}
			\label{Dplot}
		\end{minipage}
	}
	\centering
	\caption{The normalized maximum droplet diameters $D_{\max}$ at different times in Fig.~\ref{iso}, comparing with the Hinze criterion. The different times correspond to the times in Fig.~\ref{iso}. $R_0=64.0$ is the radius of the initial droplet. The red, dark green, blue vertical dash lines mark typical times for deformation ($ t^* = 0.403 $), breakup ($ t^* = 0.896$ and $5.15$), restoration  ($ t^* = 7.84$ and $14.6$), respectively.}
	\label{Dmax}
\end{figure}

Fig.~\ref{iso} shows the evolution of the droplet shape at different dimensionless times, where $t^*=t\epsilon\left( 0\right) /K\left( 0\right)$. The initial shape is a sphere located at the center of the box. This droplet may deform and break up due to the turbulent background flow.
As noted by Qian {\it et al.}~(2006)~\cite{qian2006simulation} and Albernaz {\it et al.}~(2017),~\cite{albernaz_do-quang_hermanson_amberg_2017} no simple criterion related to breakup could be identified. Hinze~(1955)~\cite{hinze1955fundamentals} argued that the competition of capillary pressure and pressure fluctuations due
to turbulent eddies 
determines the characteristic diameter $D_{c}$ of the droplets in turbulence. Assuming the inertial-subrange scaling, he obtained 
that $D_{\text{Hinze}}=C \left( \sigma/ \rho \right)^{3/5}  \epsilon^{-2/5}$ in HIT background flow field. This criterion can also be derived based on the dimensional analysis, if one assumes that only the physical quantities $\rho, \sigma, \epsilon$ determine $D_{\text{Hinze}}$.~\cite{hinze1955fundamentals} Here $C=0.725$ based on Clay~(1940)'s experimental observatons.~\cite{clay1940mechanism,hinze1955fundamentals}
While the flow Reynolds number in our simulation is too low to have the inertial subrange, we will nevertheless apply the
Hinze criterion as an estimate.
According to the Hinze criterion, at $t^*=0$,  the maximum droplet diameter that does not undergo
breakup can be estimated to be $D_{\text{Hinze}}\sim 7.20$ in our simulation (see Fig.~\ref{Dmax}). 
It is noted that $D_{\max}$ in Fig.~\ref{Dmax} is computed by first obtaining the volume of the largest inter-connected
$\phi \ge 0.5$ region, then $D_{\max}\equiv \sqrt[3]{6V_{\max}/\pi}$.
Clearly, the droplet with $D_0 = 128.0$ placed in the flow can certainly break up.
Indeed, as seen in Fig.~\ref{iso}, the droplet surface becomes wrinkled and stretched, and breaks down into many small irregular parts. According to Fig.~\ref{iso}, a spherical droplet breakup in DHIT can be divided into three stages: (1) the deformation stage, where the large spherical droplet evolves into an irregular geometric pattern with complex interface structures, (2) the breakup stage, during which many small droplets are separated from the large one, (3) the restoration stage, during which all the droplets become spherical or merge with the nearby droplets, finally reaching a quasi-stationary state. The deformation stage takes the shortest time, while the restoration stage takes the longest time, as the kinetic energy decreases with time in a power-law form in DHIT. 
Hence the logarithmic coordinates are usually used for time in the figures for this physical problem.

\begin{figure}[t!]
	\centering
	\includegraphics[width=0.7\columnwidth,trim={0cm 0cm 0cm 0cm},clip]{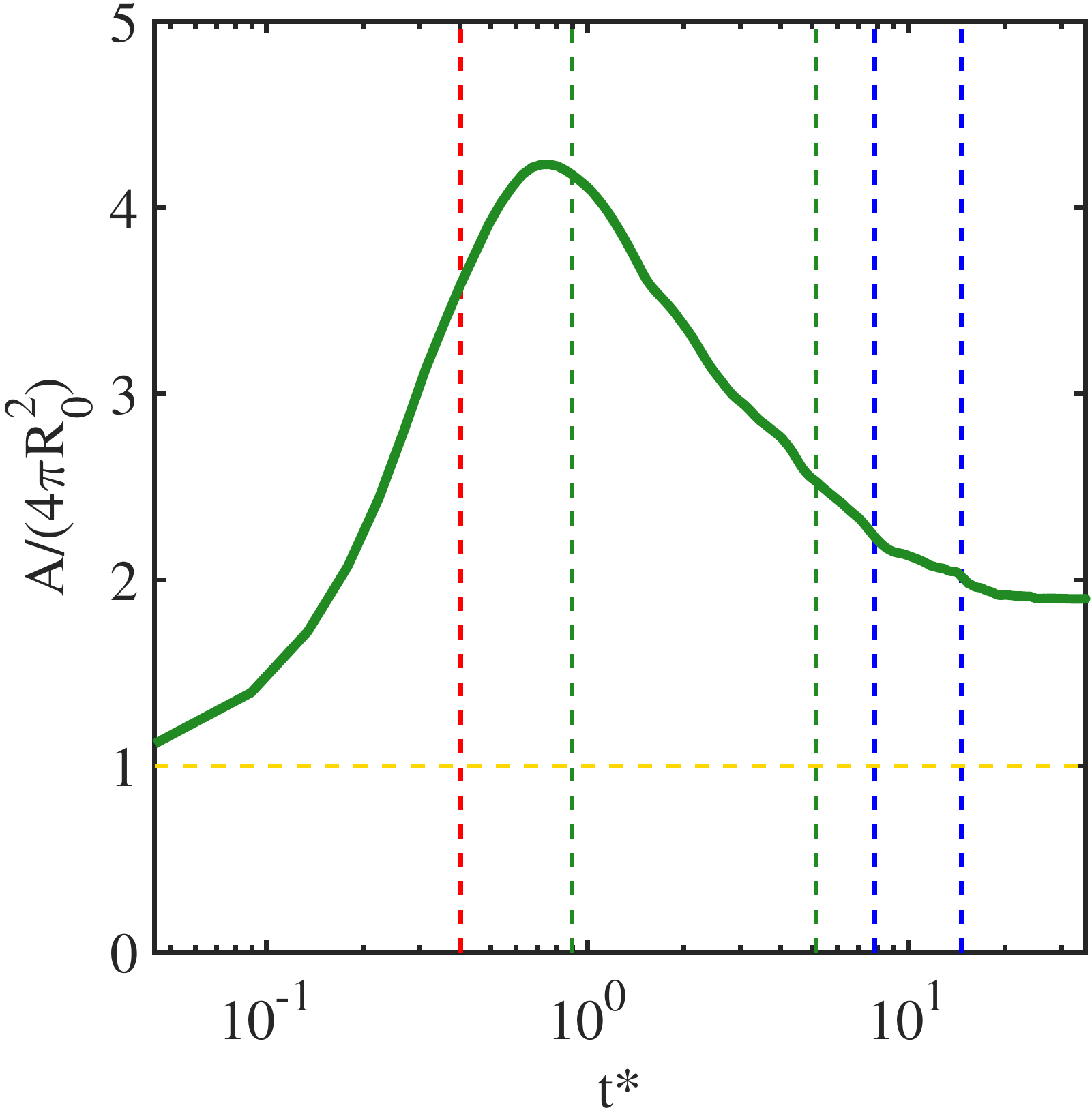}
	\centering
	\caption{Evolution of the normalized total interface area. The total area $A$ is the area of isosurface $\phi=0.5$. Initially, $A(t^*=0)=4\pi R_0^2$.
		The dashed yellow line is the constant 1. The vertical lines  mark  $t^*$=0.403,0.896,5.15,7.84,14.6, respectively.}
	\label{arcc}
\end{figure}

\begin{figure}[t!]
	\centering
	\includegraphics[width=0.55\columnwidth,trim={0cm 0cm 0cm 0cm},clip]{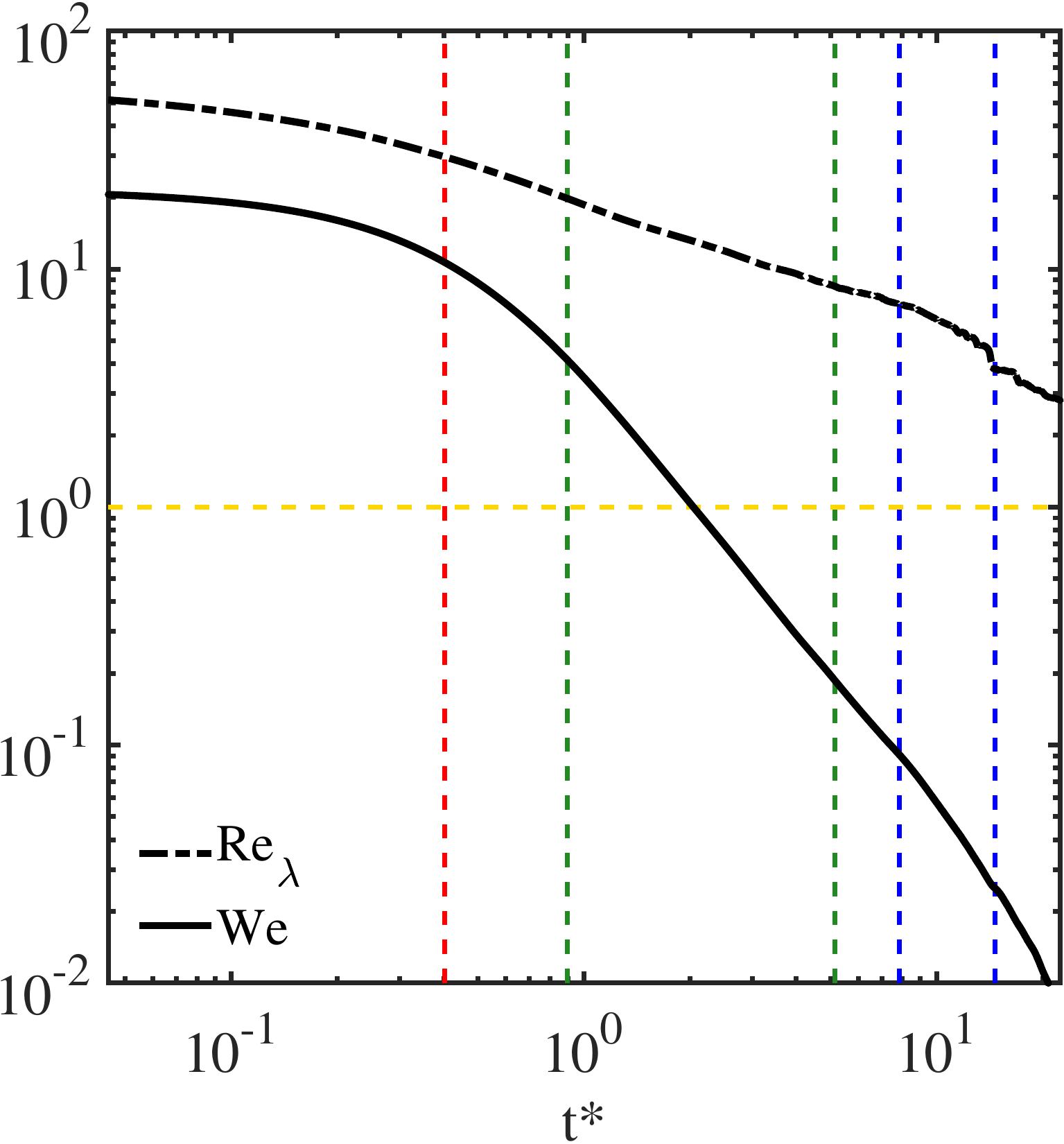}
	\centering
	\caption{Evolution of the dimensionless parameters $Re_\lambda$ and $We$. Here $We = \rho u^{\prime 2}R_0/\sigma$. The dashed yellow line marks the constant one. The red, dark green, blue vertical dashed lines mark typical times for deformation ($ t^* = 0.403 $), breakup ($ t^* = 0.896$ and $5.15$), restoration  ($ t^* = 7.84$ and $14.6$), respectively.}
	\label{ReWe}
\end{figure}

The first row in Fig.~\ref{iso} (Fig.~\ref{3000}-\ref{12000}) is at the deformation stage. 
The turbulent kinetic energy disrupts the  free energy equilibrium, augmenting the free energy of the system through increased interface area.
The initial smooth spherical droplet develops irregular wrinkles on the whole interface. Then the wrinkles gradually elongate and intertwine in all directions, sometimes creating holes and changing the topological structure. At the end of this stage, the geometric structure is very complicated and the interface area is increased by a factor about 4.2 comparing to the initial state (see Fig.~\ref{arcc}). It is noted that $\We=10.66$ 
and $Re_{\lambda}=30$  at $t^* =0.403$ (see Fig.~\ref{12000} and Fig.~\ref{ReWe}). Therefore, during the first stage  
the velocity and pressure fluctuations in turbulence usually serve as the driving mechanism for droplet deformation. The surface tension force and viscous force act as a resistance to deformation. 

The second row in Fig.~\ref{iso} (Fig.~\ref{23000}-\ref{118000}) shows the interface evolution at the breakup stage. The final state of the deformation stage provides adequate conditions for this breakup stage. The local volume distribution of the droplet phase is very nonuniform. Then regions where the cross-sectional areas of the droplet phase are small, may separate to generate daughter droplets. The factors that affect fragmentation include the following four aspects: (1) the fluid inertia developed during the deformation stage, which will continue to elongate the interface resulting in fragmentation, (2)  the local pressure
inside the droplet near the narrow neck is larger due to large local interface curvature, causing a local interfacial
instability to develop which eventually leads to neck breakup (Similar idea was proposed by Qian {\it et al.}~(2006),~\cite{qian2006simulation} where they suggested that the capillarity effect causes breakup in a highly elongated region),
(3) the velocity and pressure fluctuations also influence the breakup process, during which 
the droplet-breakup may be locally promoted or suppressed depending on the nature of
the fluctuations,
(4) viscous force acts to retard the breakup process. At the same time, the daughter droplets become convex and spherical in the breakup stage due to the influence of surface tension acting on decreasing the interface area, 
as shown in Fig.~\ref{arcc}. The competition of inertial force/viscous force and inertial force/surface tension force are presented by $Re_\lambda$ and $\We$ in Fig.~\ref{ReWe}, respectively. The dominant factor is the inertial force at the beginning of this stage, and is switched to the surface tension force at the end of this stage.
As a result, this stage is the most complicated among the three stages. We can observe many small structures in Fig.~\ref{23000}, resolved by the adequate grid resolution applied.

The third row in Fig.~\ref{iso} (Fig.~\ref{178000}-\ref{503000}) shows the interface evolution at the restoration stage. 
Take $t^*=7.84$ (see Fig.~\ref{178000} and Fig.~\ref{Dmax}), $D_{\text{Hinze}}\sim  106.33$, while the largest droplet diameter in the simulation is $D_{\max}\sim \sqrt[3]{6V_{\max}/\pi}= 104.84$. This implies that the breakup may stop around this time.
The velocity and pressure fluctuations are now
small because of the decay of kinetic energy, they then have less influence on the motion of droplets. The droplets cannot break up any more, while continue to become convex and 
spherical due to decreasing Weber numbers (Fig.~\ref{ReWe}). 
Fig.~\ref{ReWe} also shows that the surface tension force is the driving force at this stage.
The viscous force still acts as the resistance. When two droplets are close to each other, the smaller one could be merged into the larger one. Fig.~\ref{259000} shows the coalescence of a large 
droplet and a small droplet, and 
in Fig.~\ref{330000} the coalescence of two medium-sized droplets are observed, during that time the total interface area will change a little bit as the droplet shape changes.
The merging phenomenon increases the local velocity magnitude in the region between the merging droplets 
(see Fig.~\ref{330000v}). The droplets will become quasi-stationary at the end of this stage, continuing to evolve and merge slowly.   The total interface area decreases due to restoration, 
while increases in the early part of coalescence process.
The overall trend is a decreasing interface area and
minimization of the free energy (see Fig.~\ref{arcc}).

Fig.~\ref{Dmax} demonstrates that our results on the maximum droplet diameter seem to  be in quantitative agreement with the Hinze criterion. The demarcation time between breakup stage and restoration stage can be defined by $t^*(D_{\max}=D_{\text{Hinze}})$, which is roughy 7.8 according to 
Fig.~\ref{Dmax}. Furthermore, Fig.~\ref{Dmax} shows that the maximum droplet diameter does not
change much during the whole evolution in DHIT, only about 0.8 times of the initial diameter, which means the maximum droplet volume at the end of the simulation is roughly half of the initial droplet volume.

The above three stages represent the competition among three factors: the turbulent fluctuations, the viscous force, and the surface tension force. The likely contributions of these factors to the three stages are summarized in Table \ref{TabStages}.

\begin{table*}[]
	\centering
	\caption{The influence of various factors on different stages}
	\label{TabStages}
	\begin{tabular}{cccc}
		\toprule
		\multirow{2}{*}{} & \multicolumn{1}{c}{deformation stage} & \multicolumn{1}{c}{breakup stage}   & \multicolumn{1}{c}{restoration stage}\\
		\midrule
		turbulence (inertial force)
		&$+(-)$&$+(-)$& $-(+)$\\
		viscous force          &$-$&$-$&$-$\\
		surface tension force        &$-(+)$&$-(+)$&$+$\\
		\bottomrule
	\end{tabular}\\
	{'$+$' represents 'promote', '$-$' represents 'prevent'. '$+(-)$' represents 'usually promote but sometimes prevent'.}\\
\end{table*}

\subsection{Evolution of the number of droplets and droplet diameter distribution}

The variation of the total number of droplets as a function of time is shown in Fig.~\ref{Nd}.  
Here we developed an algorithm to identify individual droplets (see Appendix~\ref{sec: Droplets}). A droplet is defined 
as a set of inter-connected lattice nodes with $\phi \ge 0.5$, from which the volume and the equivalent diameter
of the droplet can be determined.
Droplets of volume one are not considered in the counting as they could not be resolved by the phase field model.
At the deformation stage, the number $N_d$ is on the order of 1, since the initial droplet is spherical and
requires time to deform and break up. During the deformation stage, small droplets may detach from the large droplet. Then the  value of $N_d$ may be larger than one at the end of this stage. $N_d=4$ at $t^*=0.403$, which means there are three small droplets being generated by this time. At the breakup stage, many small daughter droplets are formed, and $N_d$ quickly reaches the order of 10. $N_d$ continues to increase to about 60 at the end of the breakup stage.
During the restoration stage, $N_d$  tends to decrease due to the coalescence events.
Roughly speaking, $dN_d/dt^*\ge 0$ at the deformation and breakup stages, and $dN_d/dt^*\le 0$ at the restoration stage.
Perlekar~{\it et al.}~(2012)~\cite{perlekar2012droplet} also show $N_d$ varies with time in their droplet-breakup simulation at the FHIT background flow.
They point out that the volume fraction $\varphi$ and Reynolds number $Re_\lambda$ will affect $N_d$.
They found that $N_d$ was roughly around 10, between 20 and 50, between 30 and 70, for $\varphi=0.5\%,5\%,10\%$, respectively, with $Re_\lambda=15$ in their simulations.
Our result $N_d\sim 60$ with $\varphi=6.54\%$ and initial $Re_\lambda=58$ is reasonable compared to their results, 
since our flow Reynolds number is larger.

\begin{figure}[t!]
	\centering
	\includegraphics[width=0.65\columnwidth,trim={0cm 0cm 0cm 0cm},clip]{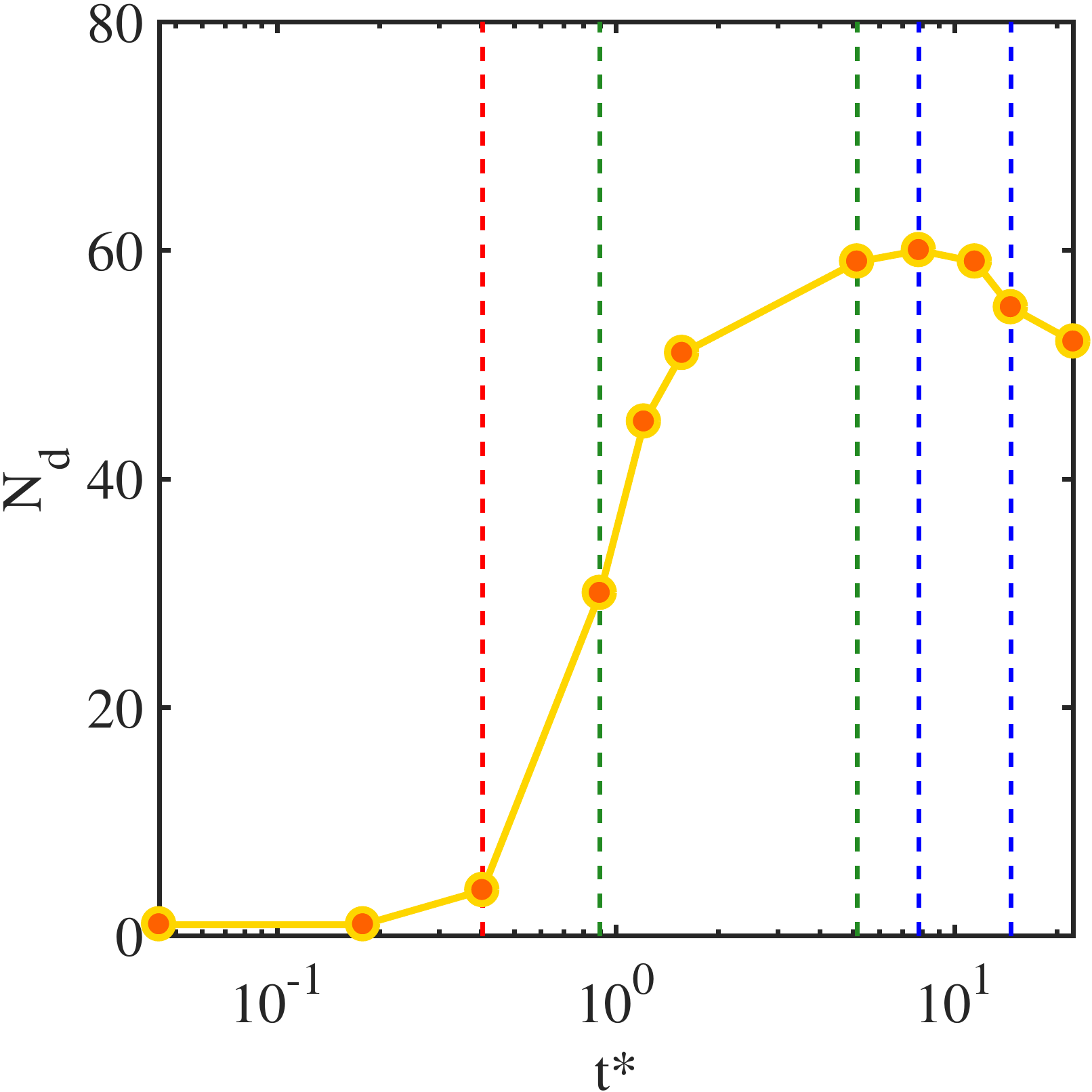}
	\centering
	\caption{The number of droplets ($N_d$) at different times in Fig.~\ref{iso}. The red, dark green, blue vertical dashed lines mark  typical times for deformation ($ t^* = 0.403 $), breakup ($ t^* = 0.896$ and $5.15$), restoration  ($ t^* = 7.84$ and $14.6$), respectively.}
	\label{Nd}
\end{figure}

\begin{figure}[t!]
	\centering    
	\subfigure[$t^*=1.21$]{
		\begin{minipage}[t]{0.4\linewidth}
			\centering
			\includegraphics[width=1.\columnwidth,trim={0cm 0cm 0cm 0cm},clip]{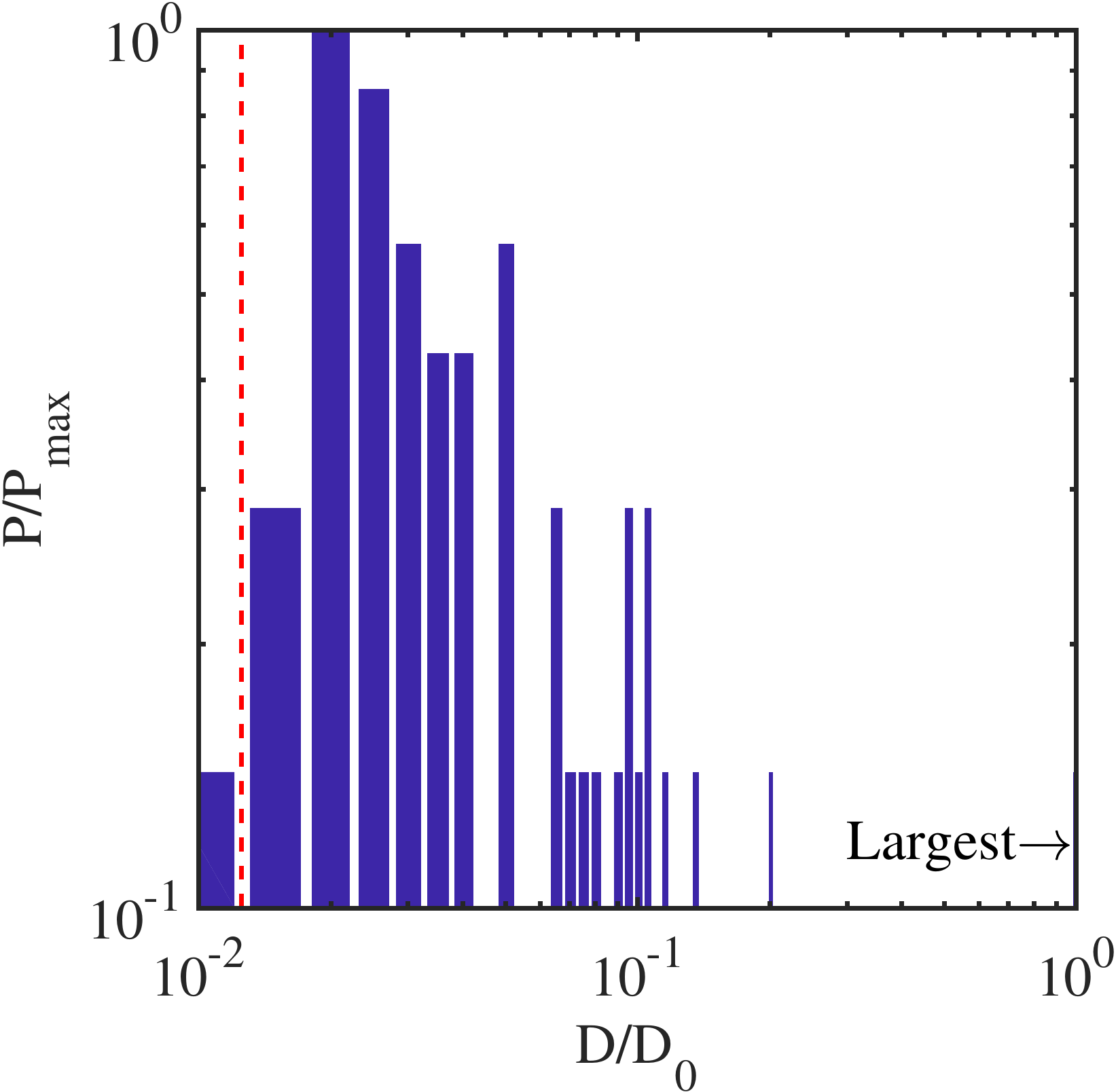}
			\label{PDFdroplets30000}
		\end{minipage}
	}
	\subfigure[$t^*=5.15$]{
		\begin{minipage}[t]{0.4\linewidth}
			\centering
			\includegraphics[width=1.\columnwidth,trim={0cm 0cm 0cm 0cm},clip]{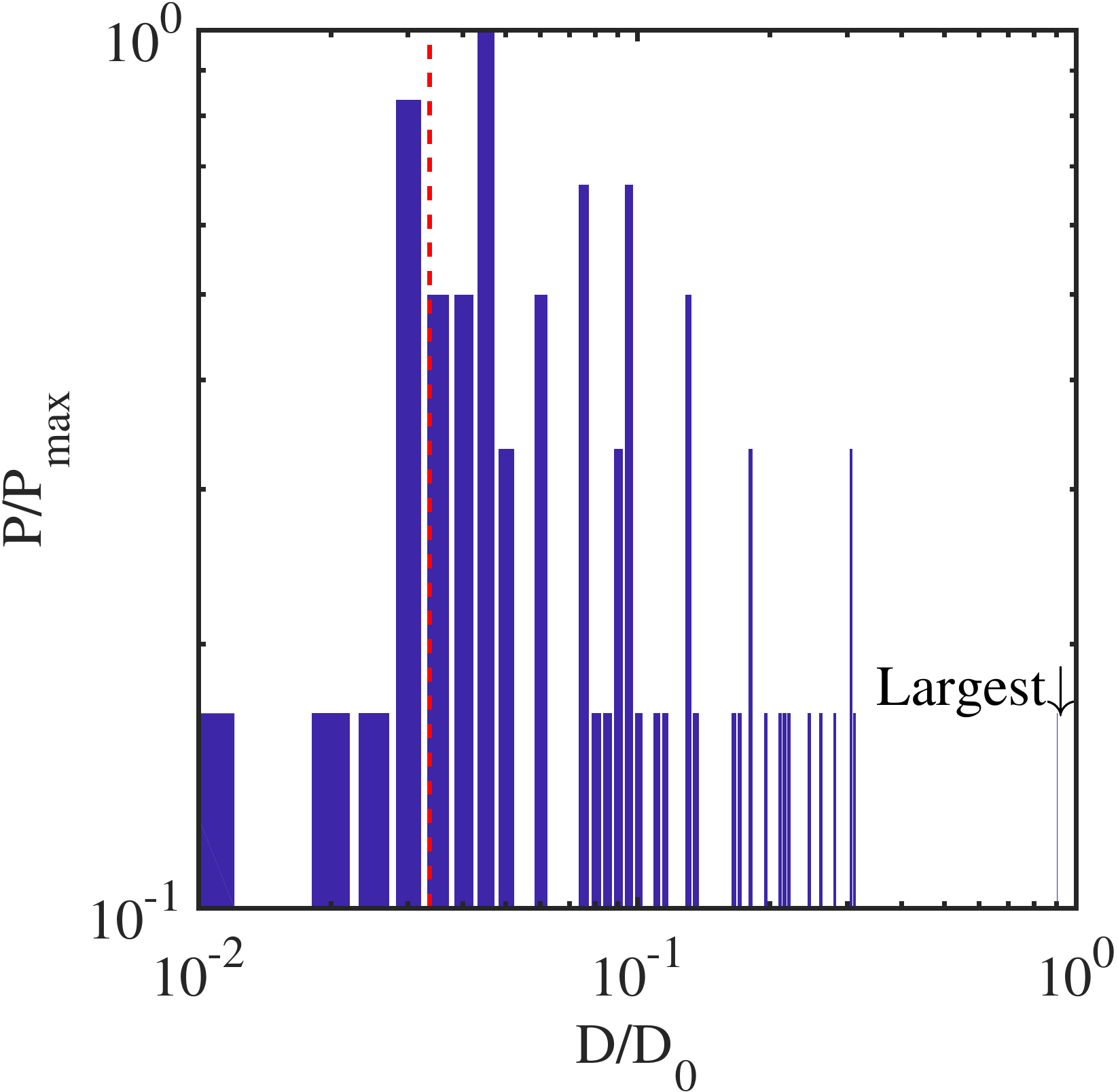}
			\label{PDFdroplets118000}
		\end{minipage}
	}\\
	\subfigure[$t^*=11.5$]{
		\begin{minipage}[t]{0.4\linewidth}
			\centering
			\includegraphics[width=1.\columnwidth,trim={0cm 0cm 0cm 0cm},clip]{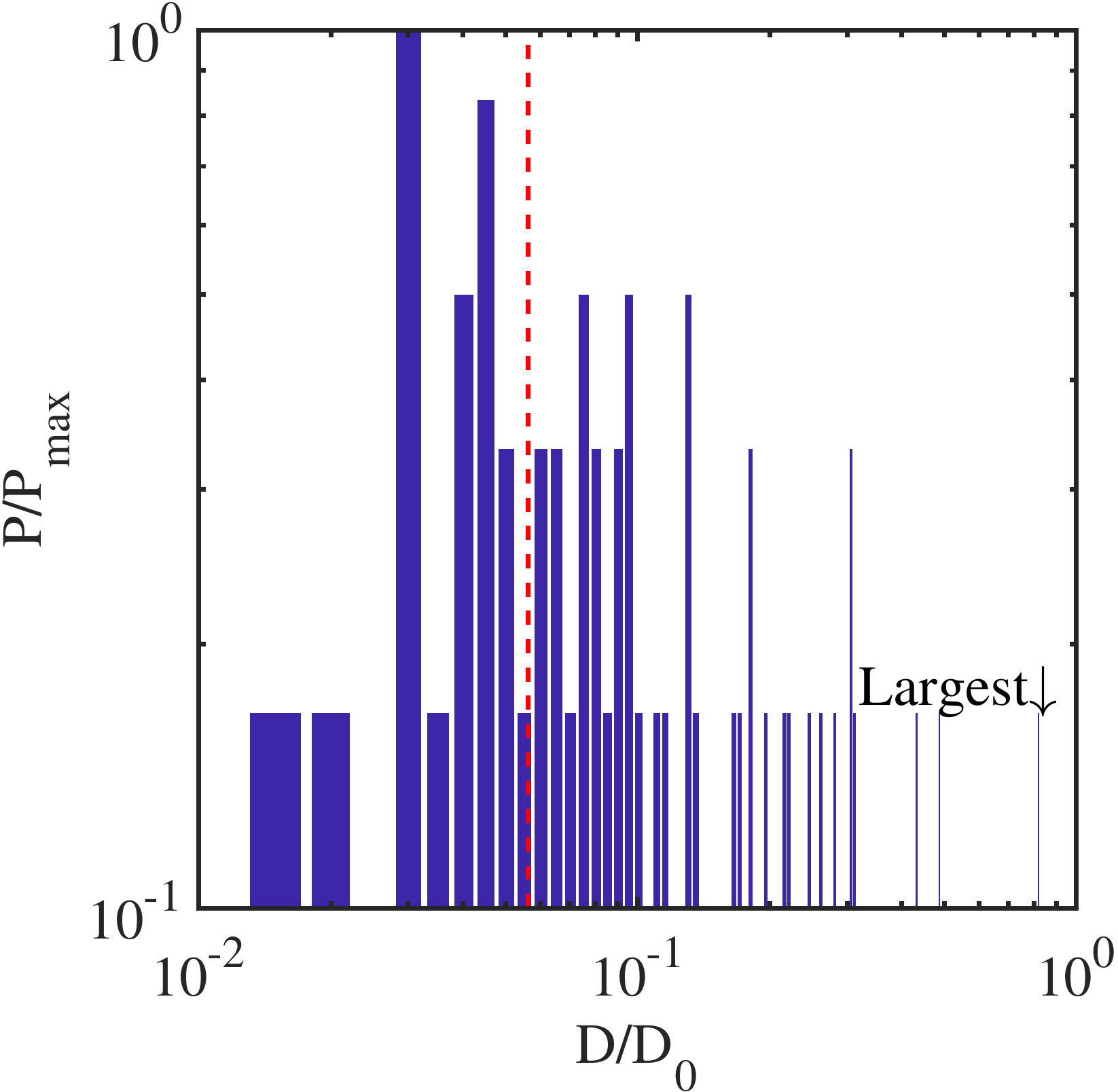}
			\label{PDFdroplets259000}
		\end{minipage}
	}
	\subfigure[$t^*=22.4$]{
		\begin{minipage}[t]{0.4\linewidth}
			\centering
			\includegraphics[width=1.\columnwidth,trim={0cm 0cm 0cm 0cm},clip]{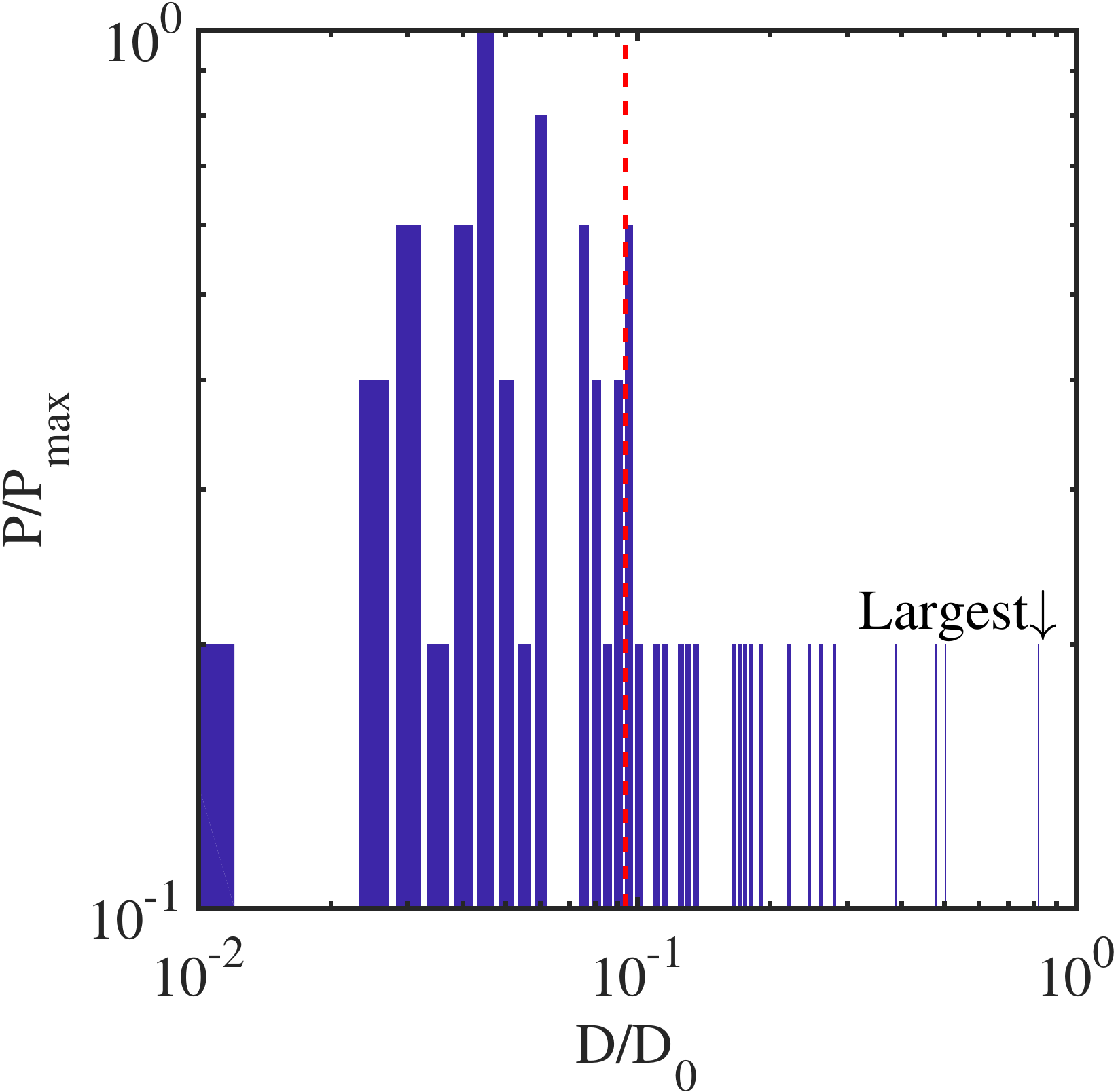}
			\label{PDFdroplets503000}
		\end{minipage}
	}
	\centering
	\caption{Probability distribution function (PDF) of droplet diameter at different times. The red dashed line represents $\eta/D_0$. The arrow points to the largest droplet.}
	\label{PDFdroplets}
\end{figure}

In order to observe the size distribution of droplets in the evolution process, we plot the probability distribution function of droplet radius in Fig.~\ref{PDFdroplets}. The diameter is calculated by $D\equiv \sqrt[3]{6V/\pi}$.
It is observed that, during the breakup stage (Figs.~\ref{PDFdroplets30000}-\ref{PDFdroplets118000}) 
and the  restoration stage (Figs.~\ref{PDFdroplets259000}-\ref{PDFdroplets503000}), most droplets have 
a diameter smaller than 10\% of the initial droplet diameter, namely, 
84.4\%, 62.7\%, 61.0\%, 63.5\% of droplets in number are with a diameter less than 
$0.1D_0$, at $t^*=1.21, 5.15, 11.5, 22.4$, respectively.
At $t^*=1.21$, the large droplet remains in one piece, and only the small droplets are formed from the
highly distorted parts of the interface, hence the ratio is quite large, over 80\%.
Later the medium-size droplets are separated from the main part of the large droplet, due to the strong deformation of the large droplet.
As a result, the ratio becomes smaller and reach a stable value around 60\%.
The diameter of the largest droplet is about 80\% of the initial $D_0$ in the end, so it is always close to 1 in the log-log plot (Fig.~\ref{PDFdroplets}) during the whole evolution.
The vertical red line marks the location of the Kolmogorov scale.
It is observed in Fig.~\ref{PDFdroplets} that more and more droplets become smaller than the Kolmogorov scale when time proceeds,
namely, 2.2\%, 13.6\%, 33.9\%, 57.7\% of the droplets are smaller than $\eta$ in diameter, respectively.

\subsection{Velocity and vorticity magnitudes on the fluid-fluid interface}

\begin{figure*}
	\centering    
	\subfigure[$t^*=0$]{
		\begin{minipage}[t]{0.22\linewidth}
			\centering
		\includegraphics[width=1.05\columnwidth,trim={3cm 0cm 3cm 0cm},clip]{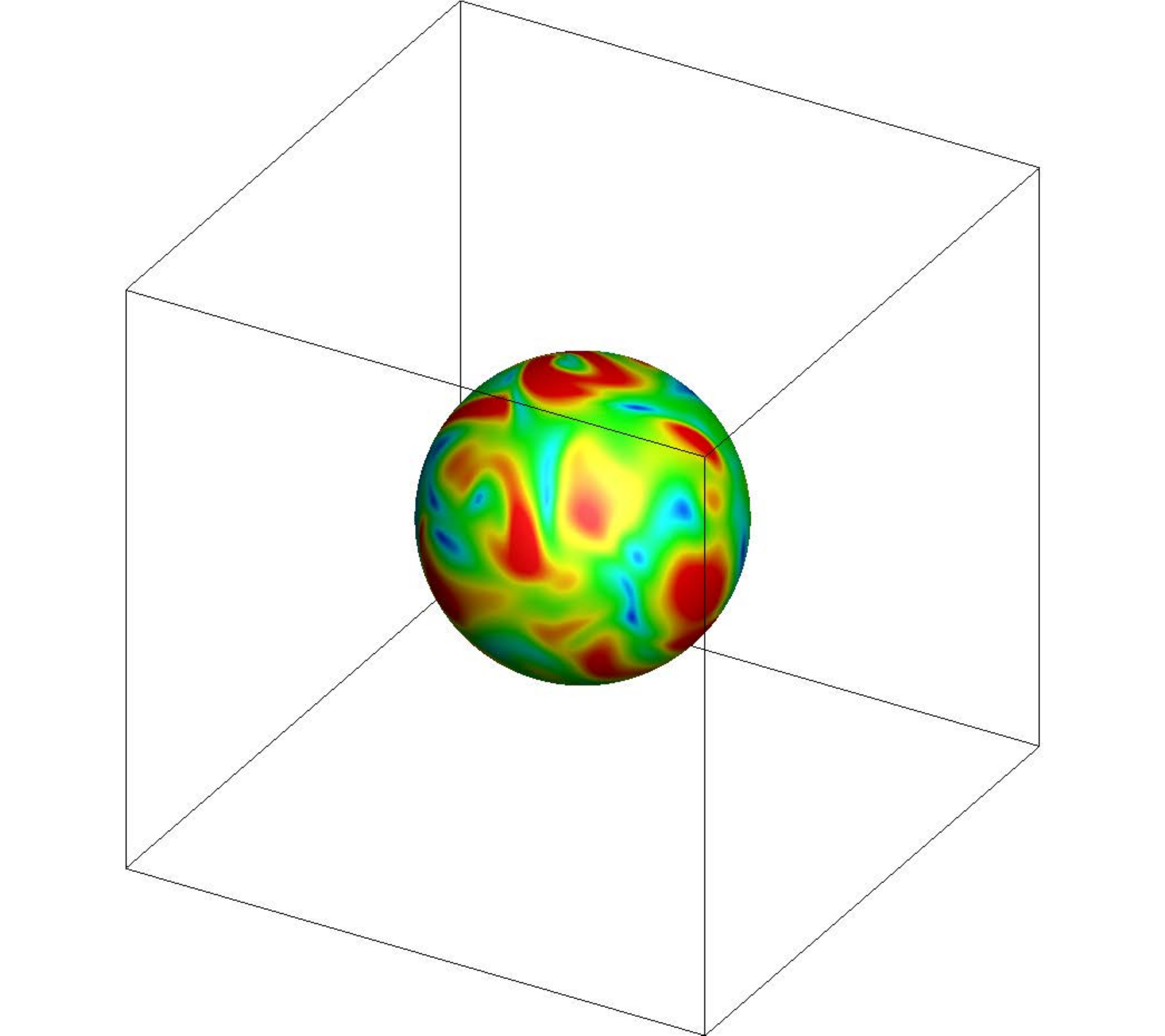}
			\label{3000v}
		\end{minipage}
	}
	\subfigure[$t^*=0.403$]{
		\begin{minipage}[t]{0.22\linewidth}
			\centering
			\includegraphics[width=1.05\columnwidth,trim={3cm 0cm 3cm 0cm},clip]{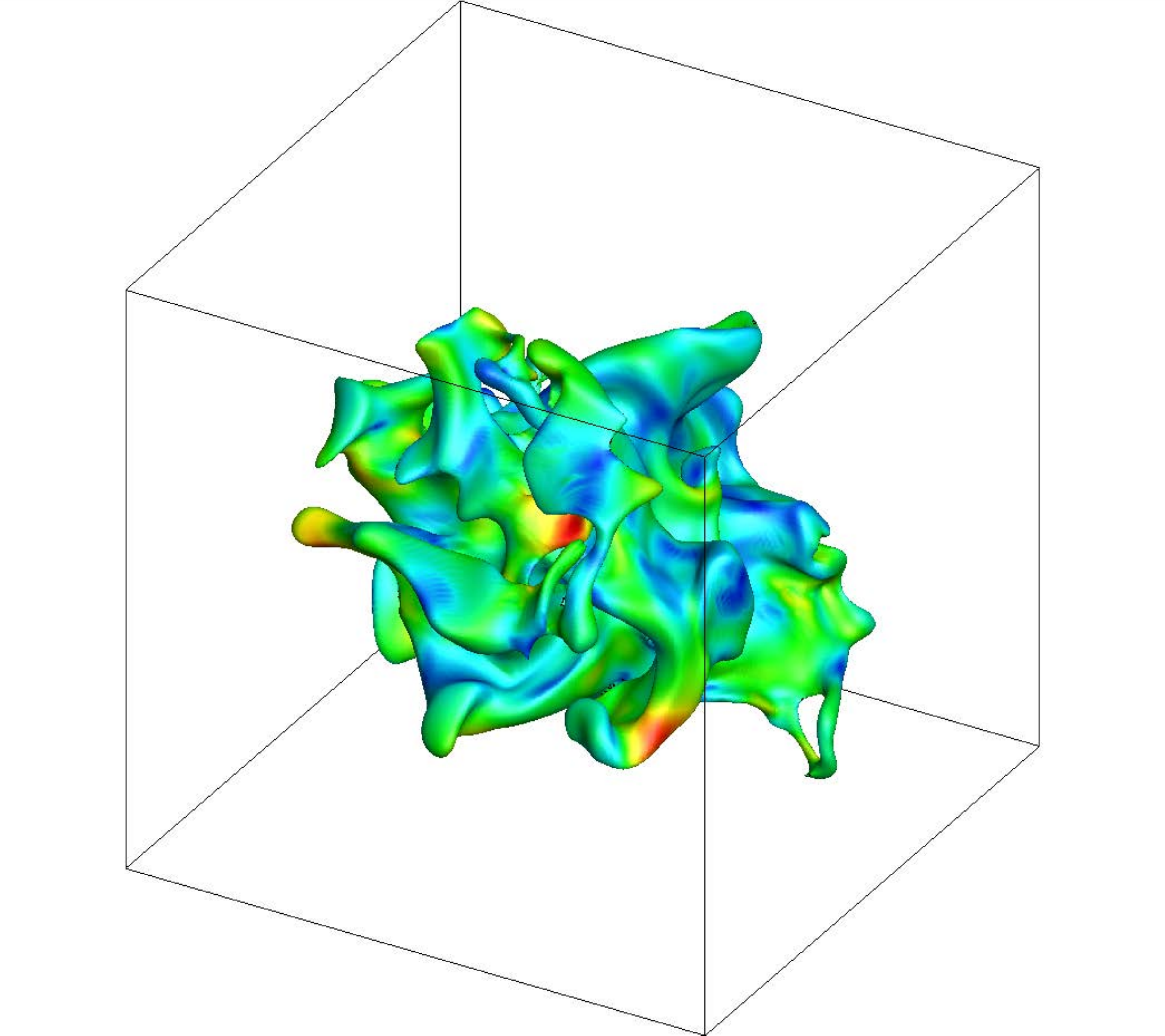}
			\label{12000v}
		\end{minipage}
	}
	\subfigure[$t^*=0.896$]{
		\begin{minipage}[t]{0.22\linewidth}
			\centering
			\includegraphics[width=1.05\columnwidth,trim={3cm 0cm 3cm 0cm},clip]{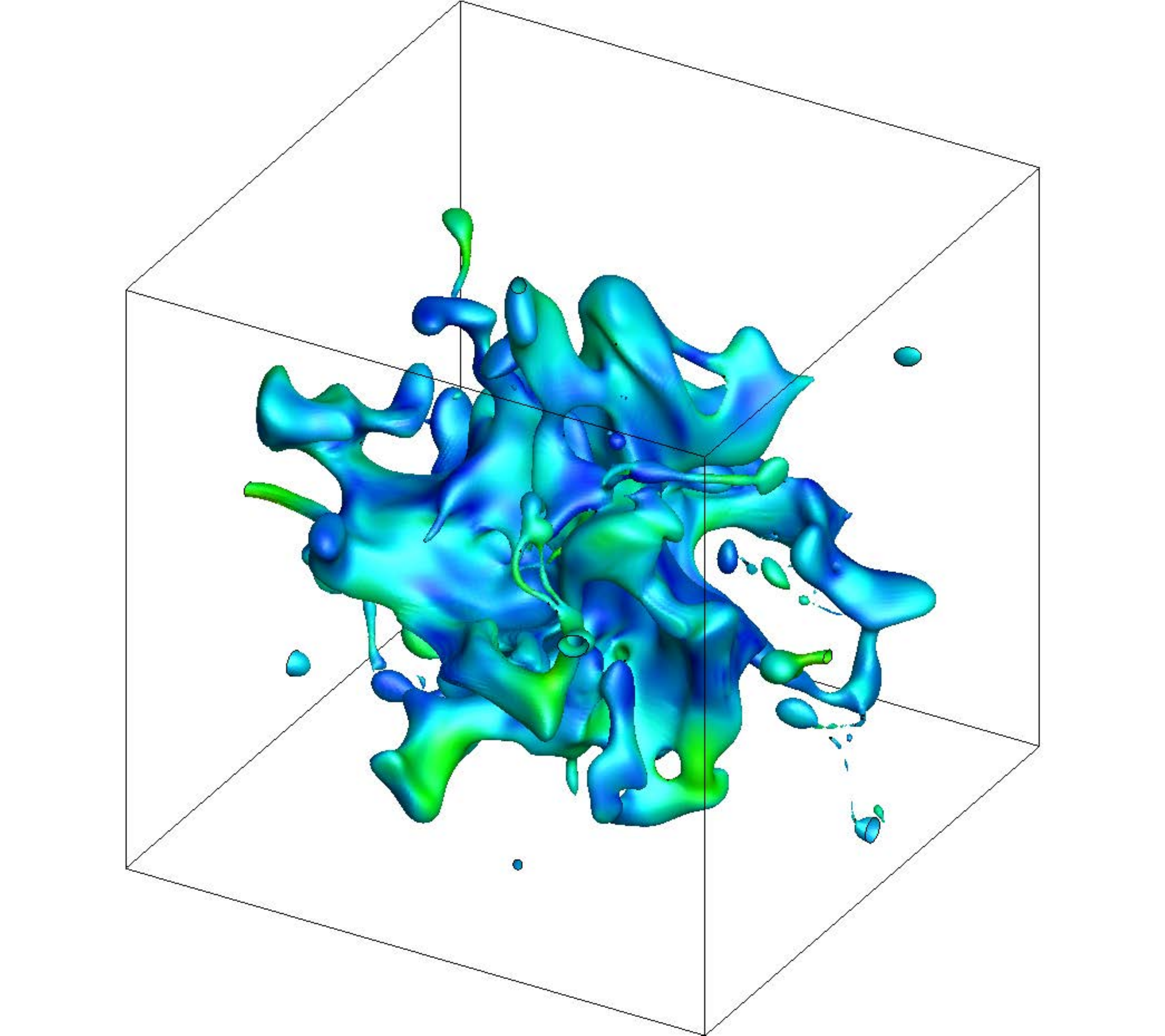}
			\label{23000v}
		\end{minipage}
	}
	\subfigure[$t^*=14.6$]{
		\begin{minipage}[t]{0.22\linewidth}
			\centering
			\includegraphics[width=1.05\columnwidth,trim={3cm 0cm 3cm 0cm},clip]{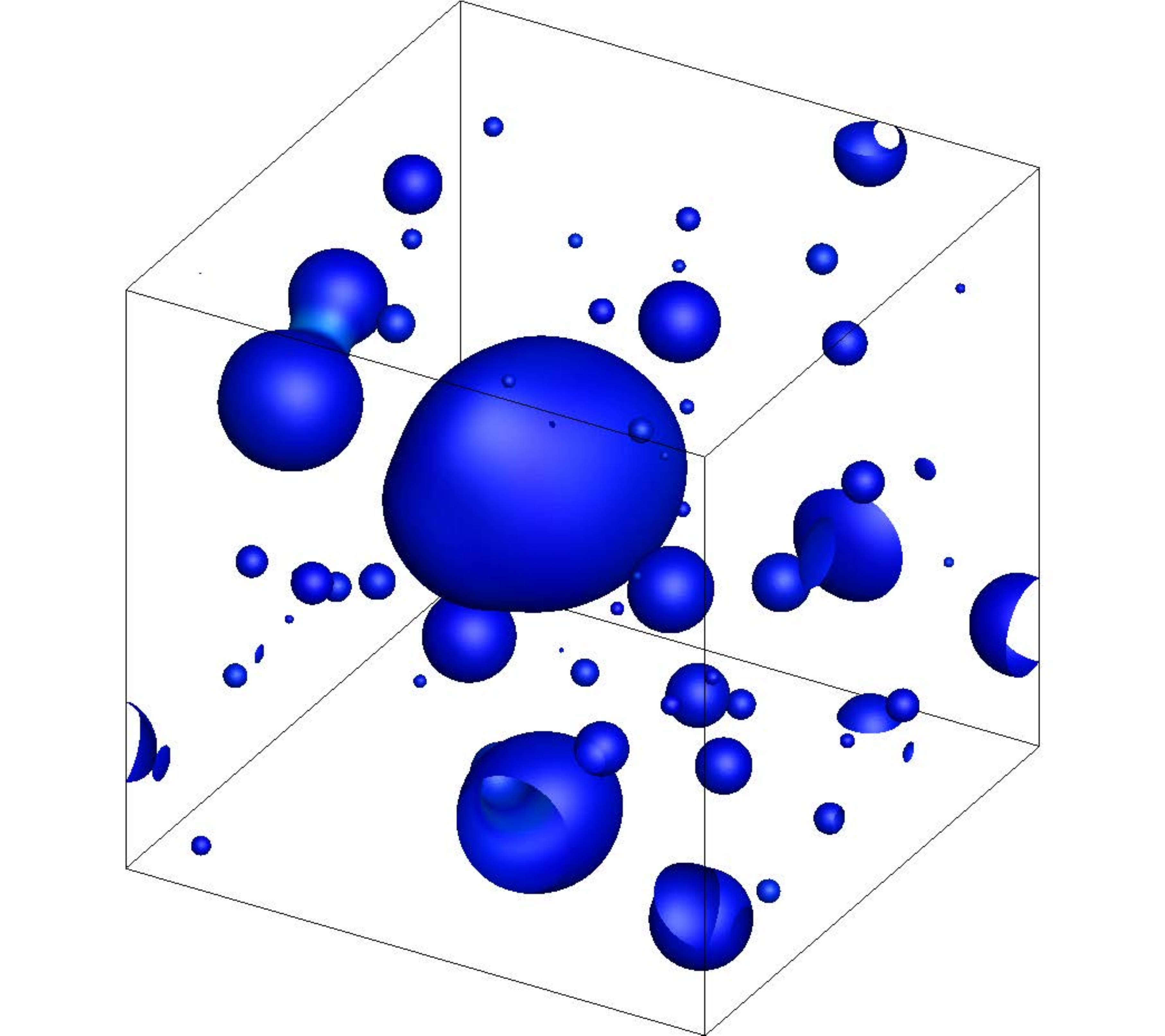}
			\label{330000v}
		\end{minipage}
	}
	\\
	\begin{minipage}[t]{0.15\linewidth}
		\centering
		\includegraphics[width=0.6\columnwidth,trim={2cm 4cm 2cm 2cm},clip]{pngpdf/xyz1.pdf}
	\end{minipage}
	\begin{minipage}[t]{0.38\linewidth}
		\centering
		\includegraphics[width=1.0\columnwidth,trim={0cm 1.5cm 0cm 1.5cm},clip]{pngpdf/u.pdf}
	\end{minipage}
	\centering
	\caption{Velocity magnitude $u^*=\left| \boldsymbol{u}\right|/\left[ u^{\prime}(t^*=0)\right] $ on the fluid-fluid interface $\phi = 0.5$ at different times.}
	\label{vel}
\end{figure*}

\begin{figure*}
	\centering    
	\subfigure[$t^*=0$]{
		\begin{minipage}[t]{0.22\linewidth}
			\centering
			\includegraphics[width=1.05\columnwidth,trim={3cm 0cm 3cm 0cm},clip]{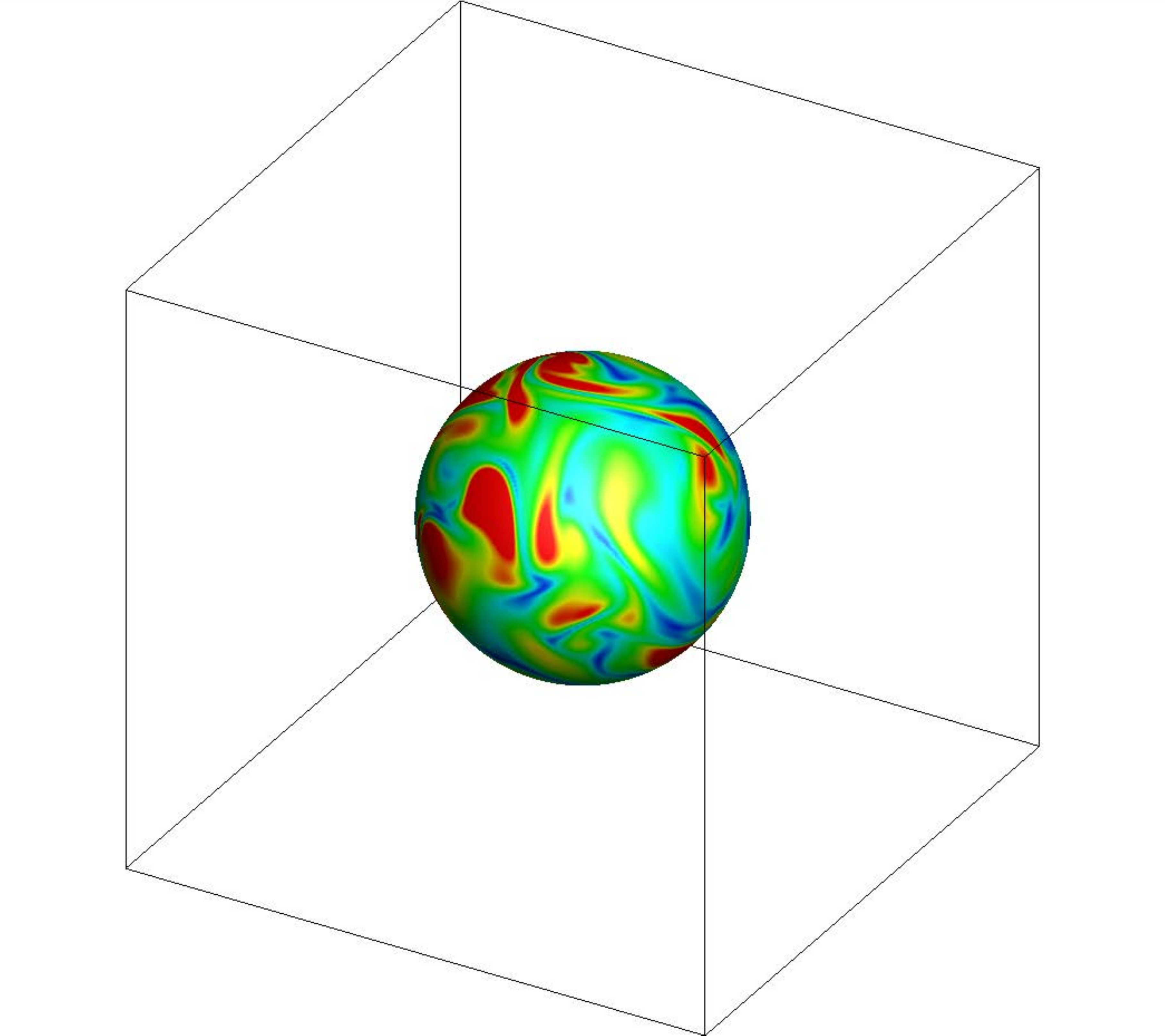}
			\label{3000wCD}
		\end{minipage}
	}
	\subfigure[$t^*=0.403$]{
		\begin{minipage}[t]{0.22\linewidth}
			\centering
			\includegraphics[width=1.05\columnwidth,trim={3cm 0cm 3cm 0cm},clip]{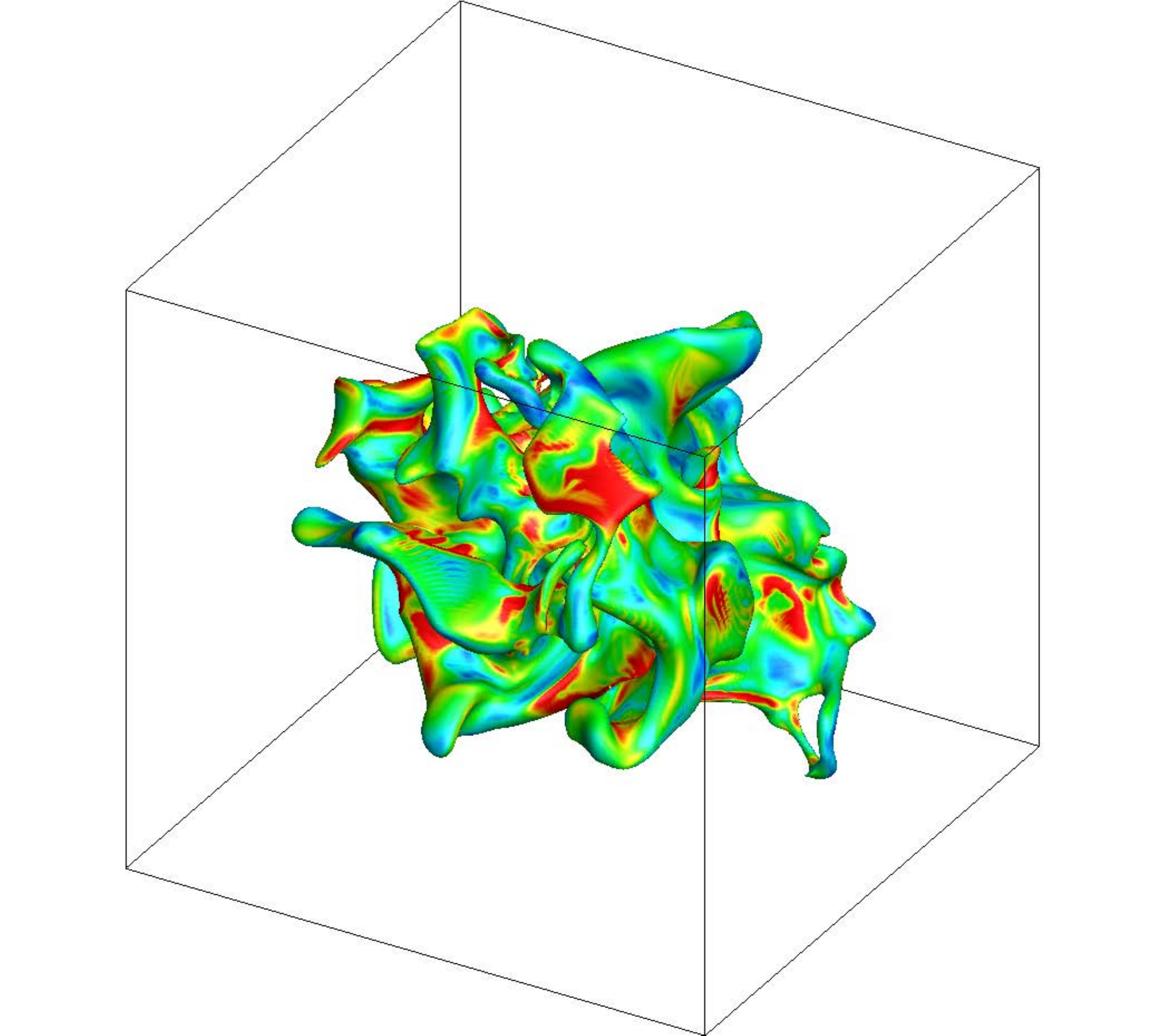}
			\label{12000wCD}
		\end{minipage}
	}
	\subfigure[$t^*=0.896$]{
		\begin{minipage}[t]{0.22\linewidth}
			\centering
			\includegraphics[width=1.05\columnwidth,trim={3cm 0cm 3cm 0cm},clip]{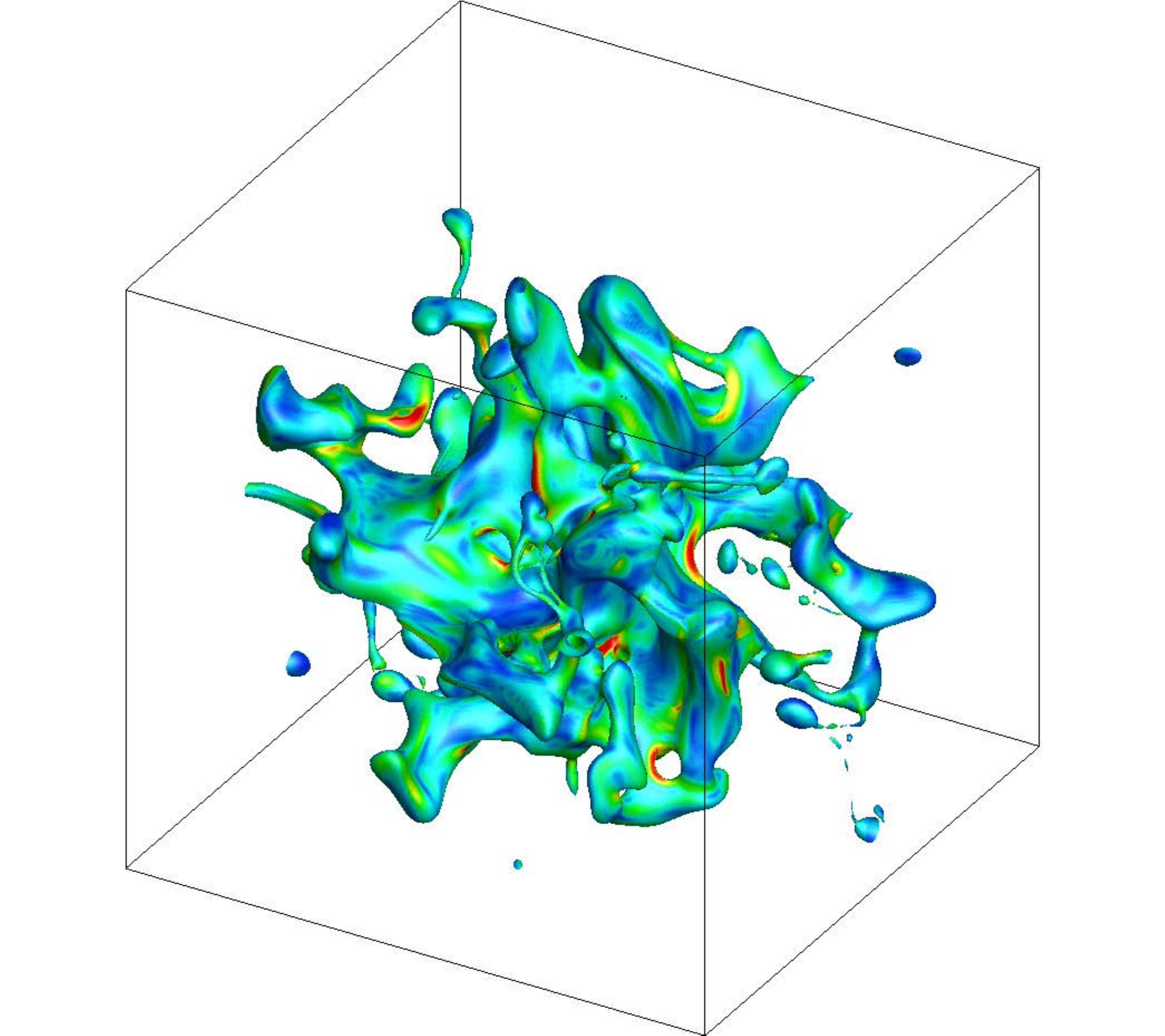}
			\label{23000wCD}
		\end{minipage}
	}
	\subfigure[$t^*=14.6$]{
		\begin{minipage}[t]{0.22\linewidth}
			\centering
			\includegraphics[width=1.05\columnwidth,trim={3cm 0cm 3cm 0cm},clip]{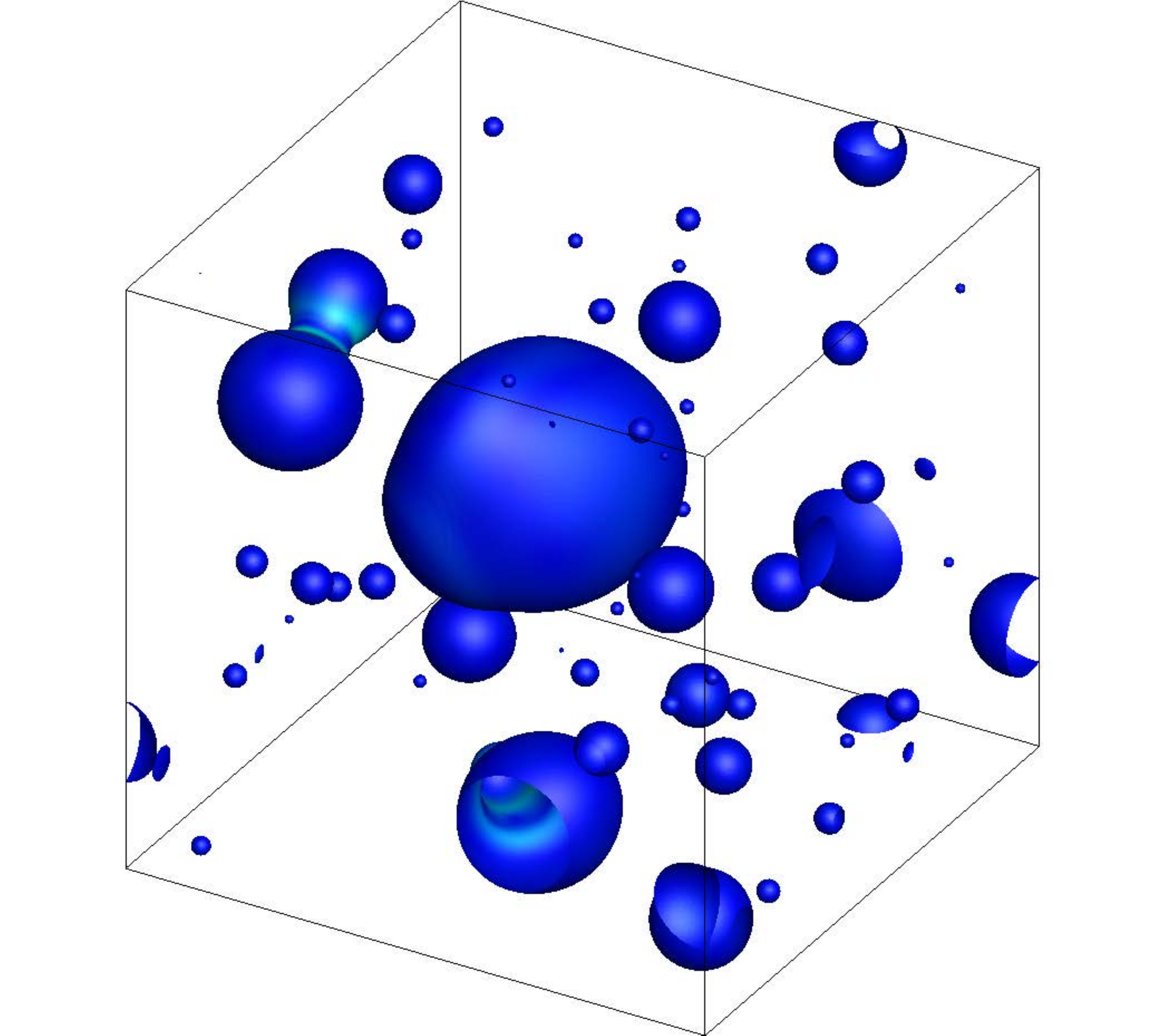}
			\label{330000wCD}
		\end{minipage}
	}
	\\
	\begin{minipage}[t]{0.15\linewidth}
		\centering
		\includegraphics[width=0.6\columnwidth,trim={2cm 4cm 2cm 2cm},clip]{pngpdf/xyz1.pdf}
	\end{minipage}
	\begin{minipage}[t]{0.38\linewidth}
		\centering
		\includegraphics[width=1.0\columnwidth,trim={0cm 1.5cm 0cm 1.5cm},clip]{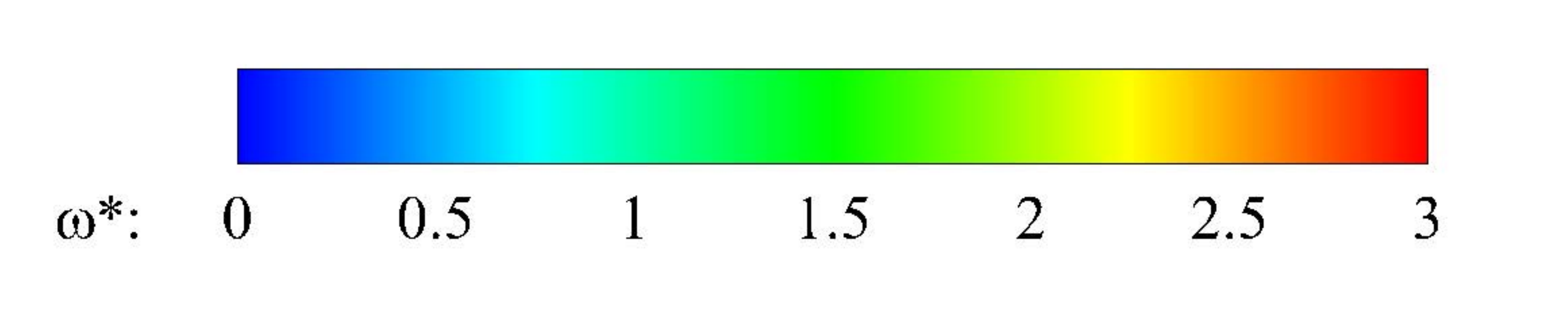}
	\end{minipage}
	\centering
	\caption{Vorticity magnitude $\omega^*=\left| \boldsymbol{\omega}\right|/\left[ \omega^{\prime}(t^*=0)\right] $ on the fluid-fluid interface $\phi = 0.5$ at different times.}
	\label{vor}
\end{figure*}

Fig.~\ref{vel} and Fig.~\ref{vor} show the dimensionless velocity and vorticity magnitudes on the droplet surface, respectively. Table \ref{TabAver} provides the average velocity and vorticity magnitudes on the isosurface $\phi=0.5$, compared with the average values in the whole computational domain. The four times shown in Table \ref{TabAver} are the same as that in Fig.~\ref{vel}-\ref{vor}. 


\begin{table*}[]
	\centering
	\caption{The average velocity and vorticity magnitudes on the interface at different times.}
	\label{TabAver}
	\begin{tabular}{ccccccccc}
		\toprule
		\multirow{2}{*}{} & \multicolumn{2}{c}{$t^*=0$} & \multicolumn{2}{c}{$t^*=0.403$} & \multicolumn{2}{c}{$t^*=0.896$} & \multicolumn{2}{c}{$t^*=14.6$}\\
		\cmidrule(r){2-3} \cmidrule(r){4-5} \cmidrule(r){6-7} \cmidrule(r){8-9}
		&  $u^*$      &  $\omega^*$   &  $u^*$      &  $\omega^*$      &  $u^*$      &  $\omega^*$
		&  $u^*$      &  $\omega^*$     \\
		\midrule
		$\left\langle \cdot\right\rangle _S $             &1.6053                          & 1.6207                    & 0.9024                   & 1.7932           & 0.5009           & 0.9797          & 0.0077           & 0.0641                     \\
		$\left\langle \cdot\right\rangle _V $             &1.5767                          & 1.5023                    & 1.1189                   & 1.4208           & 0.6944           & 0.8431          & 0.0533           & 0.0222           \\
		$\frac{\left\langle \cdot\right\rangle _S-\left\langle \cdot\right\rangle _V}{\left\langle \cdot\right\rangle _V} $             &0.0182                          & 0.0788                    & -0.1935                   & 0.2621           & -0.2786           & 0.1621          & -0.8563           & 1.8851          \\
		$\frac{\left\langle \cdot\right\rangle _S-\left\langle \cdot\right\rangle _{S,t^*=0}}{\left\langle \cdot\right\rangle _{S,t^*=0}} $             &0                          & 0                    & -0.4379                   & 0.1064           & -0.6880           & -0.3955          & -0.9952           & -0.9605          \\
		$\frac{\left\langle \cdot\right\rangle _V-\left\langle \cdot\right\rangle _{V,t^*=0}}{\left\langle \cdot\right\rangle _{V,t^*=0}} $             &0                          & 0                    & -0.2904                   & -0.0543           & -0.5596           & -0.4388          & -0.9662           & -0.9852          \\
		\bottomrule
	\end{tabular}\\
	{$\left\langle \cdot\right\rangle _S$ denotes the average value on the fluid-fluid interface at $\phi=0.5$.
		$\left\langle \cdot\right\rangle _V$ refers to the average values in the whole computational domain.
	}\\
\end{table*}

At the two-phase initialization time ($t^*=0$, Fig.~\ref{3000v} and Fig.~\ref{3000wCD}), the velocity magnitude field and vorticity magnitude field are irregularly distributed on the spherical surface, because of the 
initial homogeneous isotropic flow field. The interface averages of velocity magnitude and vorticity magnitude are slightly larger than their respective volume averaged values. It is a coincidence, because the droplets have just been put in and have not had any effect on the flow field. 
Later in the deformation stage ($t^*=0.403$), the average velocity magnitude on the interface has decreased significantly, 
only 56\% of the initial velocity magnitude on the fluid-fluid interface.
The average velocity magnitude on the interface is roughly 20\% less than that on the whole volume now, 
due to conversion of kinetic energy to the surface tension energy. 
From this point on, the average velocity magnitude on the interface continues to decrease relatively to the bulk region, showing that the surface tension force tends to damp the interface velocity magnitude, 
partially a reflection of the no-slip condition as the droplet interface have smaller velocity fluctuations.

The volume averaged vorticity magnitude decreases slowly, much slower than the reduction of velocity magnitude. It is interesting that the interface averaged vorticity magnitude is increased slightly at $t^*=0.403$, compared to the initial vorticity magnitude. 
The reasons are as follows.
First, the averaged vorticity magnitude in the whole volume does not change much. Second, the wrinkling of the interface could generate localized small-scale flows and
thus vorticity, also viscous boundary layers can develop near the interface.
The interface averaged vorticity magnitude is always larger than the volume average vorticity magnitude, 
as showed in Table \ref{TabAver}. One of the reasons is that, the fluid-fluid interface can be viewed as a deformable viscous boundary layer, and vorticity can be produced there.

Fig.~\ref{23000v} and Fig.~\ref{23000wCD} are at $t^*=0.896$, the early period of the breakup stage. The averaged velocity magnitude still decreases quickly. The averaged vorticity magnitude also decreases quickly now, compared to the previous stage. The interface averaged velocity magnitude is much smaller than the volume averaged velocity magnitude, compared to that at $t^*=0.403$. Since the surface tension force continues to restrict the interface velocity, then the relative velocity magnitude on the interface decreases monotonically in time. 

Fig.~\ref{330000v} and Fig.~\ref{330000wCD} display the results during the restoration  stage
at $t^*=14.6$.  
The velocity and vorticity magnitudes are both very small now, and the averaged values are all less than 4\% of the corresponding initial results, as showed in Table~\ref{TabAver}. The interface averaged velocity magnitude is significant smaller, less than 0.5\% of the initial value. However, the velocity magnitude is slightly larger between two colliding droplets when they merge with each other, as showed in Fig.~\ref{330000v}; when they approach each other, the interface is deformed due to local hydrodynamic interaction and 
the nearby velocity magnitude could increase. 
The vorticity magnitude near the merging interface also increases with the increasing of velocity magnitude. At that time, the vorticity magnitude in the whole domain is very small (less than 1.5\% of the initial result), while the interface boundary layers result in a relatively larger vorticity magnitude on the interface (about three times) than that in the whole volume (Table~\ref{TabAver}).

\subsection{Kinetic energy decay}

\begin{figure}[t!]
	\centering    
	\subfigure[log-log plot]{
		\begin{minipage}[t]{0.47\linewidth}
			\centering
			\includegraphics[width=1.\columnwidth,trim={0cm 0cm 0cm 0cm},clip]{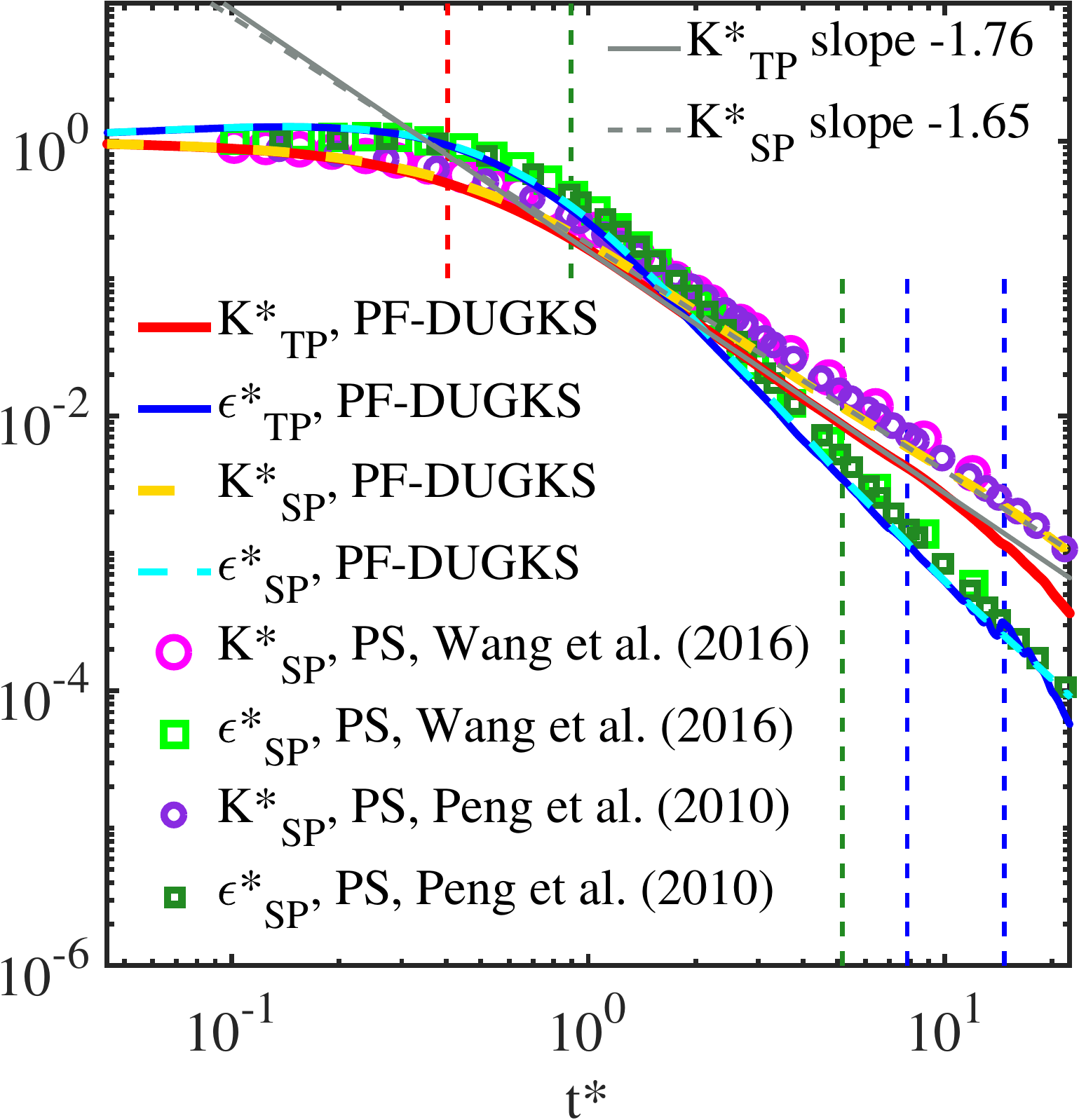}
			\label{Kelog}
		\end{minipage}
	}
	\subfigure[linear-linear plot]{
		\begin{minipage}[t]{0.47\linewidth}
			\centering
			\includegraphics[width=1.\columnwidth,trim={0cm 0cm 0cm 0cm},clip]{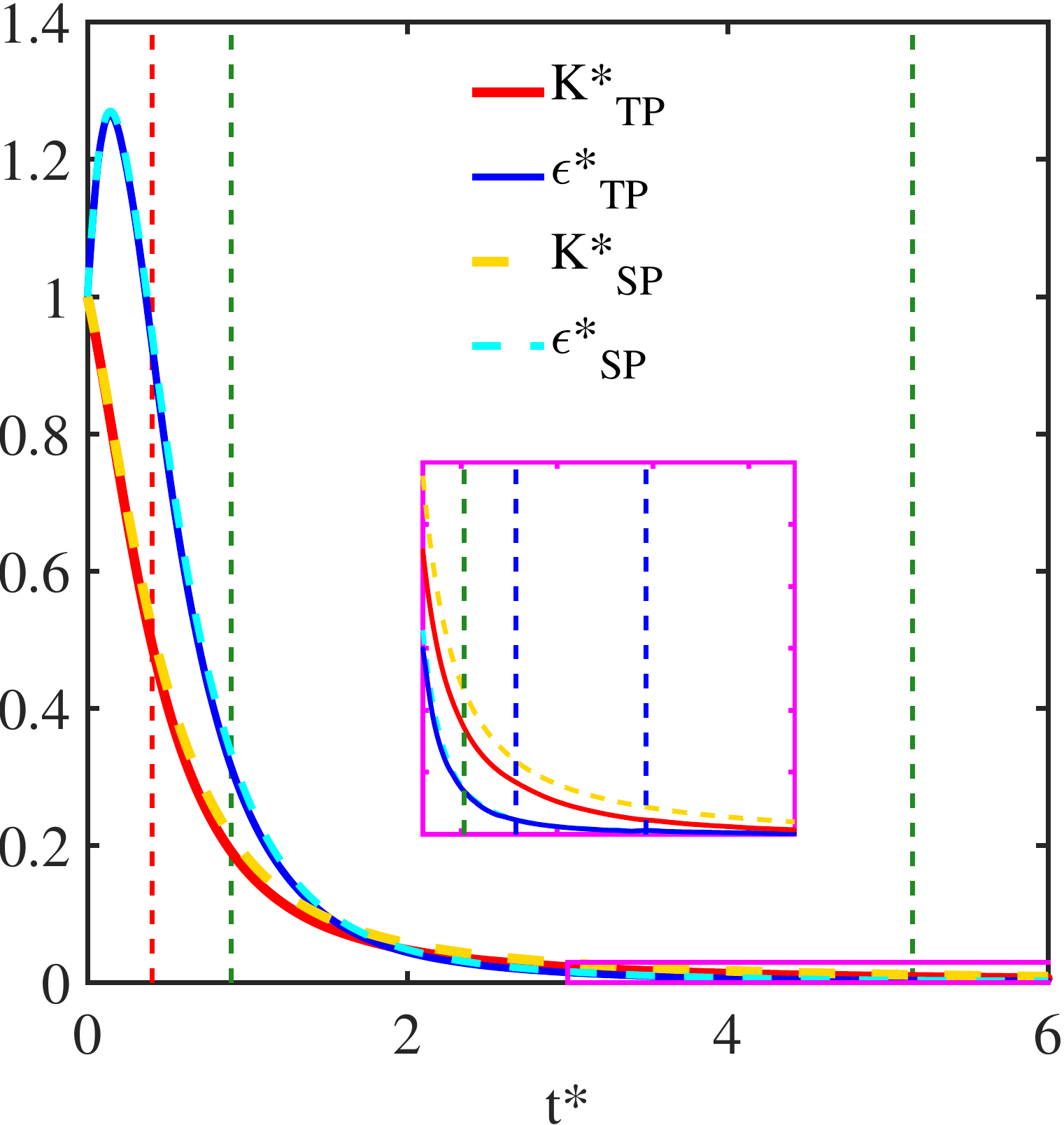}
			\label{Kelin}
		\end{minipage}
	}
	\centering
	\caption{Evolution of the normalized total kinetic energy $K^*=K(t)/K(0)$ and the normalized dissipation rate $\epsilon^*=\epsilon(t)/\epsilon(0)$. The solid lines represent results from the two-phase (TP) flow and the dashed lines represent those from the single-phase (SP) flow: (a) log-log plot, (b) linear-linear plot. The inset in  (b) shows the region $[3,22.4]\times[0,0.03]$.
		The red, dark green, blue vertical dash lines mark typical times for deformation ($ t^* = 0.403 $), breakup ($ t^* = 0.896$ and $5.15$), 
		restoration  ($ t^* = 7.84$ and $14.6$), respectively.
		The resolutions are the same ($256^3$) for our simulations and Wang's simulation,~\cite{WangPeng2016} while the resolution is lower ($128^3$) for Peng's simulation.~\cite{2010Comparison} The initial $Re_\lambda$ is about 58, 26, 24 for our simulations, Wang's simulations, and Peng's simulations, respectively.
	}
	\label{Ke}
\end{figure}

Fig.~\ref{Ke} shows the evolution of the normalized total kinetic energy $K^*=K(t)/K(0)$ and the normalized dissipation rate $\epsilon^*=\epsilon(t)/\epsilon(0)$.
In Fig.~\ref{Kelog}, the single-phase flow results from two previous studies are also added for comparison.
The slopes of our single-phase DHIT results are in good agreement with the those from the PS method reported by 
Wang {\it et al.}~(2016)~\cite{WangPeng2016} and Peng {\it et al.}~(2010).~\cite{2010Comparison} 
Our results cannot overlap with their results precisely for two reasons. First, $Re_\lambda$ are different, 
the initial $Re_\lambda$ is about 58, 26, 24 for our simulation, Wang's simulation, and Peng's simulation, respectively. 
Huang \& Leonard~(1994)~\cite{1994Power} reported that the power-law decay slope of the kinetic energy depends on $Re_\lambda$.
Second, the decay slope may depend on the shape of the initial energy spectrum used and how the flow is initialized.~\cite{2001Direct,1994Power}

Next, we compare our two-phase results with our single-phase results in log-log plot (Fig.~\ref{Kelog}) and linear-linear plot (Fig.~\ref{Kelin}).
Two general observations can be made. First, the increasing fluid-fluid interface area causes the kinetic energy for the two-phase case to decay faster, due to conversion of kinetic energy into the free energy at the deformation stage.
The power-law slopes of the kinetic energy decay at the intermediate times (from $t^*=0.896$ to $t^*=7.84$ in our simulation) are $-1.76$ and $-1.65$ for the two-phase and single-phase flow, respectively.
Second, the decay rate of the overall dissipation rate for the two-phase flow, however, is almost the same as that of the single-phase flow. Since the kinetic energy for the two-phase flow is smaller, the viscous dissipation normalized by $u'^3 /L_I $, ({\it i.e.}, $\epsilon L_I / u'^3$) for the two-phase flow is then larger,  
here $L_I=\pi/\left( 2u'^2\right) \int E\left( {k}\right)/kdk$ is the integral length scale,  meaning that the interface boundary layers tend to increase
the overall viscous dissipation. 
However, the coalescence of droplets at the restoration stage would cause transfer of free energy back to kinetic energy, which may
also affect the overall viscous dissipation rate. Because the kinetic energy and dissipation rate at that stage are so small (less than 1\% of those at the initial time), perturbations due to merging of droplets may affect their value. For example,
the dissipation rate fluctuates around $t^*=14.6$ (see Fig.~\ref{Kelog}), which could be due to the coalescence of two 
medium-size droplets at that time (see Fig.~\ref{330000}).

\begin{figure}[t!]
	\centering
	\includegraphics[width=0.65\columnwidth,trim={0cm 0cm 0cm 0cm},clip]{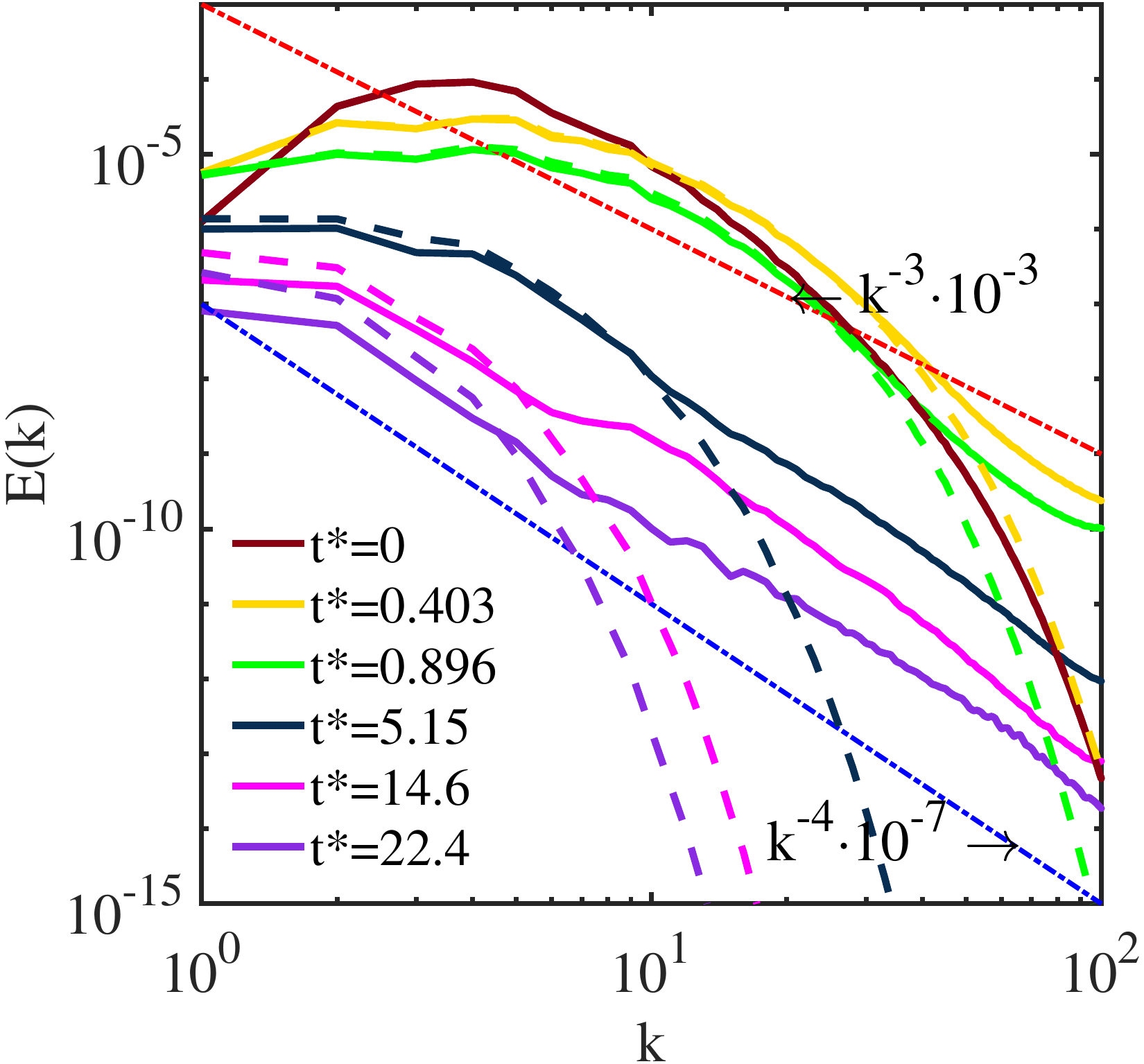}
	\centering
	\caption{The energy spectra $E(k, t)$ at different normalized times. The solid lines represent results from the two-phase flow and the dashed lines represent those from the single-phase flow.}
	\label{Ek}
\end{figure}

The energy spectrum $E(k,t)$ is also evaluated to show the energy distribution over different turbulence scales (see Fig.~\ref{Ek}). At a given time $t^*>0$, the energy contained in high wavenumbers for the two-phase flow is larger than the corresponding part in the single-phase flow, while the energy contained in small wavenumbers of two-phase flow is less than the corresponding part in the single-phase flow.
This implies that energy is redistributed among scales due to the presence of interfaces, enhancing the transfer of energy from large to small scales. Because the interfaces bring in new intermediate length scales 
which could increase the transfer rate of energy from low to high wavenumbers.~\cite{lucci2010modulation,dodd2016interaction} 
Another reason is that, surface tension force is relatively concentrated around the thin interface, locally similar to the Delta function, and acts as 
a driving mechanism for small-scale fluctuations when the interface is deformed. 

In two-phase flows containing dispersed particles and bubbles, it is well known that the energy spectrum at high wavenumber exhibits a power-law behavior due to wake flows around dispersed objects, different from the exponential decay known for a single-phase turbulence. In the pioneering work of Lance and Bataille~(1991),~\cite{lance1991turbulence} it was estimated theoretically that dispersed wakes due to bubbles can produce a power-law of slope approximately equal to $-3$ based on the experimental data.
The bubbly-flow simulation results of Bunner and Tryggvason~(2002)~\cite{bunner2002dynamics} suggested a power slope of $-3.6$ at high wavenumbers, where they applied the  front-tracking/finite difference method and did not 
allow the bubbles to coalesce. In a study of freely-decaying turbulence containing breakup of a liquid sheet applying level-set method, 
Trontin {\it et al.}~(2010)~\cite{trontin2010direct} found a power-law of slope between $-3$ and $-5$ within the breakup region.

In Fig.~\ref{Ek}, we also clearly observe power-law spectra at high wavenumbers, with a slope approximately equal to $-4$. The governing equation of $E(k,t)$ is derived in Appendix~\ref{subsec: EnergyFourier}, which
shows that there is a new term, $S(k,t)$,  due to the surface tension effect. This term 
could act to transfer the kinetic energy from  low wavenumbers to high wavenumbers.

\subsection{Kinetic energy spectra on spheres and velocity field inside/outside the droplets}

\begin{figure*}[t!]
	\centering    
	\subfigure[$t^*=0$]{
		\begin{minipage}[t]{0.46\linewidth}
			\centering
			\includegraphics[width=1.\columnwidth,trim={0cm 0cm 0cm 0cm},clip]{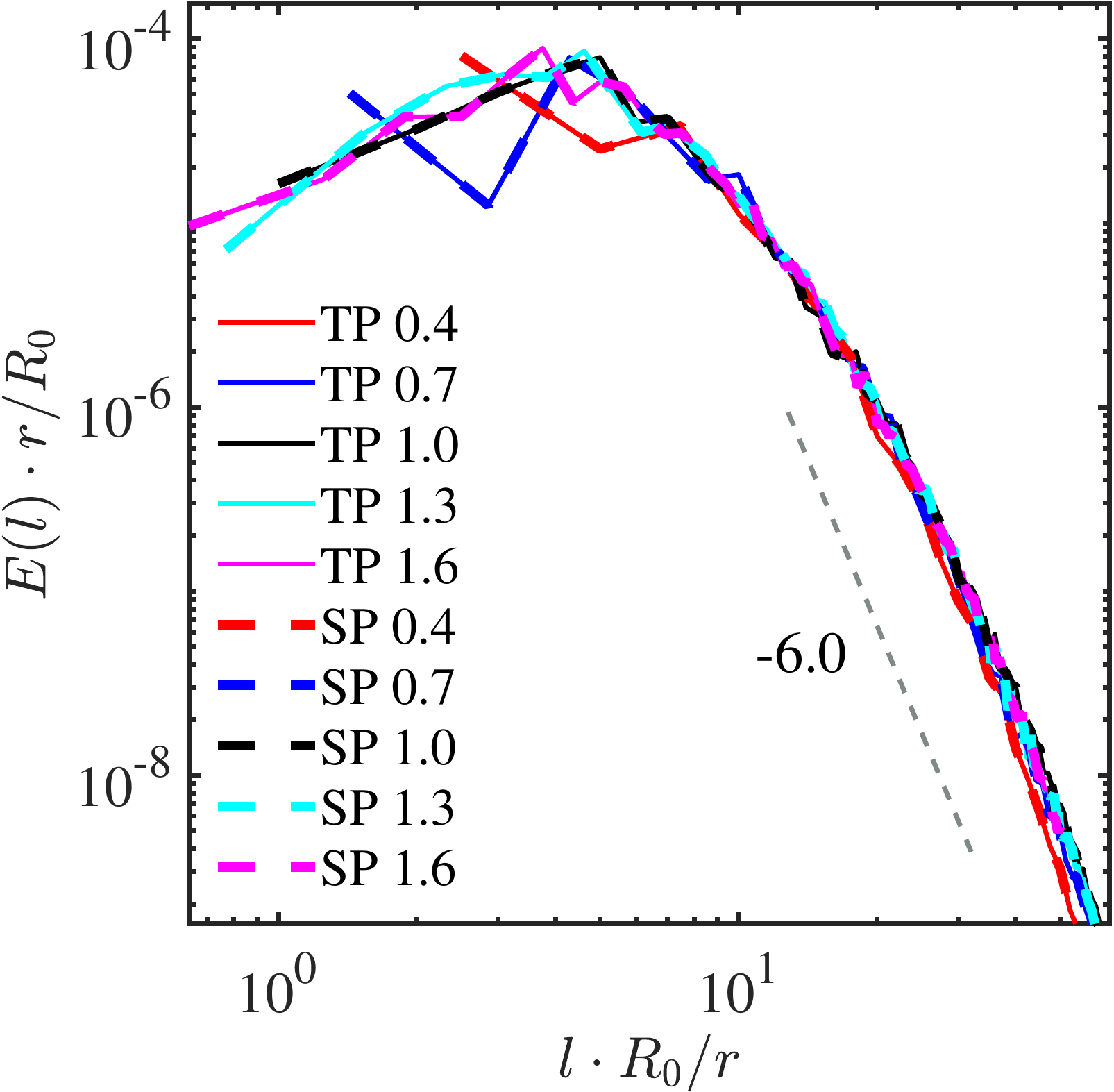}
			\label{El3000breakup256rmr}
		\end{minipage}
	}
	\subfigure[$t^*=0.403$]{
		\begin{minipage}[t]{0.46\linewidth}
			\centering
			\includegraphics[width=1.\columnwidth,trim={0cm 0cm 0cm 0cm},clip]{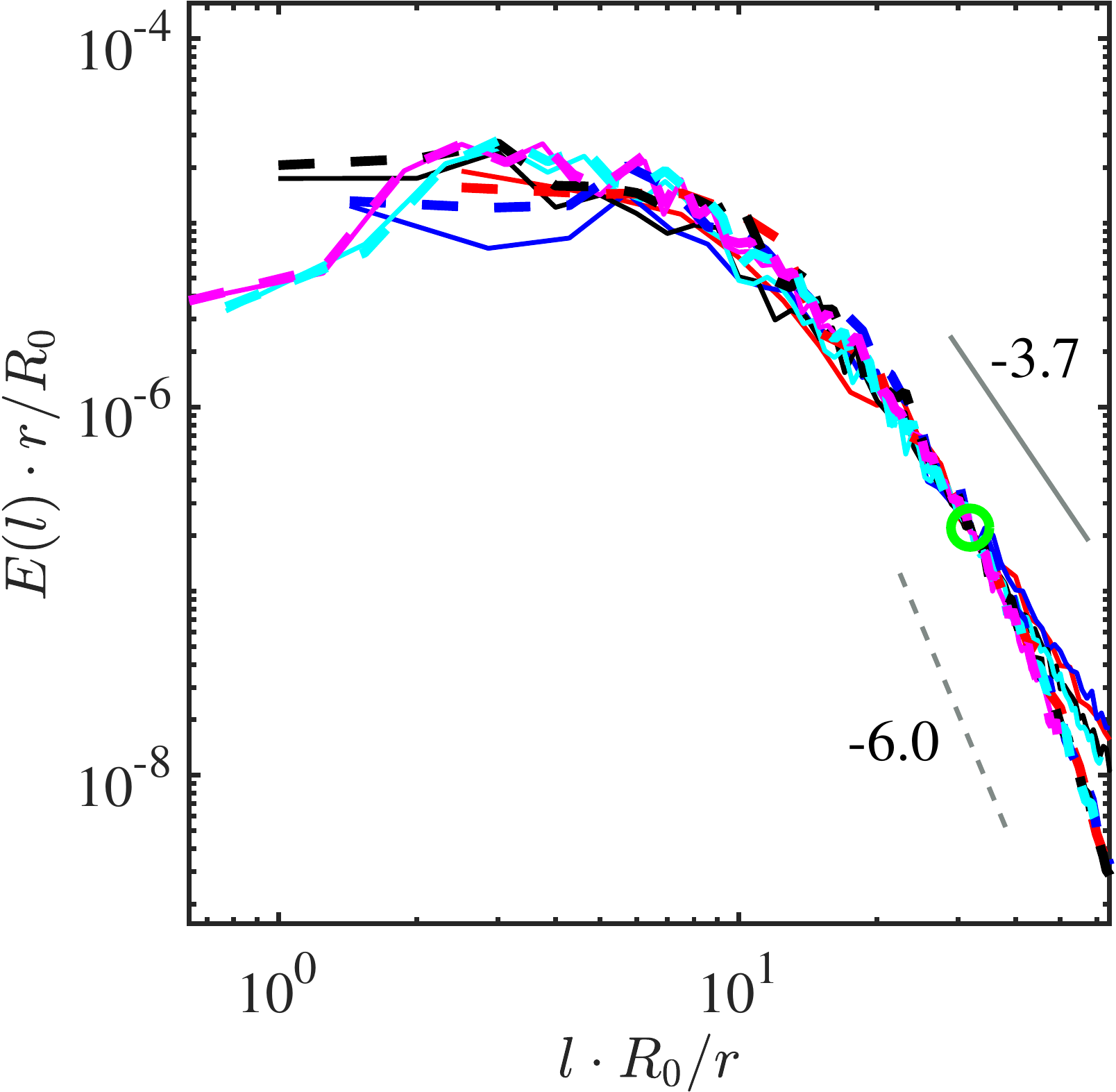}
			\label{El12000breakup256rmr}
		\end{minipage}
	}
	\\
	\subfigure[$t^*=0.896$]{
		\begin{minipage}[t]{0.46\linewidth}
			\centering
			\includegraphics[width=1.\columnwidth,trim={0cm 0cm 0cm 0cm},clip]{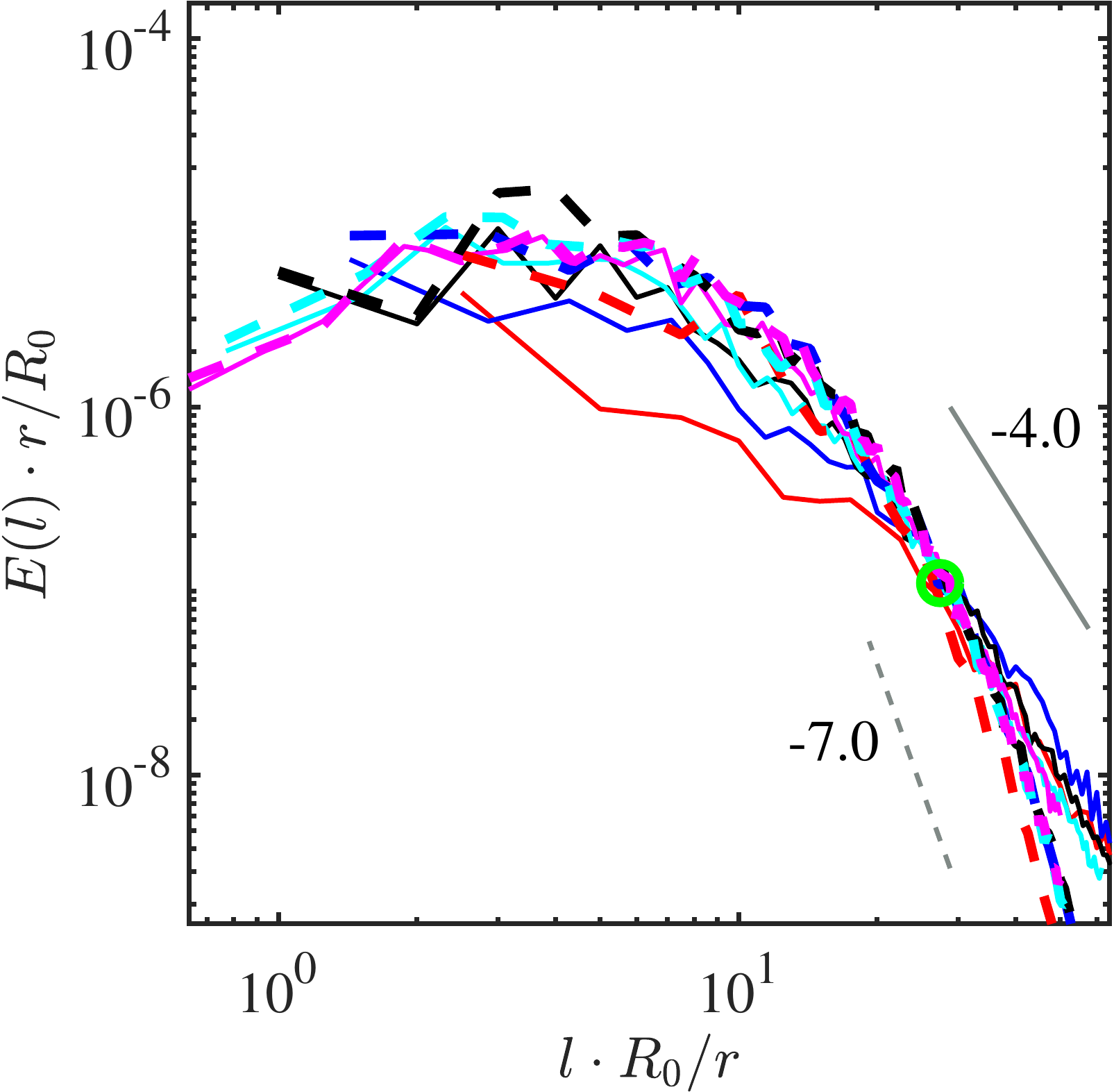}
			\label{El23000breakup256rmr}
		\end{minipage}
	}
	\subfigure[$t^*=5.15$]{
		\begin{minipage}[t]{0.46\linewidth}
			\centering
			\includegraphics[width=1.\columnwidth,trim={0cm 0cm 0cm 0cm},clip]{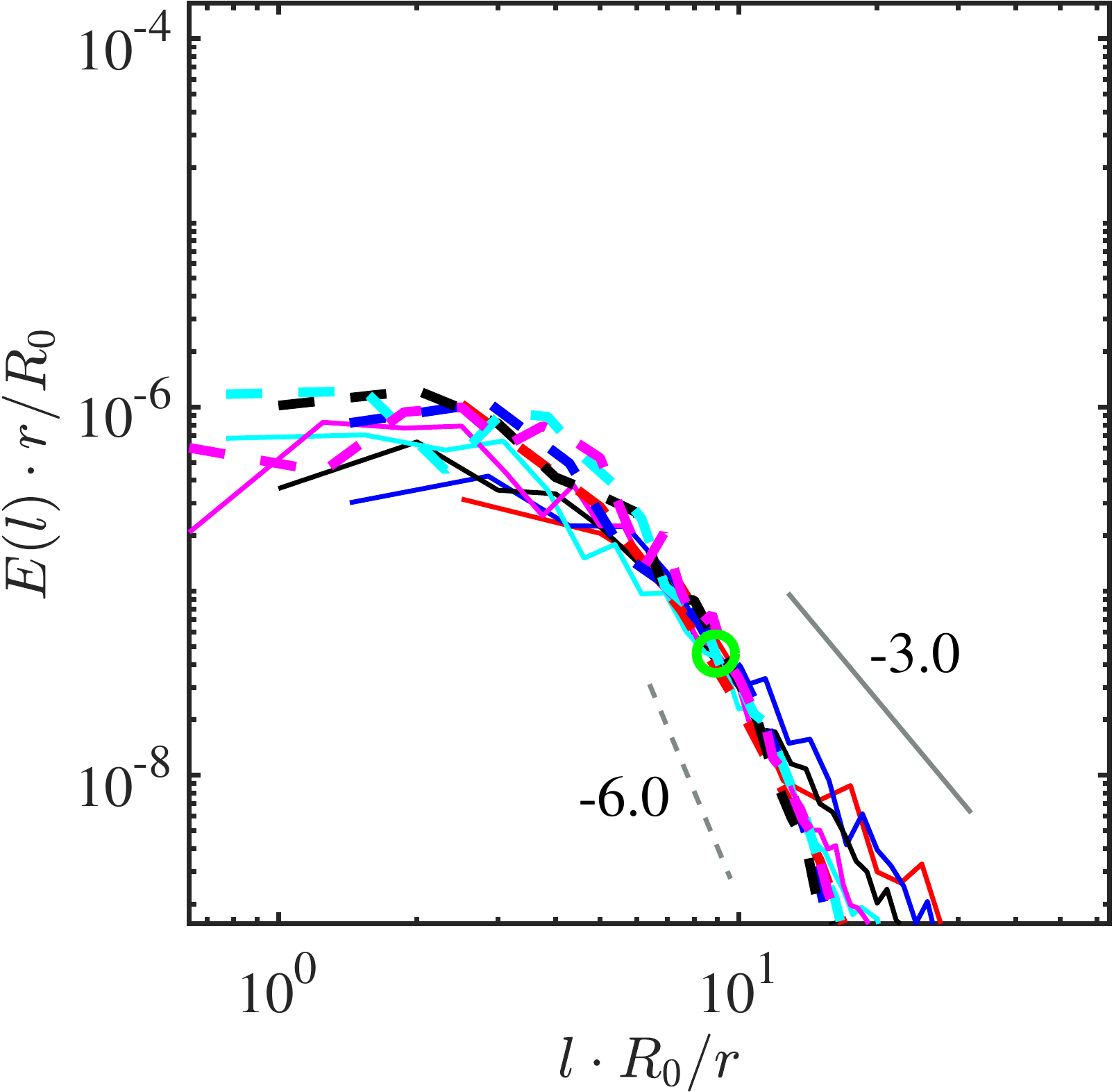}
			\label{El118000breakup256rmr}
		\end{minipage}
	}
	\centering
	\caption{Energy spectra based on the spherical harmonics at different times for the two-phase flow and the single-phase flow,
	at different radii, $r/R_0=0.4,0.7,1.0,1.3,1.6$, from the initial droplet center.
	The solid lines represent results for the two-phase (TP) flow and the dashed lines for the single-phase (SP) flow. 
	The green circles in (b), (c) and (d) indicate the approximate cross-over location, where the relative magnitudes of the
	 TP energy spectra change from below to above those of the SP flow.}
	\label{spherical}
\end{figure*}

\begin{figure*}[t!]
	\centering    
	\subfigure[$t^*=0$]{
		\begin{minipage}[t]{0.22\linewidth}
			\centering
			\includegraphics[width=1.05\columnwidth,trim={3cm 0cm 3cm 0cm},clip]{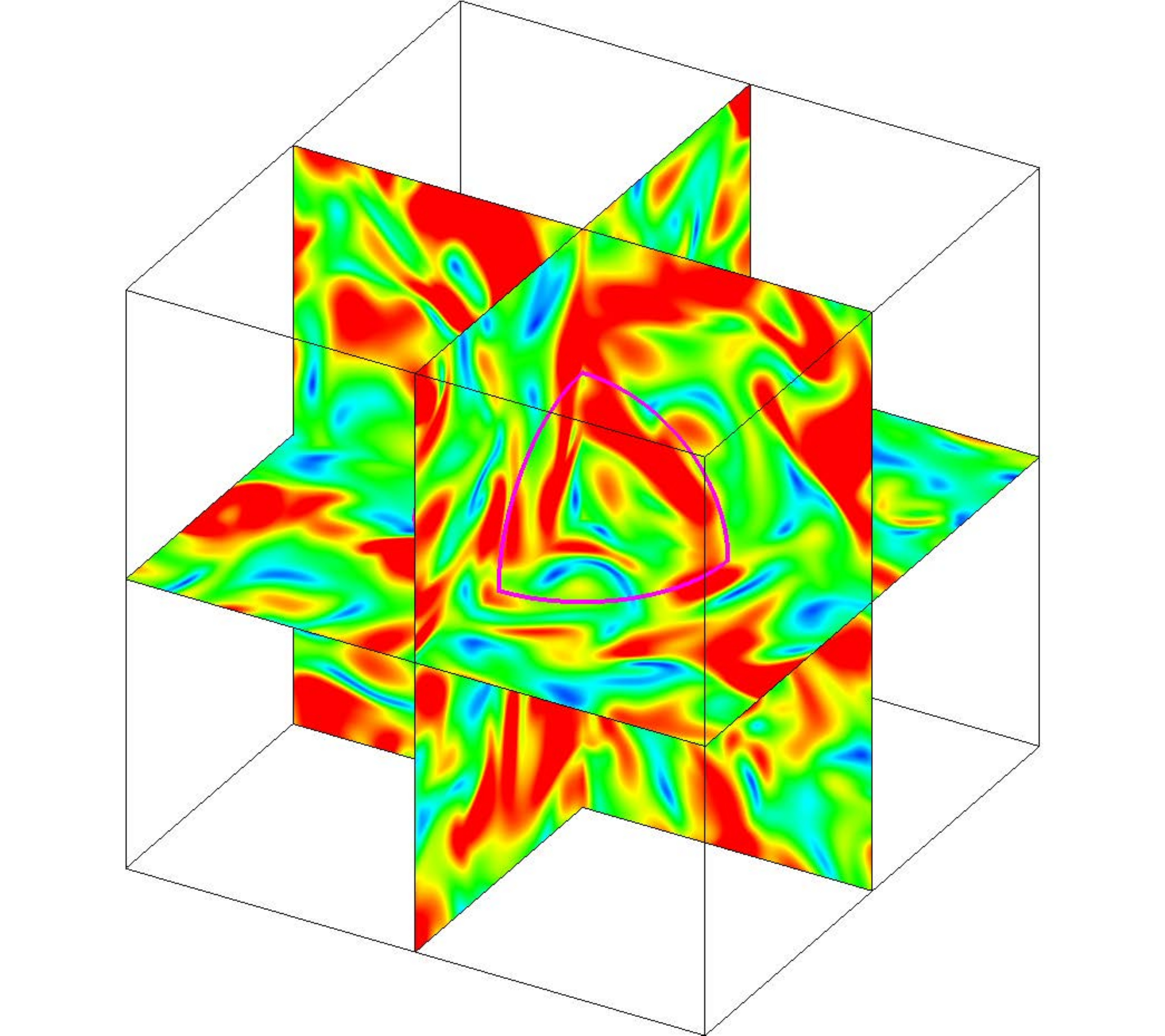}
			\label{3000vrmsCD3phi}
		\end{minipage}
	}
	\subfigure[$t^*=0.403$]{
		\begin{minipage}[t]{0.22\linewidth}
			\centering
			\includegraphics[width=1.05\columnwidth,trim={3cm 0cm 3cm 0cm},clip]{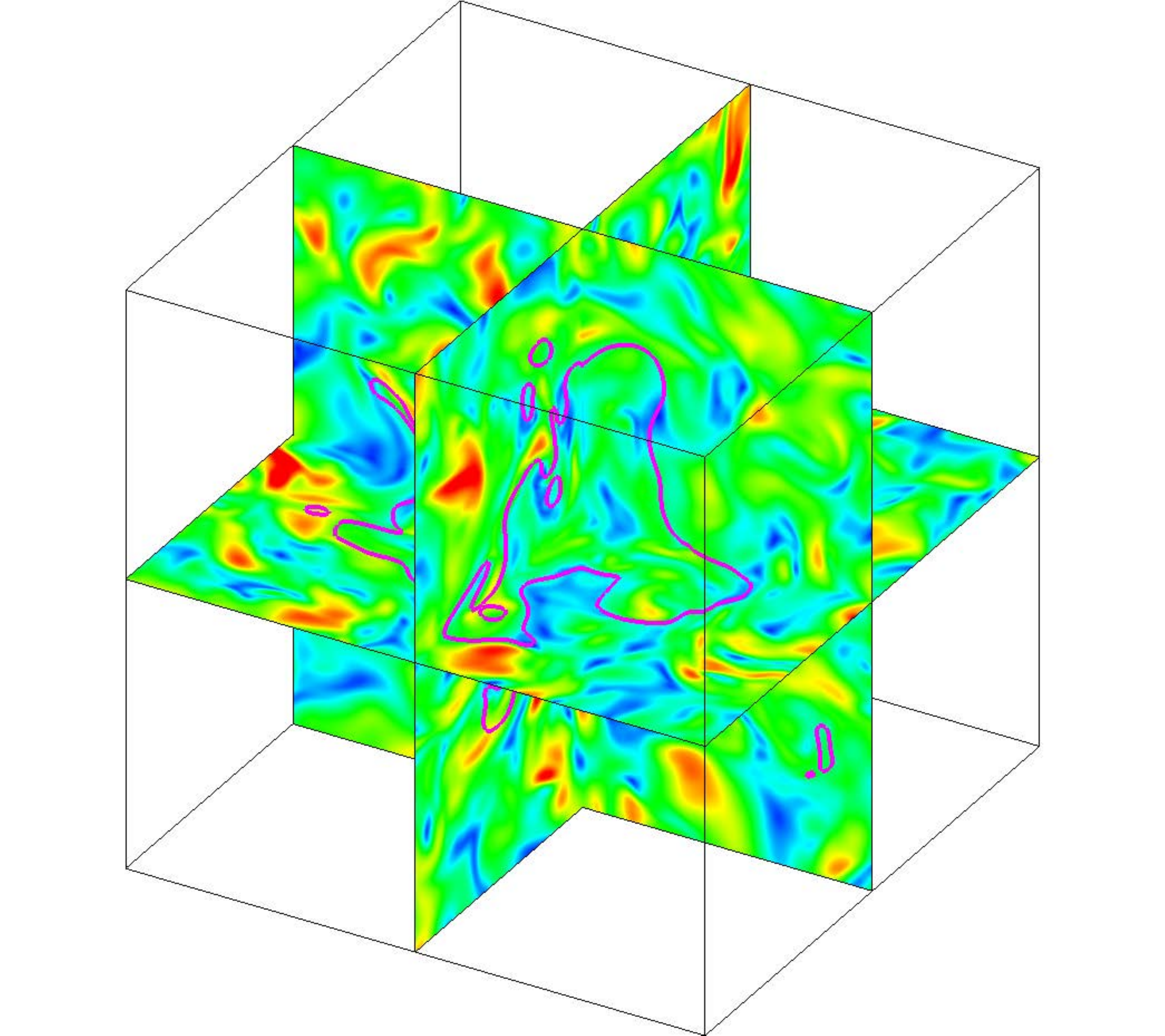}
			\label{12000vrmsCD3phi}
		\end{minipage}
	}
	\subfigure[$t^*=0.896$]{
		\begin{minipage}[t]{0.22\linewidth}
			\centering
			\includegraphics[width=1.05\columnwidth,trim={3cm 0cm 3cm 0cm},clip]{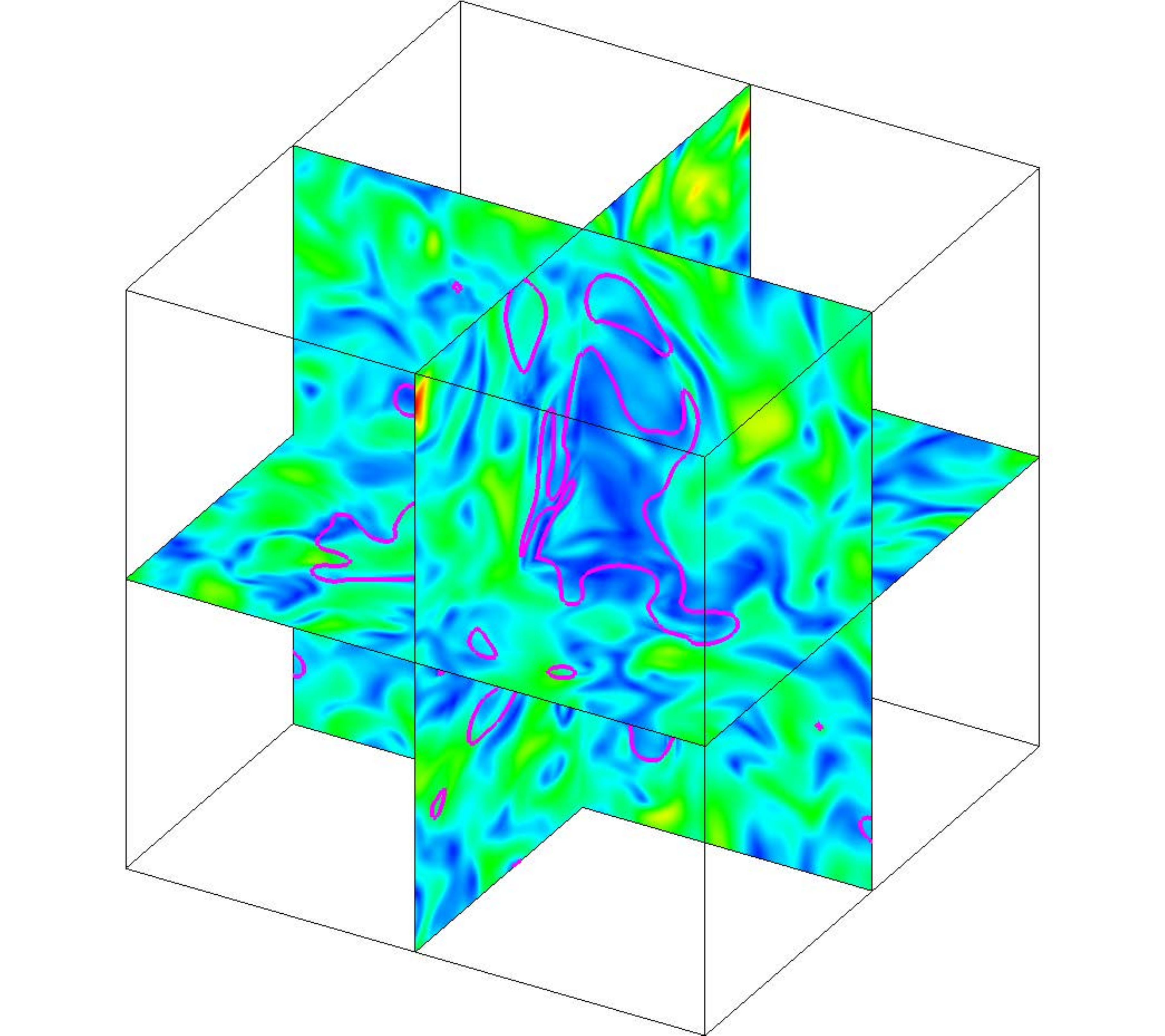}
			\label{23000vrmsCD3phi}
		\end{minipage}
	}
	\subfigure[$t^*=5.15$]{
		\begin{minipage}[t]{0.22\linewidth}
			\centering
			\includegraphics[width=1.05\columnwidth,trim={3cm 0cm 3cm 0cm},clip]{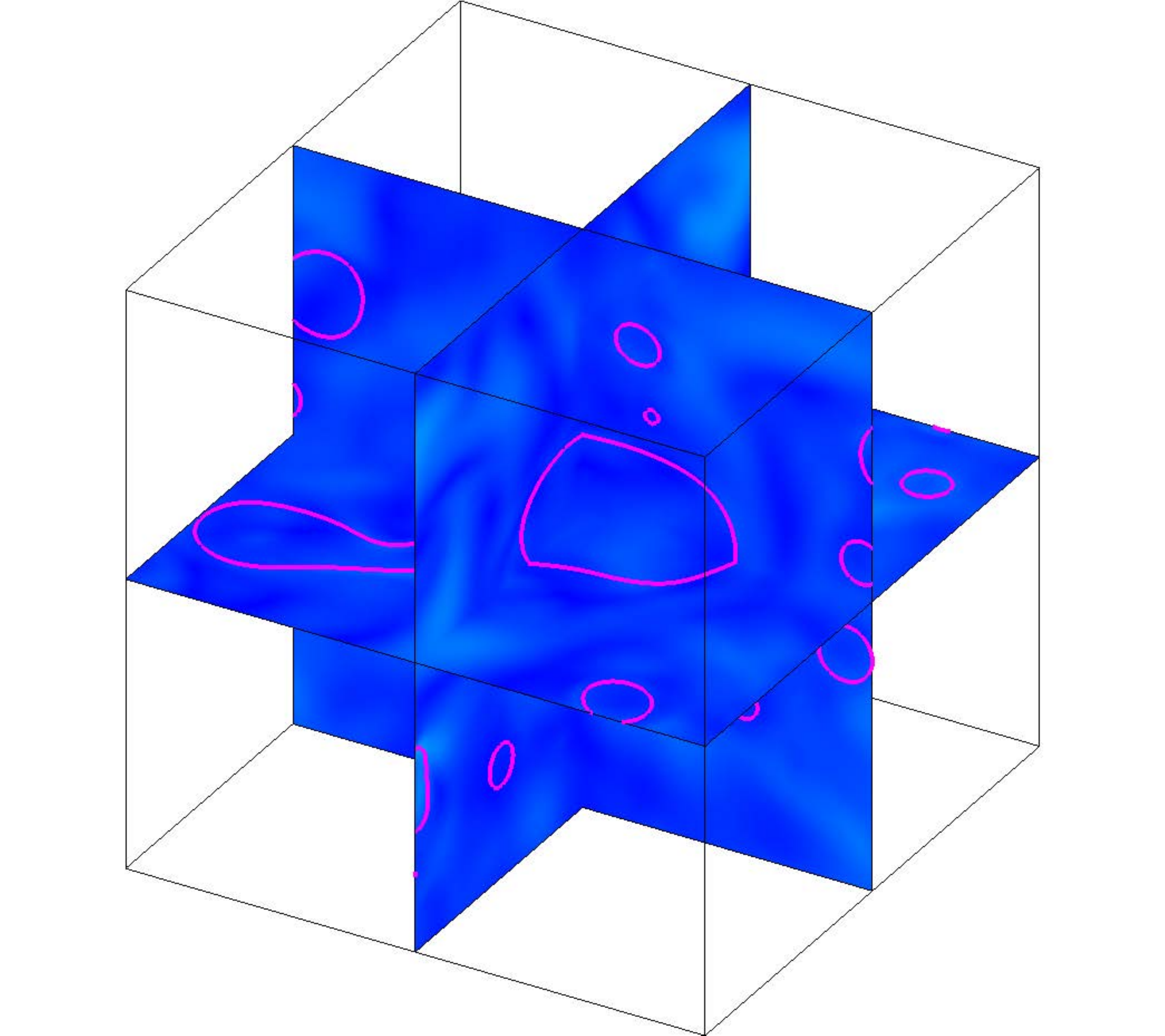}
			\label{118000vrmsCD3phi}
		\end{minipage}
	}
	\\
	\subfigure[$t^*=0$]{
		\begin{minipage}[t]{0.22\linewidth}
			\centering
			\includegraphics[width=1.05\columnwidth,trim={3cm 0cm 3cm 0cm},clip]{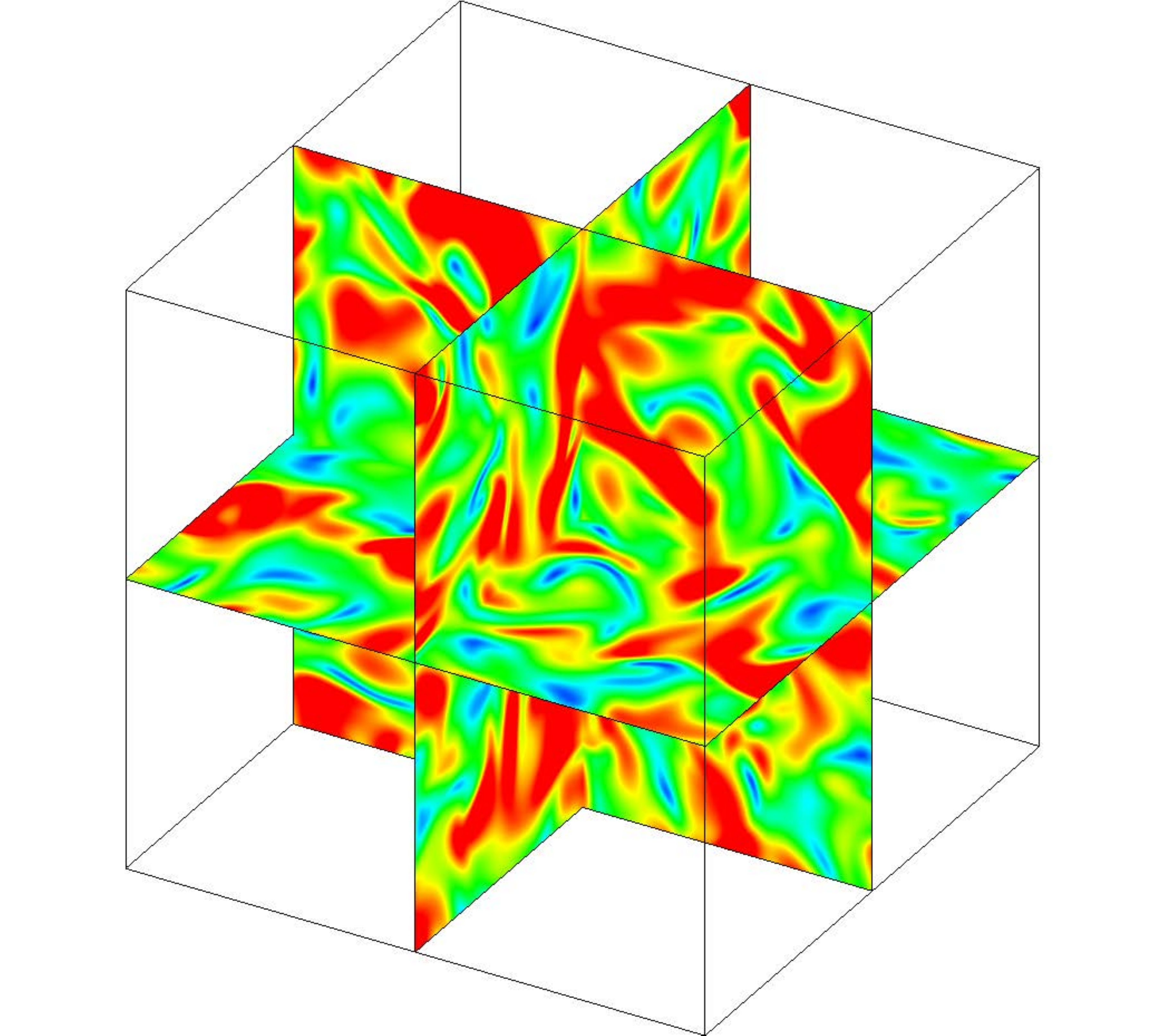}
			\label{3000vsingle}
		\end{minipage}
	}
	\subfigure[$t^*=0.403$]{
		\begin{minipage}[t]{0.22\linewidth}
			\centering
			\includegraphics[width=1.05\columnwidth,trim={3cm 0cm 3cm 0cm},clip]{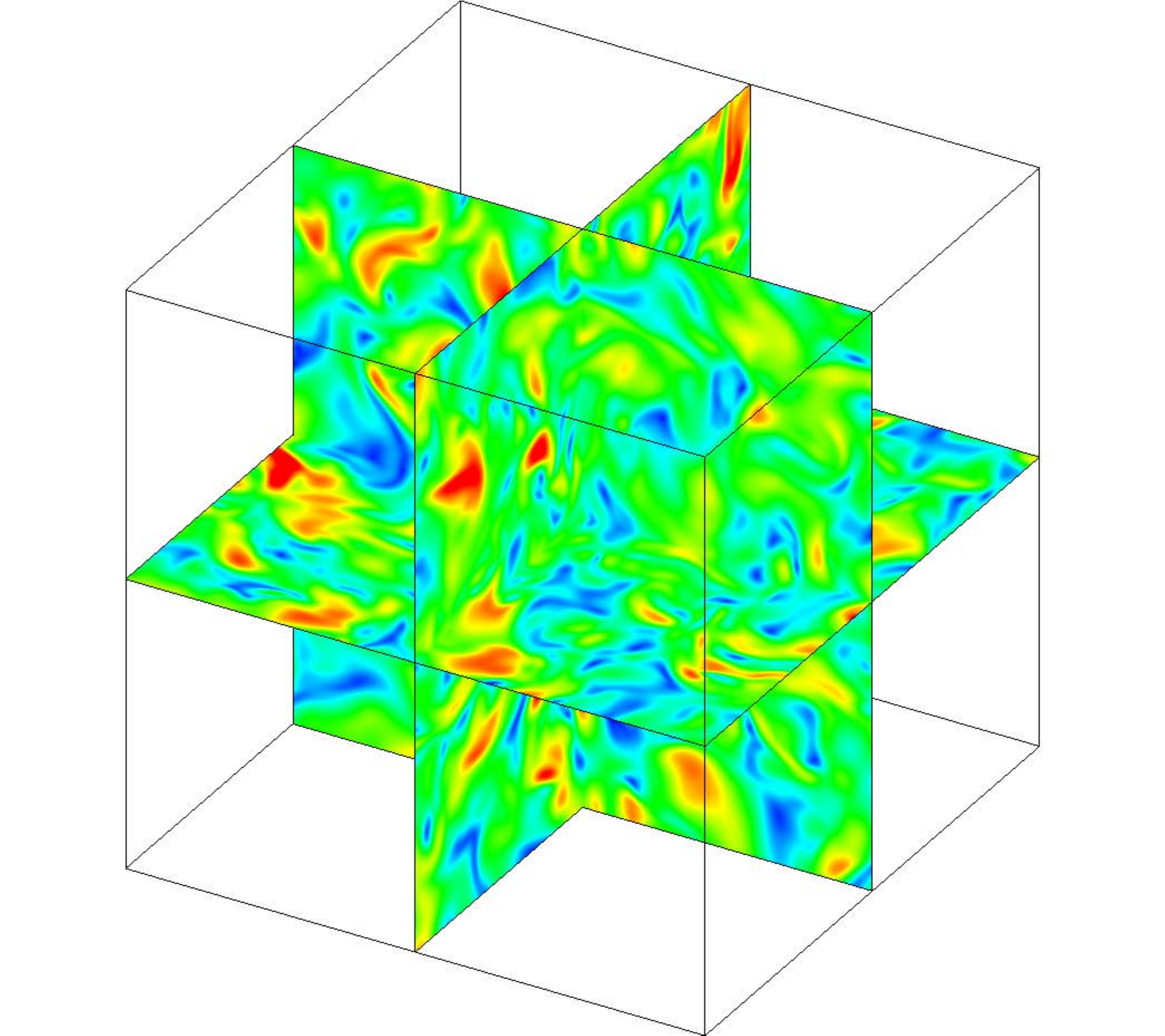}
			\label{12000vsingle}
		\end{minipage}
	}
	\subfigure[$t^*=0.896$]{
		\begin{minipage}[t]{0.22\linewidth}
			\centering
			\includegraphics[width=1.05\columnwidth,trim={3cm 0cm 3cm 0cm},clip]{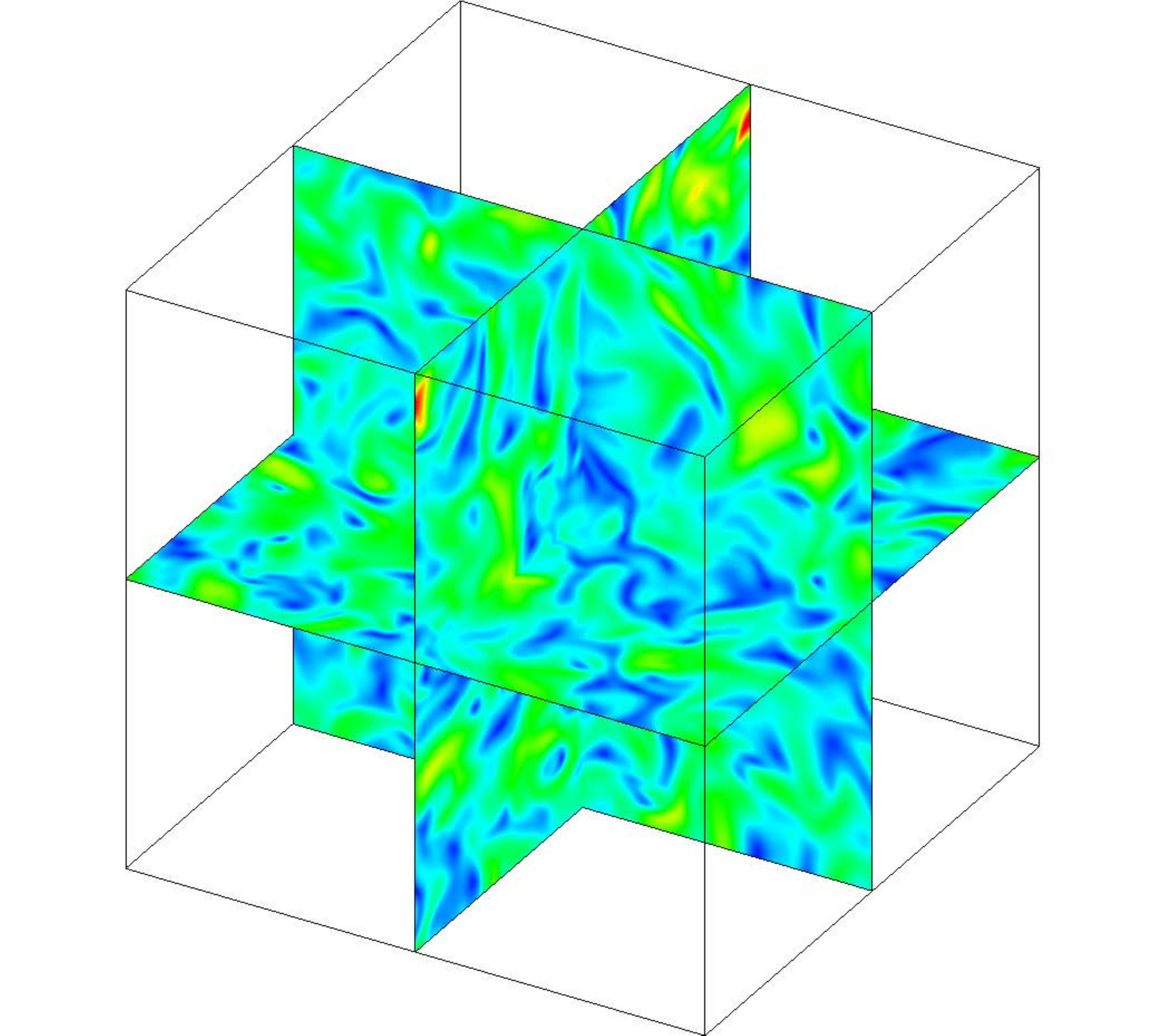}
			\label{23000vsingle}
		\end{minipage}
	}
	\subfigure[$t^*=5.15$]{
		\begin{minipage}[t]{0.22\linewidth}
			\centering
			\includegraphics[width=1.05\columnwidth,trim={3cm 0cm 3cm 0cm},clip]{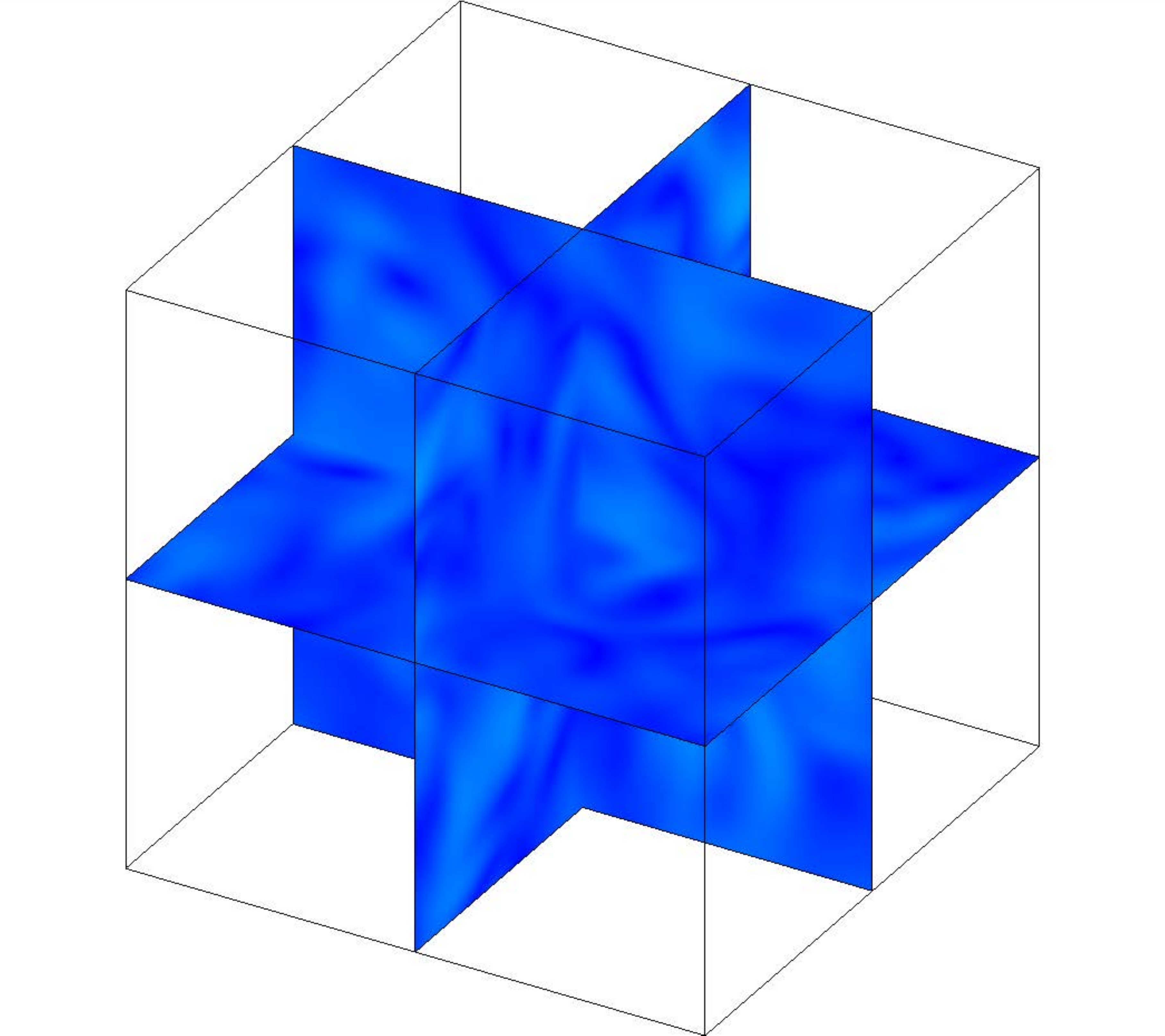}
			\label{118000vsingle}
		\end{minipage}
	}
	\\
	\begin{minipage}[t]{0.15\linewidth}
		\centering
		\includegraphics[width=0.6\columnwidth,trim={2cm 4cm 2cm 2cm},clip]{pngpdf/xyz1.pdf}
	\end{minipage}
	\begin{minipage}[t]{0.38\linewidth}
		\centering
		\includegraphics[width=1.0\columnwidth,trim={0cm 1.5cm 0cm 1.5cm},clip]{pngpdf/u.pdf}
	\end{minipage}
	\centering
	\caption{Velocity magnitude $u^*=\left| \boldsymbol{u}\right|/\left[ u^{\prime}(t^*=0)\right]$ on the central planes in the three directions at different times. The first row is for the two-phase flow and the second row for the  single-phase flow. The purple lines in the first row are the intersecting lines between fluid-fluid interface $\phi=0.5$ and the central planes.}
	\label{vel3}
\end{figure*}

Since the initial setting of the physical problem is also approximately spherically symmetric, we also show the kinetic energy on the spheres with different radii from the center of the system in Fig.~\ref{spherical}, to examine the energy distribution at different scales on the surfaces of spheres. 
The corresponding velocity fields are shown in Fig.~\ref{vel3}. 
The effects of the fluid-fluid interfaces could be more clearly shown, particularly during the deformation stage.
The main influencing factors here are the relative volume fractions of the two phases and the distribution of the fluid-fluid interface, at 
different distances from the center of the initial droplet.

At $t^*=0$ (Fig.~\ref{El3000breakup256rmr}), the kinetic energy spectra on different spherical surfaces almost overlap one another at large wavenumbers, because the initial turbulent flow field is homogeneous and isotropic.
Furthermore, the energy spectra of two-phase flow and single-phase flow are exactly the same at this time, because of the same initial velocity field,
as shown in Fig.~\ref{3000vrmsCD3phi}) and Fig.~\ref{3000vsingle}).

At an early time $t^*=0.403$, turbulence distorts the interface significantly while the interfaces slightly weaken the nearby velocity magnitude (Fig.~\ref{12000vrmsCD3phi}), compared to the corresponding regions of single-phase flow (Fig.~\ref{12000vsingle}). 
Compared to Fig.~\ref{El3000breakup256rmr}, Fig.~\ref{El12000breakup256rmr} shows that the kinetic energies decay in the large scales, while increase 
in the small scales. 
The kinetic energies of two-phase flow are smaller than those of single-phase flow in large scales, but larger in small scales. 
The reasons are the same as those for Fig.~\ref{Ek}. 
The main difference of Fourier expansion and spherical harmonics expansion is that, Fig.~\ref{Ek} is the expansion in the whole simulation region, 
while Fig.~\ref{spherical} is the expansion on the specific spherical surfaces. As a result, 
the kinetic energy in Fig.~\ref{Ek} can be viewed closed, which cannot transfer to other space, since the boundary conditions are periodic. But the kinetic energy in Fig.~\ref{spherical} is open, it can transfer from one spherical surface to other spherical surfaces. 
However, the energy spectra for spherical harmonics and Fourier appear to be similar.

When $t^*=0.896$ (Fig.~\ref{El23000breakup256rmr}), the general shape of spherical harmonics spectra are similar to those at $t^*=0.403$. However, the effect of droplets on the velocity magnitude field becomes significant at this time. The velocity magnitude inside the droplet is less than that outside the droplet (see Fig.~\ref{23000vrmsCD3phi}), and it is also less than that in the corresponding regions of single-phase flow field (compared to Fig.~\ref{23000vsingle}). 
Because there is a deformed droplet at the center of the computational domain at this time, it appears that the fluid-fluid interface or surface tension force limits the movement of the fluid inside the droplet, reducing the flow velocity fluctuations and kinetic energy.
Scarbolo \& Soldati~(2013)~\cite{scarbolo2013turbulence} also drew a similar conclusion, that the velocity magnitude
is damped inside the droplet in a turbulent channel flow simulation. Outside the droplets, the velocity magnitude contour of two-phase flow is still similar to that of single-phase flow, because a large volume fraction of the bulk region is away from the interface. 
From the point of view of spherical harmonics spectra, the large scale two-phase kinetic energy on the small spherical surface $r=0.4R_0$ is much less than that on the larger spherical surfaces, and single-phase $r=0.4R_0$ sphere, as shown in Fig.~\ref{El23000breakup256rmr}.

At $t^*=5.15$ (see Fig.~\ref{118000vrmsCD3phi} and Fig.~\ref{118000vsingle}), the velocity magnitude inside the droplets
remains smaller, but not as obvious as before, since the velocity magnitude is small in the whole flow field for the later period of DHIT. Furthermore, the velocity magnitude outside the droplets are smaller
than that in the single-phase flow, because the stored interface energy by small daughter droplets could not return
fully back to the kinetic energy, even at this late time. The spherical harmonics spectra also shows the same results (Fig.~\ref{El118000breakup256rmr}).

Quantitatively, the slopes in the high wavenumber range of kinetic energy in the spherical harmonics spaces are usually about $-6$ for single-phase flow, while the values are between $-4$ and $-3$ for the two-phase flow, which are similar to that in the Fourier spectra in Fig.~\ref{Ek}.

\subsection{Interaction between kinetic energy and free energy}
\begin{figure*}
	\centering    
	\subfigure[Deformation stage]{
		\begin{minipage}[t]{0.4\linewidth}
			\centering
			\includegraphics[width=0.8\columnwidth,trim={0cm 0cm 0cm 0cm},clip]{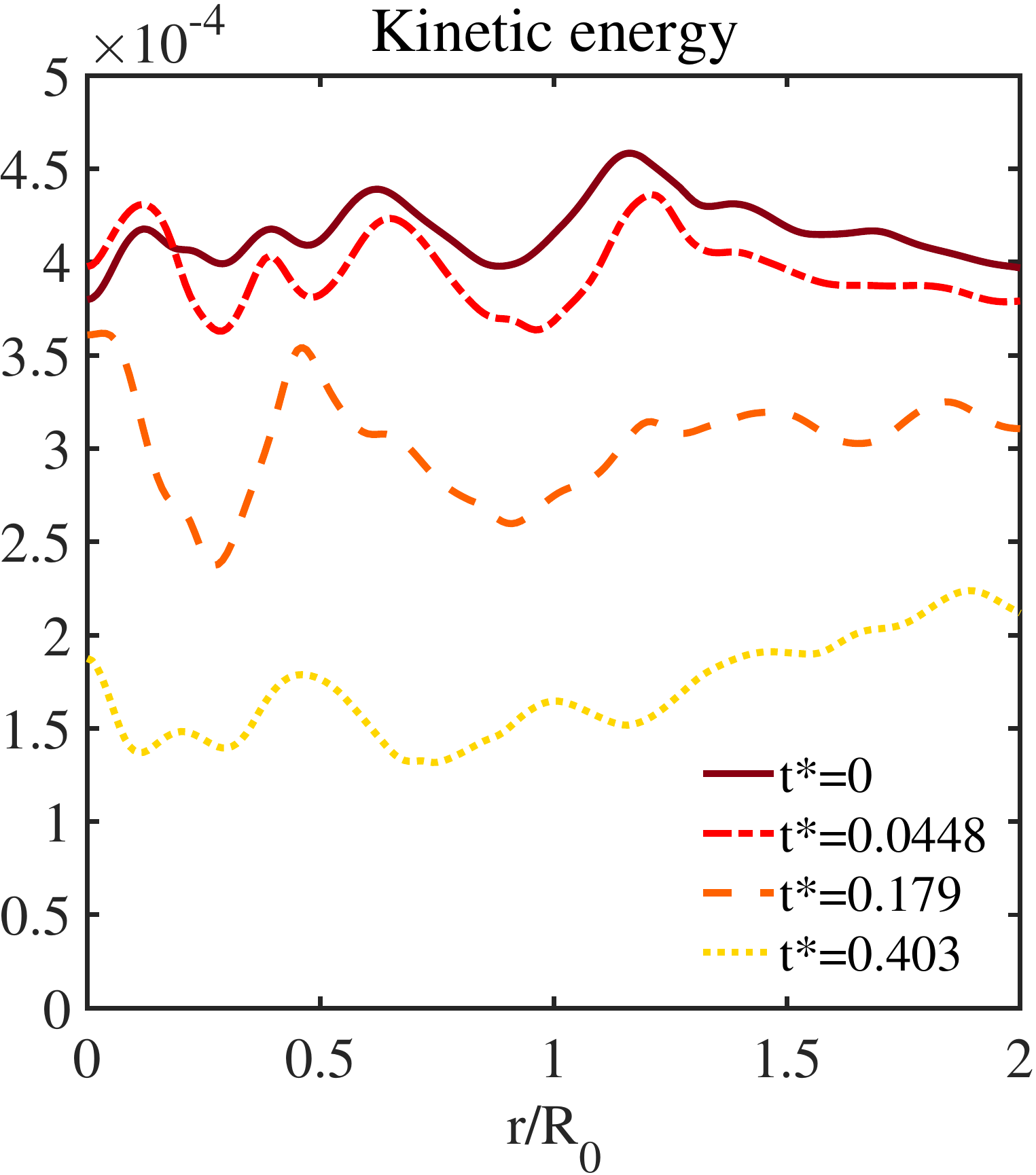}
			\label{1}
		\end{minipage}
		\begin{minipage}[t]{0.4\linewidth}
			\centering
			\includegraphics[width=0.8\columnwidth,trim={0cm 0cm 0cm 0cm},clip]{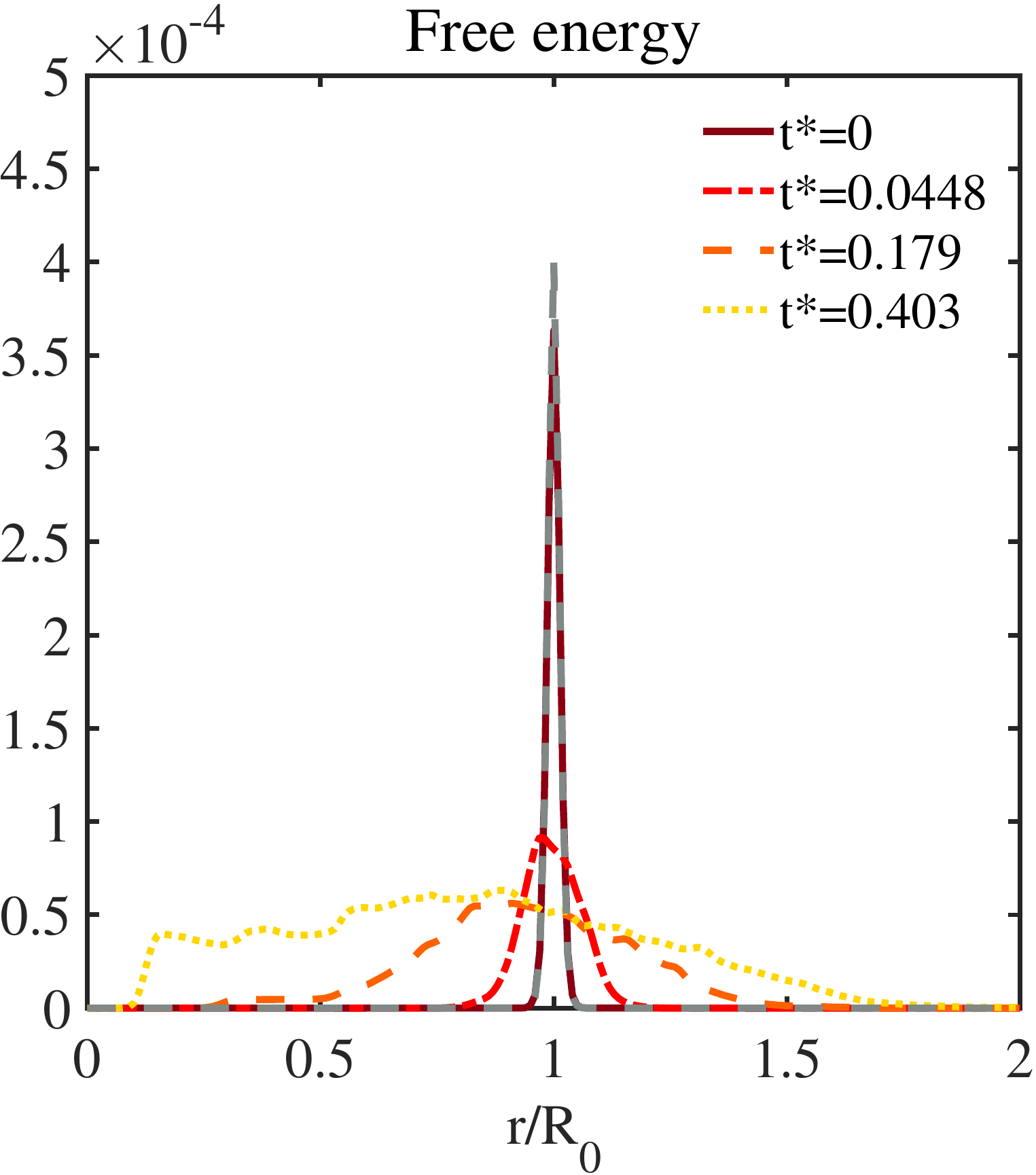}
			\label{2}
		\end{minipage}\label{Deformation}
	}
	\\
	\subfigure[Breakup stage]{
		\begin{minipage}[t]{0.4\linewidth}
			\centering
			\includegraphics[width=0.8\columnwidth,trim={0cm 0cm 0cm 0cm},clip]{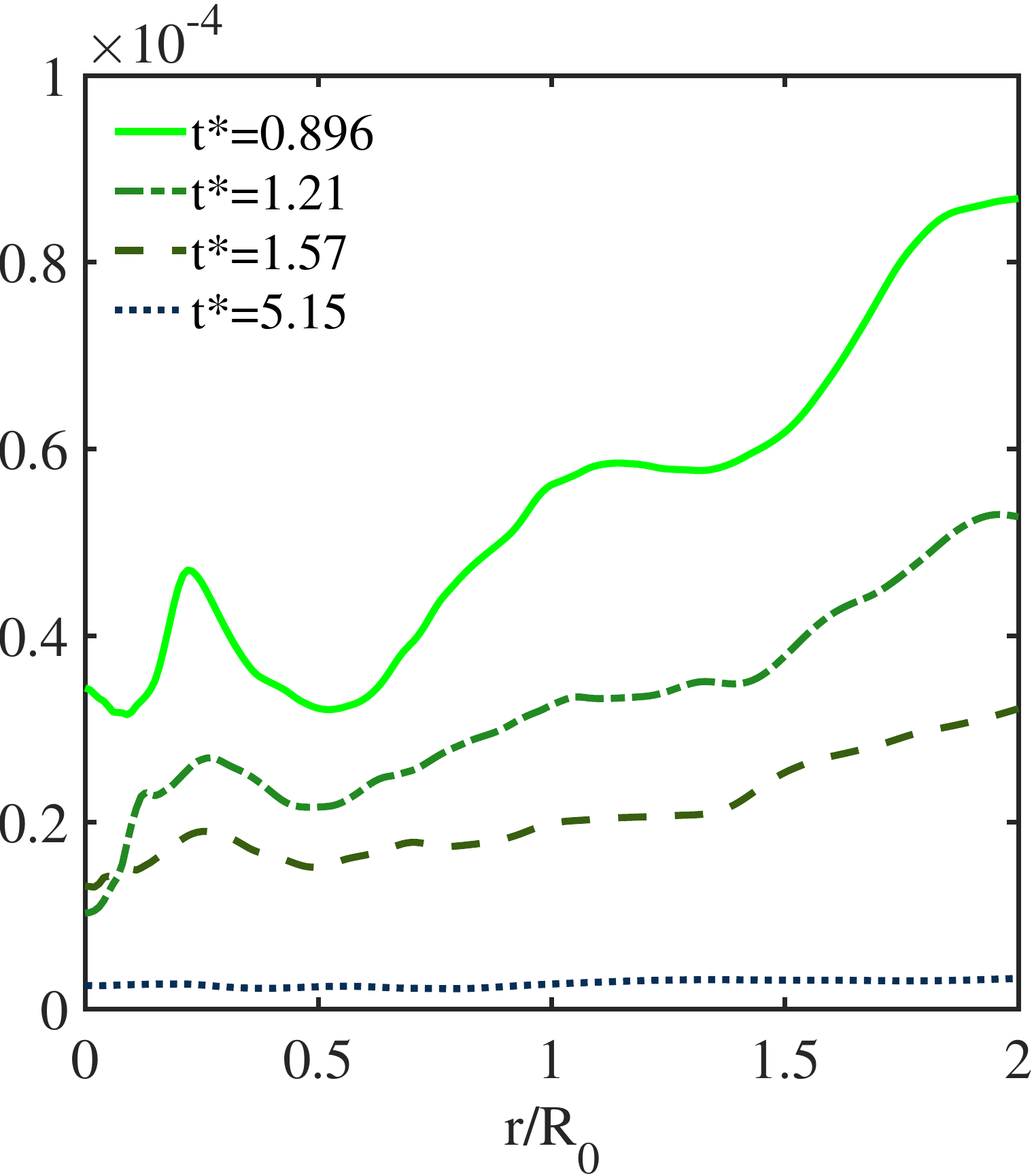}
			\label{3}
		\end{minipage}
		\begin{minipage}[t]{0.4\linewidth}
			\centering
			\includegraphics[width=0.8\columnwidth,trim={0cm 0cm 0cm 0cm},clip]{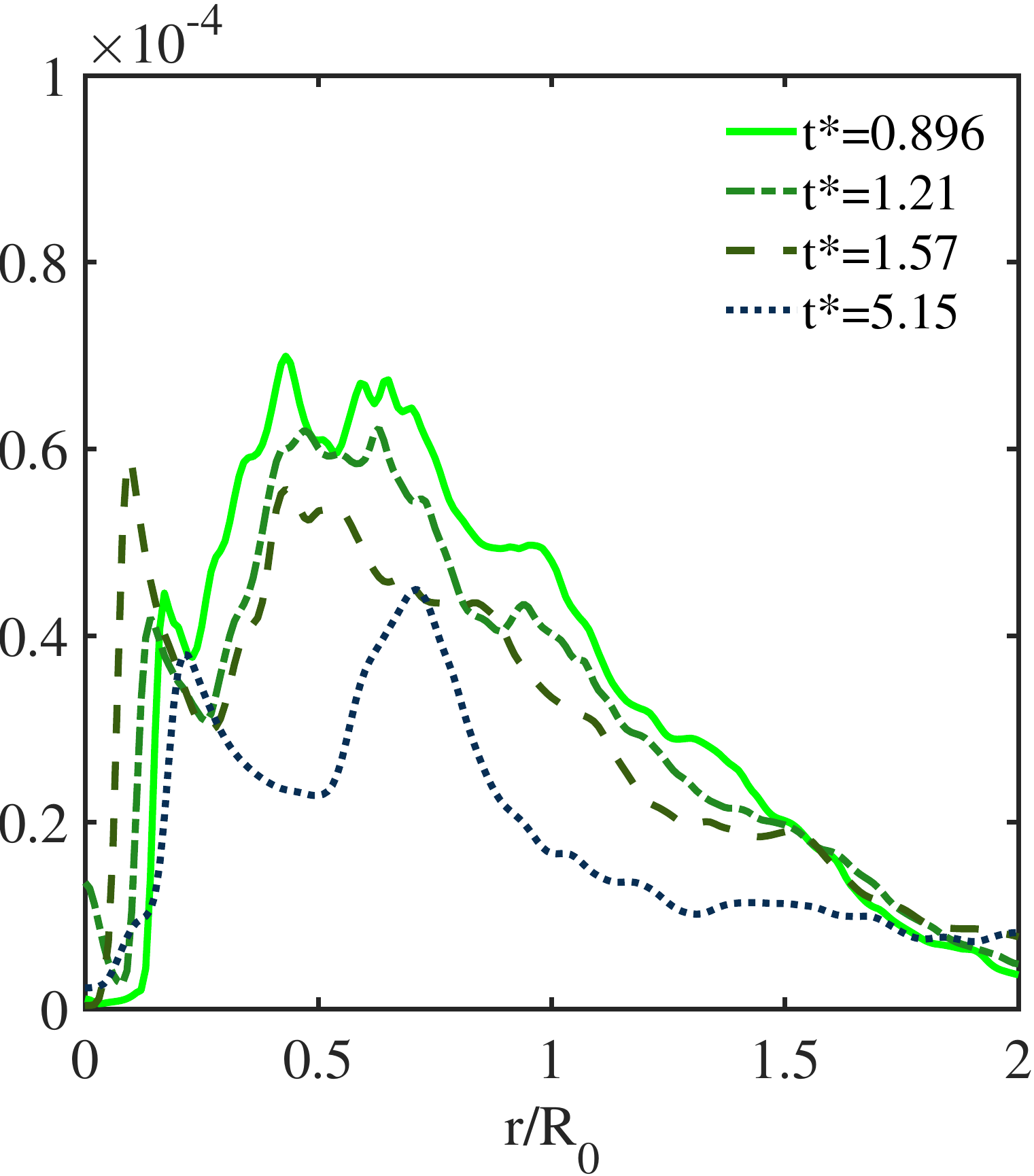}
			\label{4}
		\end{minipage}\label{Breakup}
	}
	\\
	\subfigure[Restoration stage]{
		\begin{minipage}[t]{0.4\linewidth}
			\centering
			\includegraphics[width=0.8\columnwidth,trim={0cm 0cm 0cm 0cm},clip]{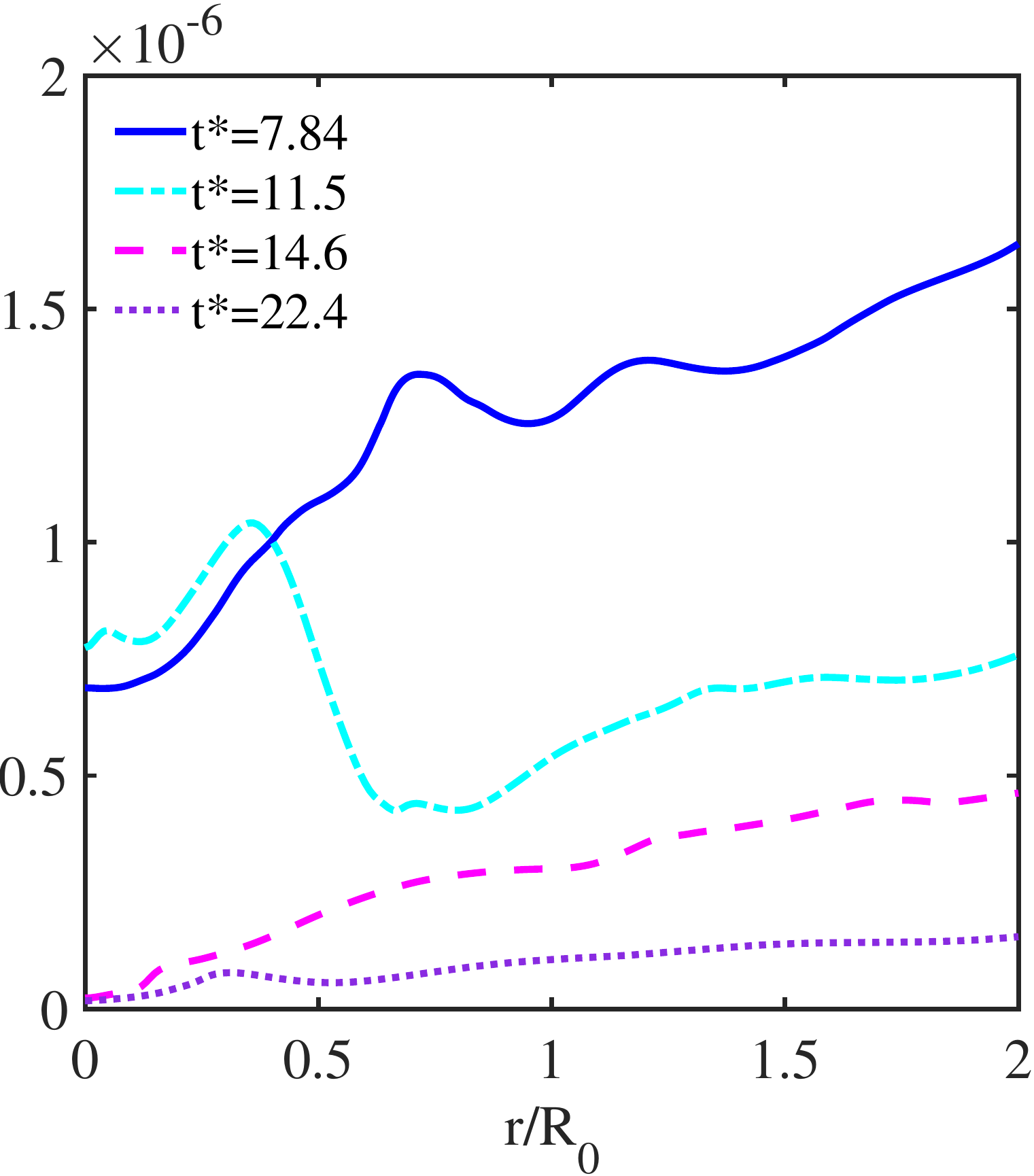}
			\label{5}
		\end{minipage}
		\begin{minipage}[t]{0.4\linewidth}
			\centering
			\includegraphics[width=0.8\columnwidth,trim={0cm 0cm 0cm 0cm},clip]{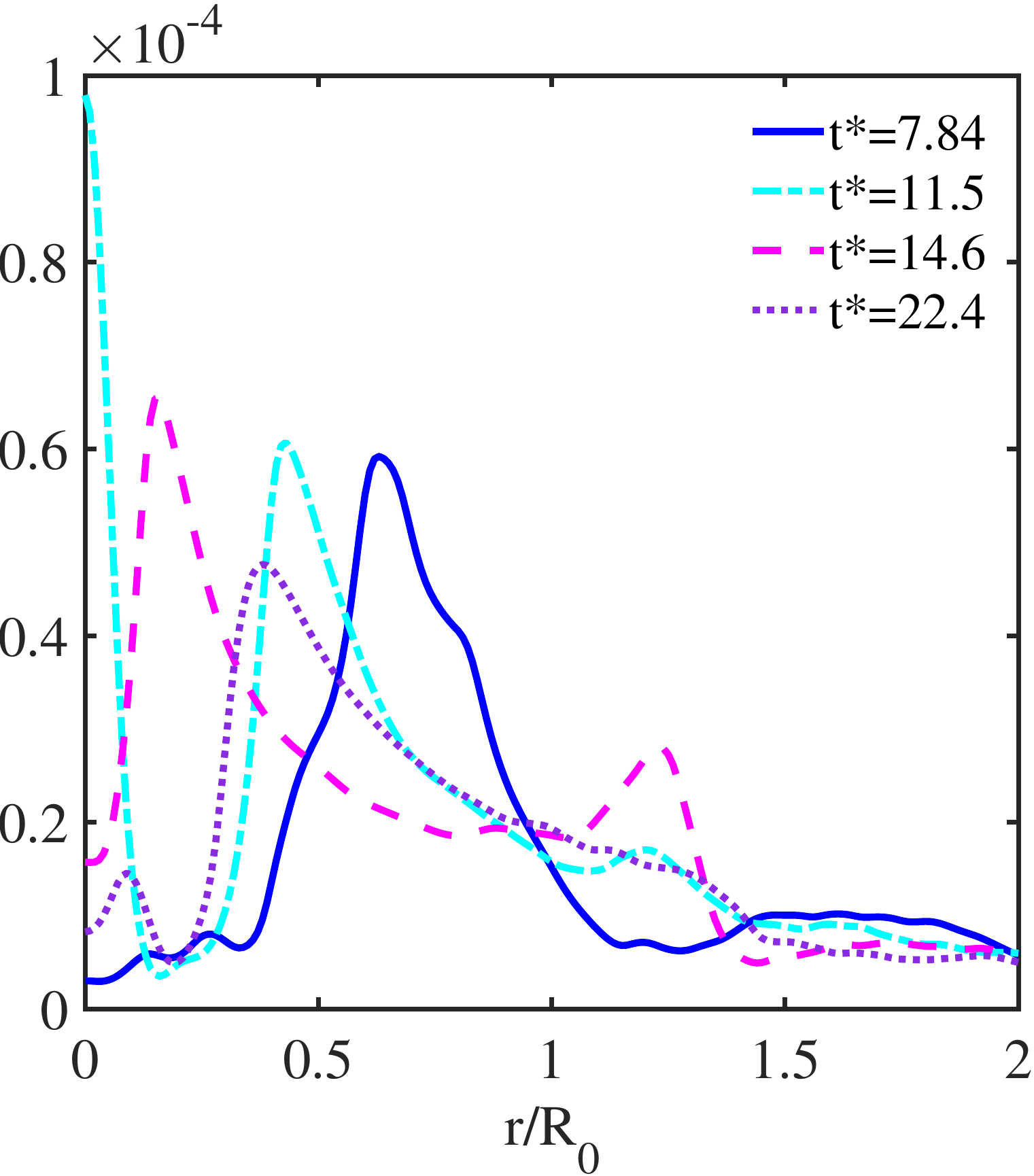}
			\label{6}
		\end{minipage}\label{Restoration}
	}
	\centering
	\caption{Spherically averaged energies as a function of radial distance from the initial droplet center at different times. 
	The gray dashed line in (a) is the analytical result of free energy at the initial time based on Eq.~(\ref{phini}).}
	\label{energystages}
\end{figure*}

To explicitly demonstrate the exchange between kinetic energy and free energy, 
Fig.~\ref{energystages} shows  the spherical averaged kinetic energy and free energy as a function of distance from the initial droplet center,
at different times during the three stages.  At $t^*=0$, the magnitude of free energy near the interface is close to kinetic energy 
because of the parameters setting, while it is 
negligible away from the interface. The averaged kinetic energy is almost independent of the radial distance from the center, because of the homogeneous isotropic flow initialization. 
Therefore, the kinetic energy is much larger than the free energy in the
integral sense  at the initial time. As time proceeds, the kinetic energy decays because of the viscous dissipation and 
the increasing free energy, while the free energy spreads over a larger region because of the droplet deformation. 
The total kinetic energy is still much larger than the free energy during the deformation stage, as shown in Fig.~\ref{Deformation}. 
During the breakup stage (Fig.~\ref{Breakup}), the kinetic energy has decreased to a magnitude which is comparable to the free energy. 
At the restoration stage (Fig.~\ref{Restoration}), the kinetic energy is much less than the free energy, 
only about 1\% of the latter. 
The above analysis, again, illustrates that (a) during the deformation stage,  kinetic energy plays a dominant role on the interface evolution with
the free energy rapidly increasing at the cost of the kinetic energy, (b) during the breakup stage,
kinetic energy and free energy are interchanged and comparable, and (c) the free energy plays a dominant role during
the restoration stage. As shown in Fig.~\ref{iso}, most of the interface region is close to the center of the computational domain. Therefore, at the restoration stage (see Fig.~\ref{Restoration}), the free energy is concentrated near the center, 
resulting in a small kinetic energy near the center partially due to the exchange between the two energy forms. 

\begin{figure}[t!]
	\centering
	\includegraphics[width=0.65\columnwidth,trim={0cm 0cm 0cm 0cm},clip]{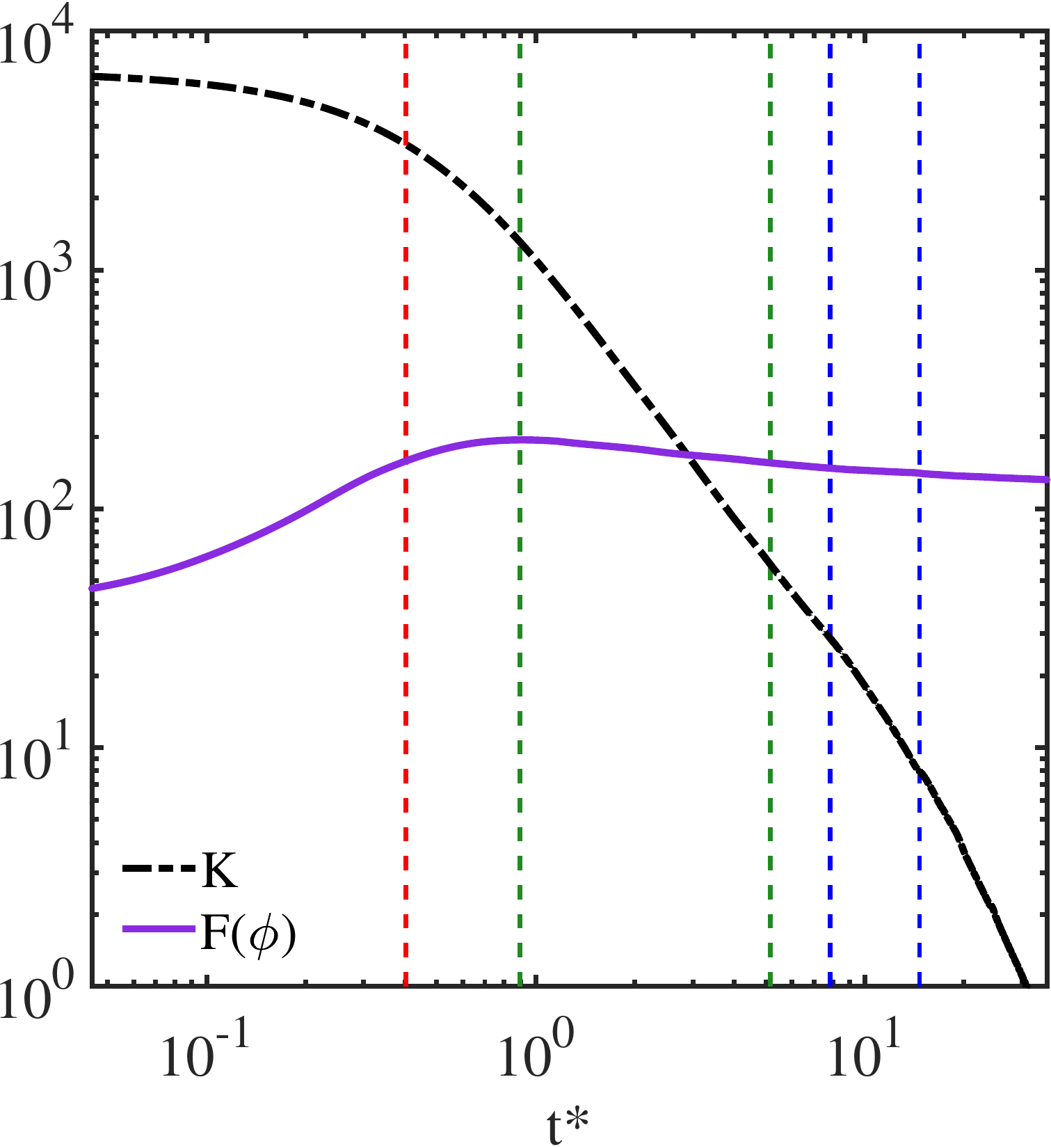}
	\centering
	\caption{Time evolution of the overall kinetic energy and free energy. The vertical dashed lines  from left to right mark $t^*$=0.403,0.896,5.15,7.84,14.6, respectively.}
	\label{E}
\end{figure}

The total kinetic energy and free energy as a function of time  are shown in Fig.~\ref{E}.  Again, the results demonstrate that the total kinetic energy is much larger than the total free energy during the deformation stage, 
becomes comparable to the total free energy during the breakup stage, and falls off to a value much smaller than the total free energy during the restoration stage. 
The total interface area and total free energy increases rapidly during the deformation stage, by a factor of about 5, then decreases
slowly with time afterwards.

\section{Conclusions}\label{sec: Con}

A three-dimensional (3D) direct numerical simulation (DNS) code is developed to simulate immiscible two-phase turbulent
flows based on the PF-DUGKS approach. This approach solves equivalently the Cahn-Hilliard-Navier-Stokes (CHNS) system by the use of two Boltzmann equations.
After a stationary droplet and a freely-deforming droplet in a turbulent background flow were simulated 
to test the PF-DUGKS  code,
we focused on applying the approach to study  the breakup process of a large spherical droplet in a DHIT background flow. 
The breakup process was observed to consist of three distinct stages. The results for the two-phase flow simulations are compared with those of the single-phase ($\phi=0$) decaying turbulence using the same code. 
The comparisons indicate that the phase-field approach can be coupled with DUGKS to simulate three-dimensional immiscible two-phase turbulent flows. 
Main results are summarized as follows.

A 3D static droplet is simulated
to show that the kinetic energy in the spurious currents are negligibly small when compared to the free energy.
In the case of a freely-deforming droplet ({\it i.e.}, with negligible surface tension effects) in decaying turbulence, 
the kinetic energy spectra on spherical surfaces based on the two-phase setting (with a negligible surface tension) are compared with
those based on the single-phase setting 
($\phi=0, \sigma \sim O(10^{-3})$). The comparisons show that 
the two limiting approaches yield the same results, implying that 
(1) the two-phase flow at very large $We$ number ($\sim O(10^{17})$) is similar to the single-phase flow because the effect of surface tension on the flow field is 
essentially  removed, (2) the use of a constant order parameter $\phi = 0$ in the whole flow field  
naturally  reduces the solved system to a single-phase flow. 
Furthermore, our results demonstrate that the spherical harmonics can help understand the simultaneous 
energy transfer process along the radial direction and across different length scales, taking advantage of the fact that
the chosen physical problem is spherically symmetric.

The breakup of a large spherical droplet in DHIT is the main focus of this study, thus we provide
an in-depth analysis of the numerical results from several different angles.  
The main parameters we used are as follows: density ratio and viscosity ratio are both set to one, the dispersed phase volume fraction $\varphi$ is 6.54\%, the initial Weber number $\We$ is 21.7, and the initial Taylor microscale Reynolds number $Re_\lambda$ is 58.
The simulation results show that the breakup of a spherical droplet in DHIT can be separated into 
three stages, namely, the deformation stage, the breakup stage and the restoration stage. 
It is shown that the dividing time between the breakup stage and the restoration stage can be roughly described by the Hinze criterion.
The deformation stage takes the shortest time, while the restoration stage takes the longest time. 

During the deformation stage, the initial spherical droplet wrinkles and elongates in every direction.
The droplet interface becomes bumpy and the interface curvature increases. 
The total interface area is increased by a factor larger than 4 at the end of this first stage.   
The kinetic energy is larger than free energy at this stage, and plays a dominant role, which can be also interpreted by the value of Weber number ($\We>>1$). 

During the breakup stage, the droplet continues to stretch and break, forming
a large numbers of daughter droplets (about 60 in the end of this stage) of different sizes. 
Roughly, over 60\% of droplets in number are with a diameter less than $0.1D_0$.
The total interface area becomes larger due to droplet breakup, 
but at the same time is reduced due to the tendency for daughter droplets to restore the spherical shape. 
The overall trend of interface area in the breakup stage is a reduction of the interface area.
The Weber number at this stage is of the order one. The kinetic energy and free energy are interchanged and comparable during this stage, and they influence the evolution of droplets, together. 

During the restoration stage, all the droplets relax towards spherical shape.
Therefore, the total area decreases. Occasionally two droplets close to each other could coalesce, disturbing the nearby flow field and the total area. It is shown that $\We <<1$ at the stage.
The restoration stage is mainly driven by surface tension force, since the kinetic energy is very small compared to the free energy.

Since the droplet is initialized in the center of the box, the free energy is distributed around the center in the whole evolution period. 
Although the kinetic energy is homogeneous isotropic at the beginning, 
its distribution can become distance-dependent, due to the reduction of velocity field by fluid-fluid interface or by surface tension force, or the conversion from kinetic energy to free energy. For the same reason, the velocity magnitude near the interface is attanuated during the whole evolution. However, the vorticity magnitude field near the interface is argumented, since the fluid-fluid interface boundary layer 
could generate vorticity.

The power-law decay of the two-phase kinetic energy at the breakup stage has an exponent $1.76$,
while it is $1.65$ for single-phase flow during the same time, as a part of the kinetic energy is converted to free energy for the two-phase flow. In the spherical harmonics space, the power-law slopes in the high wavenumber region are usually about $-6$ for single-phase flow, while the values are between $-4$ and $-3$ for two-phase flow, which are similar to the power-law slope in the Fourier spectra.

The study of interface evolution in immiscible fluids, in particular under more general conditions ({\it e.g.}, higher density ratio), is relevant to the atomization process.~\cite{canu2018does,cheron2019droplets} 
These general conditions under the PF-DUGKS approach have been addressed by combing with conservative Allen-Cahn equation.~\cite{Yangzeren2019,doi:10.1063/5.0086723}
Their applications to three-dimensional complex two-phase flows will be
a subject of further investigation.

\begin{acknowledgments}
This work has been supported by the National Numerical Wind Tunnel program, the Taizhou-Shenzhen Innovation Center, 
the National Natural Science Foundation of China (NSFC award numbers 91852205, 91741101 \& 11961131006), NSFC Basic Science Center Program (Award number 11988102), 
Guangdong Provincial Key Laboratory of Turbulence Research and
Applications (2019B21203001), Guangdong-Hong Kong-Macao Joint Laboratory for Data-Driven Fluid Mechanics and Engineering Applications (2020B1212030001) and Shenzhen Science and Technology Program (Grant No. KQTD20180411143441009). Computing resources are provided by
the Center for Computational Science and Engineering of Southern University
of Science and Technology. The authors also wish to thank Dr. Ch\'eron Victor P. Gerard, Mr. Zelong Yuan, and Mr. Mingyu Su
for helpful discussions.
\end{acknowledgments}
	
\appendix

\section{Inverse design of the model Boltzmann equations for the CHNS system based on the Chapman-Enskog analysis} 
\label{sec:inverse_design}

The purpose of this appendix is to derive all the moment-integral constraints for $f^{eq}, g^{eq}, S_{\alpha}^f$, and  $S_{\alpha}^g$  in
Eqs.~(\ref{Boltzmann}a-b),  in order to
reproduce the CHNS system, Eqs.~(\ref{macroEq}a-c). 
Two model Boltzmann equations, {\it i.e.}, Eqs.~(\ref{Boltzmann}a-b), are introduced, with  $f$ representing
 the pressure/velocity distribution function  and $g$ the order-parameter distribution function, {\it i.e.},
\begin{equation}\label{fg}
p=\int f d \boldsymbol{\xi} ,~~RT \rho {u}_j= \int f {\xi}_j d \boldsymbol{\xi} ,~~  \phi =  \int g d \boldsymbol{\xi} .
\end{equation}
As noted previously, the pressure $p$ and density $\rho$ are independent. The density field $\rho$ is
solely determined by the phase field $\phi$ according to Eq.~\eqref{rholinear}.
Since $p$ and $u_j$ share the same distribution $f$, thus $\rho RT u_j$ is viewed as the first moment of $f$.
The equilibrium $f^{eq}$ is designed by modifying the standard
Maxwellian distribution, under the Hermite expansion, as 
\begin{equation}\label{feqH}
 f^{eq}  = w( \boldsymbol{\xi} ) \left[ p + \rho RT \left(  { \boldsymbol{\xi}\cdot  \mathbf{u}  \over RT} +  {1\over 2}  
  { ( \boldsymbol{\xi}\cdot  \mathbf{u} )^2  \over (RT)^2 } -  {1\over 2}  
  { u^2  \over RT  } + {\cal O} (Ma^3 )  \right) \right],  
\end{equation}
where 
\begin{equation}\label{Hweight}
 w( \boldsymbol{\xi} )\equiv {1\over (2 \pi RT)^{D/2}} \exp \left( - {\xi^2 \over 2RT } \right) .
\end{equation}
It follows that
\begin{subequations}\label{ddt}
\begin{equation}\label{feq0}
\int f^{eq} d \boldsymbol{\xi}  =p,  
\end{equation}
\begin{equation}\label{feq1}
\quad\int f^{eq} {\xi}_j d \boldsymbol{\xi}  =RT \rho {u}_j, 
\end{equation}
\begin{equation}\label{feq2}
  \int \xi_j  \xi_m  f^{eq}  d \boldsymbol{\xi} = RT \left ( \rho u_j u _m + p \delta_{jm} \right),
\end{equation}
\begin{equation}\label{feq3}
  \int \xi_j  \xi_m  \xi_n  f^{eq}  d \boldsymbol{\xi} = \rho  (RT)^2 \left (  u_j \delta_{mn} +   u_m \delta_{jn}  + u_n \delta_{mj}  \right).
\end{equation}
\end{subequations}
It is noted that the above Hermite expansion is consistent with the D3Q19 lattice velocity model used in this paper.


Eqs.~\eqref{feq0} and~\eqref{feq1} imply the collision term in Eq.~(\ref{Boltzmannf}) makes no contribution to the zeroth-order
and first-order moment equations. Similarly, we require that the collision term in Eq.~(\ref{Boltzmanng}) conserves the phase field
by setting 
\begin{equation}\label{fgeq}
\quad \int g^{eq} d \boldsymbol{\xi}  =\phi.
\end{equation}

The continuity equation is derived from the zeroth-order moment of Eq.~(\ref{Boltzmannf}), {\it i.e.},
\begin{equation}\label{intfBoltzmann}
\int\left[\frac{\partial f}{\partial t}+{\xi}_m\frac{\partial f}{\partial x_m} =-\frac{f-f^{e q}}{\tau_{f}}+S^{f}\right] d \boldsymbol{\xi}.
\end{equation}
Substituting Eq.~(\ref{fg}), Eq.~(\ref{fgeq}) into Eq.~(\ref{intfBoltzmann}),
yields
\begin{equation} \label{CEcontinuity}
\frac{\partial p}{\partial t}+\frac{\partial }{\partial x_m}\left(RT \rho {u}_m\right)=0+
\int S^f d\boldsymbol{\xi}.
\end{equation}
Comparing this to the continuity equation Eq.~(\ref{EqMa}),
we need to set the zeroth-order moment of $S^f$ to
\begin{equation}\label{intSf}
\int S^f d\boldsymbol{\xi}=RT\left[ {u}_m \frac{\partial \rho}{\partial x_m}-\rho \gamma \frac{\partial }{\partial x_m} \left( M_{CH} \frac{\partial \mu_{\phi}}{\partial x_m} \right)\right].
\end{equation}

To proceed further, we introduce the Chapman-Enskog expansion of the distribution functions
\begin{subequations}
\begin{equation}\label{CE1}
f = f^{eq} - \tau_f \left( {\partial f^{eq} \over \partial t} + \xi_j {\partial f^{eq} \over \partial x_j} - S^{f}   \right) + {\cal O} (\tau_f^2 ), 
\end{equation}
\begin{equation}\label{CE2}
g = g^{eq} - \tau_g \left(   {\partial g^{eq} \over \partial t} + \xi_j {\partial g^{eq} \over \partial x_j}  - S^{g}   \right) + {\cal O} (\tau_g^2 ).
\end{equation}
\end{subequations}
Note that the continuity equation remains the same at the different orders in the Chapman-Enskog expansion. 

To the leading order, the first-order moment equation of Eq.~\eqref{Boltzmannf}  and 
 the zeroth-order moment equations of Eq.~\eqref{Boltzmanng}  yield
 	\begin{subequations} 
		\begin{equation} \label{NS0}
		RT \frac{\partial (\rho u_{j}) }{\partial t} + RT  {\partial  (\rho u_m u_j  + p\delta_{mj} )  \over \partial x_m}  = 0 +   \int S^f \xi_j d \boldsymbol{\xi} + {\cal O} (\tau_f),
		\end{equation}
		\begin{equation} \label{CH0}
		\frac{\partial \phi}{\partial t} +  {\partial  \over \partial x_m} \int   \xi_m  g^{eq}  d \boldsymbol{\xi} =  \int S^g d \boldsymbol{\xi}  +  {\cal O} (\tau_g).
		\end{equation}
\end{subequations}
These two equations should reproduce the momentum equation and the CH equation to the same leading order, namely, without the diffusion terms. A convenient choice for the three involved moments is
 	\begin{subequations} 
	         \begin{equation} \label{intSf1}
		  \int S^f \xi_j d \boldsymbol{\xi}  =  RT F_j,
		\end{equation}
		\begin{equation} \label{intgeq1}
		 \int   \xi_m  g^{eq}  d \boldsymbol{\xi} =  \phi u_m,
		\end{equation}
		\begin{equation} \label{intSg}
		   \int S^g d \boldsymbol{\xi} =  0.
		\end{equation}
\end{subequations}

The leading-order equations, Eqs.~\eqref{EqMa}~\eqref{NS0}~\eqref{CH0}, together with Eq.~\eqref{rholinear}, allow us to express all the time derivatives,  to the leading-order, as
	\begin{subequations}\label{ddt}
		\begin{equation}\label{dphidt}
		\frac{\partial \phi}{\partial t}=-\frac{\partial }{\partial x_m}\left(\phi u_{m} \right) 
		+{\cal O}\left(\tau_{g} \right) ,
		\end{equation}
		\begin{equation}\label{drhodt}
		\frac{\partial \rho}{\partial t}= -\frac{\partial }{\partial x_m}\left(\rho u_{m} \right)+\frac{\left(\rho_{A}-\rho_{B}\right)}{\left(\phi_{A}-\phi_{B}\right) } {1\over \gamma } \frac{\partial u_{m}}{\partial x_m}
		+{\cal O}\left(\tau_{g} \right),
		\end{equation}
		\begin{equation}\label{dpdt}
		\frac{\partial p}{\partial t}=-\rho R T \frac{\partial u_{m}}{\partial x_m}-\rho R T \gamma \frac{\partial }{\partial x_m} \left( M_{CH} \frac{\partial \mu_{\phi}}{\partial x_m} \right),
		\end{equation}    	
		\begin{equation}\label{dudt}
		\frac{\partial u_{j}}{\partial t}=-u_{m} \frac{\partial u_{j}}{\partial x_{m}}- \frac{\left(\rho_{A}-\rho_{B}\right)u_{j}}{\left(\phi_{A}-\phi_{B}\right)\gamma\rho}  \frac{\partial u_{m}}{\partial x_m}-\frac{1}{\rho} \frac{\partial p}{\partial x_{j}}+\frac{F_{j}}{\rho}
		+{\cal O}\left(\tau_{f} \right)
		+{\cal O}\left(\tau_{g}Ma \right).
		\end{equation}
	\end{subequations}
	where $Ma = U/\sqrt{RT}$ denotes the Mach number with $U$ being the velocity scale.

Now we proceed to keep the ${\cal O} (\tau_f,\tau_g)$ terms. According to Eq.~\eqref{CE1}, the momentum equation becomes
 \begin{equation}
\begin{aligned}
&\frac{\partial}{\partial t}\left(\rho u_{j}\right)+  \frac{\partial (\rho u_j u_m ) }{\partial x_m} 
\\=& - {\partial p \over \partial x_j } +  F_j  + {1\over RT} \frac{\partial}{\partial x_m} \tau_f  \int \left(  {\partial f^{eq} \over \partial t} + \xi_n {\partial f^{eq} \over \partial x_n} - S^{f}     \right)  \xi_{m}\xi_{j} d \boldsymbol{\xi} + {\cal O}(\tau_f^2) ,
\end{aligned}
\end{equation}
where the $O(\tau_{f})$ term should approximate the viscous term, {\it i.e.},
\begin{equation}\label{fxixiviscous}
\begin{aligned}
\tau_f   \int \left(  {\partial f^{eq} \over \partial t} + \xi_n {\partial f^{eq} \over \partial x_n} - S^{f}     \right)  \xi_{m}\xi_{j} d \boldsymbol{\xi} =
  RT\mu\left(\frac{\partial u_j}{\partial x_{m}}+\frac{\partial u_m}{\partial x_{j}}\right) +  {\cal O}(\tau_f^2) ,
\end{aligned}
\end{equation}
where $\mu = \rho RT \tau_f $.
 
Using Eqs.~(\ref{ddt}b-d) and Eq.~\eqref{fxixiviscous}, we can obtain
\begin{equation}\label{fxixiCE}
\begin{aligned}
& \tau_f   \int   S^{f}      \xi_{m}\xi_{j} d \boldsymbol{\xi}  
\\ = & \tau_{f}\left[\frac{\partial}{\partial t} \int f^{eq} \xi_{j}\xi_{m} d \boldsymbol{\xi}+\frac{\partial}{\partial x_{n}} \int f^{e q} \xi_{j} \xi_{m}\xi_{n} d \boldsymbol{\xi} \right] -  RT\mu\left(\frac{\partial u_j}{\partial x_{m}}+\frac{\partial u_m}{\partial x_{j}}\right)
\\  =& \tau_{f} RT
\left[ 
\frac{\partial}{\partial t}\left(\rho u_{m} u_{j}+p \delta_{m j}\right)
+ RT \frac{\partial}{\partial x_{n}}   \left (   \rho u_j \delta_{mn} +    \rho u_m \delta_{jn}  +  \rho u_n \delta_{mj}  \right)    \right] 
\\ = &  \tau_f RT  \left\{  u_{j} F_m +u_{m} F_j
+ R T \left( u_n
\frac{\partial  \rho }{\partial x_n}   \delta_{mj} + u_j {\partial \rho \over \partial x_m } +  u_m {\partial \rho \over  \partial x_j } \right) - R T \rho    \gamma \frac{\partial }{\partial x_n} \left( M_{CH} \frac{\partial \mu_{\phi}}{\partial x_n} \right)
   \delta_{mj}
\right\} 
\\   & \hspace{1cm}   + {\cal O} \left(\tau_{f}^{2}, \tau_f\tau_g Ma^2, \tau_f Ma^3 \right).
\end{aligned}
\end{equation}   
Namely, the requirement for the second-order moment of $S^f$ is
\begin{equation}\label{intSf2}
\begin{aligned}
  \int   S^{f}      \xi_{m}\xi_{j} d \boldsymbol{\xi}   = & RT  \left[ u_{j} F_m +u_{m} F_j \right]
\\  & +   (RT)^2  \left[   
 u_n
\frac{\partial  \rho }{\partial x_n}   \delta_{mj} + u_j {\partial \rho \over \partial x_m } +  u_m {\partial \rho \over  \partial x_j } -  \rho    \gamma \frac{\partial }{\partial x_n} \left( M_{CH} \frac{\partial \mu_{\phi}}{\partial x_n} \right)
   \delta_{mj}
\right]. 
\end{aligned}
\end{equation}  

 
Similarly, by retaining the ${\cal O} (\tau_g)$ terms in  the zeroth-order moment of Eq.~(\ref{Boltzmanng}), 
we have 
\begin{equation}
\frac{\partial \phi}{\partial t}+\frac{\partial ( \phi u_m ) }{\partial x_{m}} = 
 \frac{\partial}{\partial x_{m}} \left[  \tau_g  \int \left(   {\partial g^{eq} \over \partial t} + \xi_j {\partial g^{eq} \over \partial x_j}  - S^{g}   \right) \xi_{m} d \boldsymbol{\xi} \right] + {\cal O} (\tau_g^2 ).
\end{equation}
By matching the first-order term with the diffusion term in the CH equation, we have
\begin{equation}\label{intgxi}
\begin{aligned}
 \tau_{g} \int S^{g} \xi_{j} d \boldsymbol{\xi} =& \tau_{g}\left[\frac{\partial}{\partial t} \int g^{eq} \xi_{j} d \boldsymbol{\xi}+\frac{\partial}{\partial x_{m}} \int g^{e q} \xi_{m} \xi_{j} d \boldsymbol{\xi} \right] -    M_{CH} \frac{\partial\mu_{\phi}}{\partial x_{j}}     +{\cal O}\left(\tau_{g}^{2}\right)
\\=& \tau_{g}\left[ 
\frac{\partial}{\partial t}\left(\phi u_{j}\right)+\frac{\partial}{\partial x_{m}} \int g^{e q} \xi_{m} \xi_{j} d \boldsymbol{\xi} 
\right]  -    M_{CH} \frac{\partial\mu_{\phi}}{\partial x_{j}}  +{\cal O}\left(\tau_{g}^{2}\right),
\end{aligned}
\end{equation}
where $M_{CH} = \tau_g \eta RT$.
From Eqs.~(\ref{dphidt}) and (\ref{dudt}), the time derivative term can be expressed as 
\begin{equation}\label{dphiudt}
\frac{\partial}{\partial t}\left(\phi u_{j}\right)
=-\frac{\partial}{\partial x_{m}}\left(\phi u_{m}u_{j}\right)
+\frac{\phi}{\rho}\left( F_j-\frac{\partial p}{\partial x_j}\right)
+{\cal O} \left(\tau_g, \tau_f, Ma^3\right).
\end{equation}
Then 
\begin{equation}
\begin{aligned} \label{Sggeq}
 \tau_{g} \int S^{g} \xi_{j} d \boldsymbol{\xi} &= \tau_{g}
\frac{\partial}{\partial x_{m}}  \left( \int g^{e q} \xi_{m} \xi_{j} d \boldsymbol{\xi}  - \phi u_m u_j - \eta RT \mu_{\phi} \delta_{mj}      \right) 
+ \tau_g  \frac{\phi}{\rho}\left( F_j-\frac{\partial p}{\partial x_j}\right)    \\
& \hspace{1cm}  +{\cal O}\left(\tau_{g}^{2}, \tau_f \tau_g, \tau_g Ma^3\right).
\end{aligned}
\end{equation}
Since two moments can be assigned, the most convenient choice would be 
\begin{subequations}
		\begin{equation}\label{intgeq2}
		 \int g^{e q} \xi_{m} \xi_{j} d \boldsymbol{\xi}  =  \phi u_m u_j + \eta RT \mu_{\phi} \delta_{mj}, 
		\end{equation}
		\begin{equation}\label{intSg1}
		  \int S^{g} \xi_{j} d \boldsymbol{\xi} =  \frac{\phi}{\rho}\left( F_j-\frac{\partial p}{\partial x_j}\right).
		\end{equation}
\end{subequations}
 
To summarize, the integral constraints for the model Boltzmann equations, Eqs.~(\ref{Boltzmannf}) and~(\ref{Boltzmanng}), include:
\begin{enumerate}
\item The continuity equation sets the zeroth-order moment for $S^f$, Eq.~(\ref{intSf});

\item The momentum equation provides the first-order and second-order moments for $S^f$,
  Eq.~(\ref{intSf1}), and Eq.~(\ref{intSf2});

\item The CH equation leads to the first- and second-order moments of $g^{eq}$, Eqs.~\eqref{intgeq1} and~\eqref{intgeq2}, and the zeroth- and first-order moments for $S^g$, 
Eqs.~\eqref{intSg} and~\eqref{intSg1}. 

\end{enumerate}

The above analyses have outlined that the moments up to the third order for $f^{eq}$,  up to the second order for $g^{eq}$ and $S^f$,
and up to the first order for $S^g$ are involved in the inverse design.  Rigorously speaking, a Gauss-Hermite 
quadrature of a sixth order is needed for $f$, and a Gauss-Hermite 
quadrature of a fourth order is needed for $g$. The D3Q19 lattice velocity model provides only a fifth-order Gauss-Hermite 
quadrature, which is the reason that we require the ${\cal O}(Ma^3)$ terms must be negligible.

We emphasize that the above integral conditions are designed as a convenient choice, but not the only choice. 
In other words, they represent a set of sufficient conditions, not the necessary conditions. For example, one equation, Eq.~\eqref{Sggeq},
is used to assign two conditions, Eqs.~\eqref{intgeq2} and~\eqref{intSg1}, there are many possibilities.

Furthermore, even with the convenient design stated by the above integral constraints, there are many ways to specify the
precise forms for $g^{eq}$, $S^f$, and $S^g$.  We can confirm that the specific forms given in Eqs.~(\ref{feqgeq}) and~(\ref{SfSg}) 
do meet all the requirements stated above. We can also introduce other specific forms, for example, 
utilizing the Hermite expansion formulae, as done in~\cite{XiaowenShan2006,2021Inverse}.
Since the stability and accuracy of the forms in Eqs.~(\ref{feqgeq}) and~(\ref{SfSg}) have been verified by the previous studies,~\cite{zhang2018discrete,chen2019simulation} we apply them for the droplet breakup simulation here.

\section{The algorithm for computing the volume of individual droplets} \label{sec: Droplets}

During the breakup process, we wish to identify individual droplets and compute their volumes. We developed an algorithm
for this purpose. 
In this appendix, we briefly describe the algorithm to isolate the region for an individual droplet.
Here an individual droplet is defined as a set of connected nodes with $\phi \ge 0.5$.
Two nodes with $\phi \ge 0.5$ are viewed to be associated with a same droplet if they are connected by a set of nodes with $\phi \ge 0.5$. 

The procedure is as follows:
\begin{enumerate}
\item Simplify the field data. All nodes with $\phi\ge 0.5$ are reassigned a value of $\phi = 1$, and all with $\phi< 0.5$ are
set to $\phi = 0$. After this procedure, $\phi= 1$ represents the droplet phase (the dispersed phase), while $\phi= 0$ represents the background phase (the continuous phase).

\item  Find and mark the seed position for one droplet. 
We first find a node location $(i_0,j_0,k_0)$ with $\phi= 1$, which means that there is a droplet around $(i_0,j_0,k_0)$. 
Then we change the $\phi$ value from 1 to 50 for this specific position, {\it i.e.}, $\phi(i_0,j_0,k_0)=50$. 
The reason why we set $\phi(i_0,j_0,k_0)=50$ is to isolate it and other connected droplet nodes 
from the un-connected droplet nodes with $\phi= 1$. 
In fact, we only need a value larger than 27 ($=3^3$), and 50 is a convenient choice.
We refer to this location as the seed position for the current droplet.

\item Find all the droplet node points connected to $(i_0,j_0,k_0)$. 
We then check the immediate neighborhood, {\it i.e.},  the $\phi(i,j,k)$ values for nodes $i_0-1\le i\le i_0+1, j_0-1\le j\le j_0+1, k_0-1\le k\le k_0+1$.
If any of these points is outside the computational domain, then periodic boundary condition is employed to provide the $\phi$ value.
If $\phi(i,j,k)=1$ is found in this region, {\it i.e.}, $\phi(i_1,j_1,k_1)=1$, then compute the sum
 $\sum\phi(i,j,k)$ with $i_1-1\le i\le i_1+1, j_1-1\le j\le j_1+1, k_1-1\le k\le k_1+1$. 
 If this sum is larger than 50, which implies that  the two nodes $(i_1,j_1,k_1)$ and $(i_0,j_0,k_0)$ 
are connected nodes; then $\phi(i_1,j_1,k_1)$ is set to 50. 
If at least one new connected node is found in this previous search, the process continues to the region
$\phi(i,j,k)$ with $i_0-2\le i\le i_0+2, j_0-2\le j\le j_0+2, k_0-2\le k\le k_0+2$, to see if any new connected nodes can
be identified. Otherwise, all droplet nodes connected to the seed node $(i_0,j_0,k_0)$  have been found.

\item  Calculate the volume of this current droplet. Simply count the number of grid points with $\phi= 50$ encountered in Step 2 and Step 3.

\item Repeat the process to find other droplets.
We remove all the node points with $\phi= 50$ by changing the $\phi$ value to 0 for these nodes.
Then go back to Step 2,  repeat Steps 2 to 4.
\end{enumerate}

\section{Kinetic energy in the spectral space} \label{sec: Energy} 

We focus on the isothermal two-phase flow with same density and same viscosity for the two phases. The density is $\rho_0$, the dynamic viscosity is $\mu_0$, the kinematic viscosity is $\nu_0$. It is also assumed that no other forcing term exists.

\subsection{Kinetic energy evolution in the Fourier space}\label{subsec: EnergyFourier}

Converting the energy balance evolution into the Fourier space allows us to study 
the energy transfer across different scales for the multiphase flow system.

Eq.~(\ref{EqMo}) 
can be written equivalently as
\begin{equation}
\frac{\partial(\rho \boldsymbol{u})}{\partial t}+\nabla \cdot(\rho \boldsymbol{u} \boldsymbol{u})=-\nabla P+\nabla \cdot\left[\mu\left(\nabla \boldsymbol{u}+ \boldsymbol{u}\nabla\right)\right]-\kappa \nabla\cdot\left(  \nabla\phi\nabla \phi\right) ,\label{EqMo2}
\end{equation}
where $P=p+\phi\mu_{\phi}-\left( \psi+\kappa|\nabla \phi|^{2} / 2\right) $.

Taking the Fourier transform of Eq.~(\ref{EqMo2}) yields the momentum equation in the wavenumber space
\begin{equation}
\begin{aligned}
&\left( \frac{\partial }{\partial t}+\nu_0 k^{2}\right)  \hat{\boldsymbol{u}}\left(\boldsymbol{k}, t\right)
\\=&\left(\boldsymbol{I}-\frac{\boldsymbol{k} \boldsymbol{k}}{k^{2}}\right)\cdot \\&
\left\{\begin{array}{l}
-\sqrt{-1}  \sum_{\boldsymbol{k}^{\prime}+\boldsymbol{k}^{\prime \prime}=\boldsymbol{k}}\left[\boldsymbol{k}\cdot\hat{\boldsymbol{u}}\left(\boldsymbol{k}^{\prime}, t\right) \hat{\boldsymbol{u}}\left(\boldsymbol{k}^{\prime \prime}, t\right)\right]\\
+{\kappa \over \rho_0} \sqrt{-1}  \sum_{\boldsymbol{k}^{\prime}+\boldsymbol{k}^{\prime \prime}=\boldsymbol{k}}\left[\boldsymbol{k}\cdot\boldsymbol{k}^{\prime}  \boldsymbol{k}^{\prime \prime} \hat{\phi}(\boldsymbol{k}^{\prime}, t) \hat{\phi}\left(\boldsymbol{k}^{\prime \prime}, t\right)\right] 
\end{array}\right\},
\end{aligned}
\end{equation}
where $\boldsymbol{k}$ is the wavenumber, $k=\left| \boldsymbol{k}\right| $ is the magnitude of $\boldsymbol{k}$, $\boldsymbol{I}$ is the second-order unit tensor. $\left(\boldsymbol{I}-{\boldsymbol{k} \boldsymbol{k}}/{k^{2}}\right)$ is the projection tensor.

Then the kinetic energy of Fourier mode, $\hat{E}\left(\boldsymbol{k}, t\right)= \hat{\boldsymbol{u}}_*\left(\boldsymbol{k}, t\right)\cdot\hat{\boldsymbol{u}}\left(\boldsymbol{k}, t\right)/2 $, is
\begin{equation}
\frac{\partial }{\partial t}\hat{E}\left(\boldsymbol{k}, t\right)=\hat{T}\left(\boldsymbol{k}, t\right)+\hat{S}\left(\boldsymbol{k}, t\right)-2\nu_0 k^{2}\hat{E}\left(\boldsymbol{k}, t\right),
\end{equation}
where
\begin{subequations}
	\begin{equation}
	\begin{aligned}
	&\hat{T}\left(\boldsymbol{k}, t\right)
	=\left(\boldsymbol{I}-\frac{\boldsymbol{k} \boldsymbol{k}}{k^{2}}\right): \\&
	\mathcal{R}
	\left\lbrace 
	-\sqrt{-1} 
	\sum_{\boldsymbol{k}^{\prime}+\boldsymbol{k}^{\prime \prime}=\boldsymbol{k}}\left[ 
	\boldsymbol{k}\cdot \hat{\boldsymbol{u}}\left(\boldsymbol{k}^{\prime}, t\right) \hat{\boldsymbol{u}}\left(\boldsymbol{k}^{\prime \prime}, t\right)
	\hat{\boldsymbol{u}}_*\left(\boldsymbol{k}, t\right)
	\right]  
	\right\rbrace 
	\end{aligned}
	\end{equation}
	and
	\begin{equation}
	\begin{aligned}
	&\hat{S}\left(\boldsymbol{k}, t\right)
	=\left(\boldsymbol{I}-\frac{\boldsymbol{k} \boldsymbol{k}}{k^{2}}\right): \\&
	\mathcal{R}
	\left\{
	\frac{\kappa}{\rho_0} \sqrt{-1} 
	\sum_{\boldsymbol{k}^{\prime}+\boldsymbol{k}^{\prime \prime}=\boldsymbol{k}}\left[ 
	\boldsymbol{k}\cdot \boldsymbol{k}^{\prime}  \boldsymbol{k}^{\prime \prime}\hat{\boldsymbol{u}}_*\left(\boldsymbol{k}, t\right) \hat{\phi}(\boldsymbol{k}^{\prime}, t) \hat{\phi}\left(\boldsymbol{k}^{\prime \prime}, t\right)\right]    
	\right\}
	\end{aligned}
	\end{equation}
\end{subequations}
are the transfer rate due to nonlinear triadic interactions and the surface tension effect, respectively. Here a hat $\hat{\cdot} $ represents a Fourier transform (the hat in the main text  is dropped for simplicity),  
the subscript asterisk $\left( \cdot\right)_* $ represents the complex conjugate, 
$\mathcal{R}\left\{ \cdot\right\}$ is the real part.
It is well known that $\sum_{\boldsymbol{k}} \hat{T}\left(\boldsymbol{k}, t\right)=0$.~\cite{pope2001turbulent} However, the sum for $\hat{S}\left(\boldsymbol{k}, t\right)$ may not
be zero, namely, $\sum_{\boldsymbol{k}}\hat{S}\left(\boldsymbol{k}, t\right)\neq 0$, because the kinetic energy could
be transferred to the free energy if the interface area is increased.

\subsection{Kinetic energy expansion in the spherical harmonics space}\label{subsec: Energyspherical}
The standard Fourier spectrum is more suitable for analyzing HIT, in particular, 
the full-developed forced homogeneous isotropic two-phase flow containing many small droplets.
For the problem of evolution of a spherical droplet located in decaying turbulent flow, 
it is a spherical symmetric problem if we ignore the influence of periodic boundary conditions. Therefore, we may
introduce a generalized Fourier analysis, namely, the spectral method on a spherical surface,~\cite{2014Spectral,2010Representation,courty2006oscillating,groemer1996geometric} 
to analyze the kinetic energy evolution.

At a given time $t$, the expansion of velocity on a spherical surface of radius $r_0$ under spherical harmonics can be written as
\begin{subequations}
	\begin{equation}
	\boldsymbol{u}(r_0, \Omega)=\sum_{l=0}^{\infty} \sum_{m=-l}^{l} \tilde{\boldsymbol{u}}(l, m ; r_0) Y_{l}^{m}(\Omega) 
	\end{equation}
	with the coefficient
	\begin{equation}
	\tilde{\boldsymbol{u}}(l, m ; r_0)=\int \boldsymbol{u}(r_0, \Omega) Y_{l*}^{m}(\Omega) d \Omega ,
	\end{equation}		
\end{subequations}
where $\Omega$ is the solid angle, $Y_{l}^{m}(\Omega)$ ($l$ = 0, 1, 2, ...; $m$ = 0, $\pm$ 1, $\pm$ 2, ..., $\pm l$) 
denotes the 
well-known 
spherical harmonics function of degree $l$ and order $m$.~\cite{2014Spectral,2010Representation,courty2006oscillating,groemer1996geometric} 
The average kinetic energy on the spherical surface of radius $r_0$ is
\begin{equation}
\begin{aligned}
K\left( r_0\right) =&\frac{1}{2}\frac{\int|\boldsymbol{u}(r_0, \Omega)|^{2} d \Omega}{\int d \Omega}
\\=&\frac{1}{8\pi}\sum_{l=0}^{\infty} \sum_{m=-l}^{l}  \tilde{\boldsymbol{u}}_{*}(l, m ; r_0)\cdot \tilde{\boldsymbol{u}}(l, m ; r_0) 
\\=&\sum_{l=0}^{\infty}\hat{E}(l; r_0),
\end{aligned}
\end{equation}
where $\hat{E}(l; r_0)=\sum_{m=-l}^{l}  \tilde{\boldsymbol{u}}_{*}(l, m ; r_0)\cdot \tilde{\boldsymbol{u}}(l, m ; r_0)/\left( 8\pi\right) $ 
is the average kinetic energy of degree $l$ on the spherical surface.

\bibliography{refs}
\newpage

\end{document}